\documentclass[usenatbib]{mn2e}

\usepackage{amsmath, mathrsfs, amssymb}
\usepackage{graphicx}
\usepackage{rotating}
\usepackage{longtable,lscape}
\usepackage{morefloats}
\usepackage{color}
\usepackage[usenames,dvipsnames,svgnames,table]{xcolor}
\usepackage{enumitem}
\usepackage{caption}

\title[Spectral hardening in state transitions]{Spectral hardening as a viable alternative to disc truncation in black hole state transitions}

\author[Salvesen et al.]{Greg~Salvesen$^{1,2}\thanks{E-mail: salvesen@colorado.edu}\thanks{National Science Foundation Graduate Fellow.}$, Jon~M.~Miller$^{3}$, Rubens~C.~Reis$^{3}$\thanks{Einstein Postdoctoral Fellow.}\thanks{Michigan Society Fellow.}, \& Mitchell~C.~Begelman$^{1,2}$ \\
$^{1}${JILA, University of Colorado and National Institute of Standards and Technology, 440 UCB, Boulder, CO 80309-0440, USA.} \\
$^{2}${Department of Astrophysical and Planetary Sciences, University of Colorado, 391 UCB, Boulder, CO 80309-0391, USA.} \\
$^{3}${Department of Astronomy, University of Michigan, 500 Church Street, Ann Arbor, MI 48109, USA.}}

\begin{document}

\label{firstpage}

\maketitle

% ABSTRACT
\begin{abstract}
Constraining the accretion flow geometry of black hole binaries in outburst is complicated by the inability of simplified multi-colour disc models to distinguish between changes in the inner disc radius and alterations to the emergent spectrum, parameterised by the phenomenological colour correction factor, $f_{\rm col}$.  We analyse {\it Rossi X-ray Timing Explorer} observations of the low mass Galactic black hole X-ray binary, GX 339--4, taken over seven epochs when the source was experiencing a state transition.  The accretion disc component is isolated using a pipeline resulting in robust detections for disc luminosities, $10^{-3} \lesssim L_{\rm disc} / L_{\rm Edd} \lesssim 0.5$.  Assuming that the inner disc remains situated at the innermost stable circular orbit over the course of a state transition, we measure the {\it relative} degree of change in $f_{\rm col}$ required to explain the spectral evolution of the disc component.  A variable $f_{\rm col}$ that increases by a factor of $\sim 2.0-3.5$ as the source transitions from the high/soft state to the low/hard state can adequately explain the observed disc spectral evolution.  For the observations dominated by a disc component, the familiar scaling between the disc luminosity and effective temperature, $L_{\rm disc} \propto T_{\rm eff}^{4}$, is observed; however, significant deviations from this relation appear when GX 339--4 is in the hard intermediate and low/hard states.  Allowing for an evolving $f_{\rm col}$ between spectral states, the $L_{\rm disc}-T_{\rm eff}^{4}$ law is recovered over the full range of disc luminosities, although this depends heavily on the physically conceivable range of $f_{\rm col}$.  We demonstrate that physically reasonable changes in $f_{\rm col}$ provide a viable description for multiple state transitions of a black hole binary without invoking radial motion of the inner accretion disc.
\end{abstract}

% KEYWORDS
\begin{keywords}
accretion, accretion discs - black hole physics - X-rays: binaries - X-rays: individual (GX 339--4)
\end{keywords}

%===========================================================================
% INTRODUCTION
\section{Introduction}
\label{section:intro}
The nature of the accretion flow for black hole binaries (BHBs) undergoing state transitions remains unclear and is an active topic of observational and theoretical pursuit.  The phenomenology describing the predominant spectral states experienced by BHBs include the high/soft state (HSS), characterised by high luminosity and a standard thin accretion disc as the dominant spectral component; the low/hard state (LHS), characterised by low luminosity with a dominant hard power-law spectral component that is likely produced by inverse Compton scattering of soft photons by a relativistic electron corona \citep{Shapiro1976}; and the intermediate state (IS), which is itself further divided into the hard intermediate state (HIS) and soft intermediate state (SIS) (see, e.g., \citeauthor{Belloni2010} \citeyear{Belloni2010} for an extensive description of these phenomenological states).  

Currently, the picture of a black hole accreting at a significant fraction of the Eddington mass accretion rate, $\dot{M}_{\rm Edd}$, from a geometrically thin, optically thick disc extending to the innermost stable circular orbit (ISCO) is the accepted description of the HSS.  However, there is a lack of consensus regarding a universal geometry of the disc and innermost accretion flow in the LHS when the black hole feeds at a low fraction of $\dot{M}_{\rm Edd}$ and/or the flow has a low radiative efficiency.  According to the paradigm put forth by \citet{Esin1997}, the accretion disc in the LHS is radially recessed (i.e., truncated) with the interior region being composed of a radiatively inefficient accretion flow (RIAF) \citep{NarayanYi1994}.  Alternatively, models where the inner disc remains at the ISCO in the LHS have been proposed by invoking magnetically reconnecting flares driving a coronal outflow \citep[e.g.,][]{Beloborodov1999}.  Another model advocating against disc truncation is a standard thin disc that generates a magnetically dominated corona where the amount of accretion energy liberated in the corona increases with decreasing mass accretion rate \citep{MerloniFabian2002}.  Recently, \citet{Reis2012} proposed that the intermediate states may differ from one another only in the geometrical extent of the X-ray corona, with the SIS being the result of a collapse of the previously extended corona as the system moves into the disc-dominated HSS.  The notion of an accretion disc, where the magnetorotational instability governs dissipation and angular momentum transport \citep{BalbusHawley1991}, producing a magnetically dominated corona is supported by sophisticated numerical simulations of local shearing boxes \citep[e.g.,][]{Hirose2006, Hirose2009} and global discs \citep[e.g.,][]{Beckwith2009}.  By varying only the mass accretion rate, the spectral signatures of the HSS and LHS were reproduced in the global disc simulations of \citet{Schnittman2012}, without the need for radial variation of the inner accretion disc.  The physical mechanism responsible for triggering a state transition is poorly understood, making it difficult to conceive what would drive a sudden disc recession or a drastic change in coronal activity required to explain the observed spectral evolution; thus, lending credence to the need for a focused study of disc evolution during state transitions.

Observational studies of BHB spectra have successfully applied a multi-colour disc (MCD) blackbody model \citep{Mitsuda1984, Makishima1986} to the observed quasi-thermal soft X-ray spectral component in the HSS and even in the LHS in some cases \citep[e.g.,][]{Miller2006, Reis2010}.  Provided the observational parameters are well-constrained, the radial location of the inner disc boundary can be measured from the best-fit MCD model parameters.  However, this technique makes the critical assumption that the deviations to the emergent disc spectrum caused by the disc vertical structure can be parametrized by a frequency-independent constant, known as the colour-correction, or spectral hardening, factor, $f_{\rm col}$ \citep{ShimuraTakahara1995b}.  

Recently, investigations into the possibility of a variable $f_{\rm col}$ experienced a strong revival in the observational community.  Studies of black hole X-ray binary accretion discs across various luminosity states using data from the {\it Rossi X-ray Timing Explorer} ({\it RXTE}) \citep[e.g.,][]{Nowak2002, Ibragimov2005, Wilms2006, Dunn2011} and the {\it Swift} observatory \citep{ReynoldsMiller2011} provide extensive systematic analyses of the properties of disc evolution in BHBs.  The global studies of \citet{Dunn2011} and \citet{ReynoldsMiller2011} examined all of the available archived data for all of the well-known, confirmed BHBs.  For nearly all BHBs where an outburst was observed, \citet{Dunn2011} found that $f_{\rm col}$ remained relatively constant in the most disc-dominated states (i.e., the HSS), closely following the theoretical $L_{\rm disc} \propto T_{\rm eff}^{4}$ relation between the disc luminosity and disc effective temperature.  For observations with a relatively weaker disc component, characteristic of the IS, deviations from the luminosity-temperature relationship arose.  Supposing that the inner disc radius does not change and allowing for a variable colour correction factor, $1.6 \lesssim f_{\rm col} \lesssim 2.6$, where $f_{\rm col}$ increased as the disc fraction decreased, the $L_{\rm disc} \propto T_{\rm eff}^{4}$ scaling was recovered.  This finding sets the stage for a thorough consideration of an increasing $f_{\rm col}$ when a BHB transitions from the HSS to the LHS as a possible alternative to a truncated disc geometry in the LHS.  Performing a similar analysis, \cite{ReynoldsMiller2011} speculate that $f_{\rm col}$ may attain values as high as 5 as the BHB spectrum becomes less disc dominated.  The main differences between our work and these recent observational studies are: (1) We focus exclusively on a single source, the well-known BHB transient, GX 339--4, over multiple state transitions.  (2) We systematically explore different Comptonisation models.\footnote{\citet{ReynoldsMiller2011} also consider both phenomenological and physical Comptonisation models.}  (3) We introduce a novel approach for making relative measurements in order to reduce uncertainties.

The methodology of this paper is to systematically investigate to what extent variations in the disc structure and/or disc truncation can explain the observed spectral evolution for multiple state transitions of a BHB.  Most importantly, we will demonstrate that a colour correction factor that dynamically evolves within a physically reasonable range provides an adequate alternative to a truncated disc for explaining the changing disc spectrum.

%===========================================================================
% Observations and Data Reduction
\section{Observations and Data Reduction}
\label{section:obsdata}
Motivated by the lack of consensus on accretion disc evolution over the course of a BHB state transition, we aim to address the question: Can physically reasonable changes in the vertical disc structure alone explain the observed evolution of disc properties during a state transition without invoking a truncated disc geometry?  In light of this question, we sought a source that met the following criteria: (1) The source was observed over multiple state transitions, permitting comparison of disc variations between different transitions.  (2) Given that our methodology hinges on the presence of a disc, a disc component must have been robustly detected in the LHS.  (3) Conflicting claims for both a severely recessed disc and a disc extending to the ISCO in the LHS are published for the source.  (4) Variable low-frequency quasi-periodic oscillations (LFQPOs), which may be linked to the disc evolution, have been observed coincident with state transitions.  (5) Reasonable constraints are known on mass, distance, and inclination; however, this is of ancillary importance.  Based on these criteria, we selected GX 339--4 as the focus of our analysis.

% TABLE 1
\begin{table*}
\addtolength{\tabcolsep}{-2pt}
\centering
\begin{tabular}{c c c c c c c c c c c}
\hline
\hline
Trans. ID & Start & End & Type & $N_{\rm obs}$ & $N_{\rm disc}^{\texttt{pow/bknpow}}$ & $N_{\rm disc}^{\texttt{simpl}}$ & $N_{\rm disc}^{\texttt{comptt}}$ & $\nu_{\rm QPO}^{\rm min}$ & $\nu_{\rm QPO}^{\rm max}$ & $R_{\rm Kep}^{\rm max} / R_{\rm Kep}^{\rm min}$ \\
 & dd/mm/yy (MJD) & dd/mm/yy (MJD) & & & & & & (Hz) & (Hz) & \\
\hline
R02 & 18/04/02 (52382) & 23/05/02 (52417) & Rise & 33 & 32 & 22 & 31 & 0.159(1)$^{a}$ & 7.9(1)$^{a}$ & 14 \\
D03 & 15/02/03 (52685) & 11/05/03 (52770) & Decay & 53  & 33 & 30 & 23 & 0.29(1)$^{a}$ & 8.8(3)$^{a}$ & 9.7 \\
R04 & 05/07/04 (53191) & 25/08/04 (53242) & Rise & 30 & 21 & 21 & 21 & 0.306(7)$^{a}$ &  8.0(2)$^{a}$ & 8.8 \\
R07 & 17/01/07 (54117) & 23/02/07 (54154) & Rise & 34 & 28 & 17 & 29 & 0.142(3)$^{a}$ & 6.67(2)$^{a}$ & 13 \\
D07 & 02/05/07 (54222) & 06/06/07 (54257) & Decay & 46 & 40 & 39 & 40 & 0.070(3)$^{a}$ & 4.05(9)$^{a}$ & 15 \\
R10 & 26/03/10 (55281) & 20/05/10 (55336) & Rise & 59 & 11 & 9 & 10 & 0.22(1)$^{b}$ & 6.7(2)$^{b}$ & 9.8 \\
D11 & 07/01/11 (55568) & 15/02/11 (55607) & Decay & 28 & 12 & 12 & 12 & 0.9(1)$^{c}$ & 4.5(3)$^{b}$ & 2.9 \\
\hline
\end{tabular}
\caption{Table summarising the key features of each GX 339--4 state transition.  From {\it left} to {\it right}, the columns are the state transition ID, start time of the transition, end time of the transition, type of transition, number of archived {\it RXTE} observations, number of observations requiring a disc component for Comptonisation models \texttt{pow/bknpow}, \texttt{simpl}, and \texttt{comptt}, minimum/maximum LFQPO frequency observed, and fractional change in inner disc radius implied if $\nu_{\rm QPO}^{\rm min}$ and $\nu_{\rm QPO}^{\rm max}$ are associated with the Keplerian frequency and assumed to trace the inner disc location.  $^a$\citet{ShaposhnikovTitarchuk2009}; $^b$\citet{Motta2011}; $^c$\citet{Stiele2011}}
\label{tab:transitions}
\end{table*}

\subsection{GX 339--4}
\label{sec:gx339}
GX 339--4, discovered by \citet{Markert1973},  is a low-mass X-ray binary and recurring transient hosting a dynamically confirmed black hole with a mass $M \ge 5.8 M_{\odot}$ (\citeauthor{Hynes2003} \citeyear{Hynes2003}; see also \citeauthor{MunozDarias2008} \citeyear{MunozDarias2008}) and distance $D = 8 \pm 2 {\rm~kpc}$ (\citeauthor{Zdziarski2004} \citeyear{Zdziarski2004}; see also \citeauthor{Hynes2004} \citeyear{Hynes2004}).  Supposing that the inner disc inclination, $i$ ($0^{\circ}$ for face on, $90^{\circ}$ for edge on), can be assumed to align with the binary inclination, then the lack of observed eclipses constrain $i \le 60^{\circ}$ \citep{Cowley2002}, while radio observations imply $i \le 30^{\circ}$ \citep{Gallo2004}.  Results from modelling the X-ray reflection features of GX 339--4 in the SIS also suggest a lower disc inclination of $i \sim 10^{\circ}-30^{\circ}$ \citep{Miller2004b, Miller2009} and in the LHS \citep{Reis2008}.  Owing to the uncertainties on the intrinsic parameters (namely, $M$, $D$, $i$) of GX 339--4 and the limitations of existing X-ray data, seeking absolute measurements that depend on these parameters presents a serious challenge.

GX 339--4 experienced multiple outbursts since its discovery, making it an appealing source for temporal evolution studies.  The seven state transitions we select to study are listed in Table \ref{tab:transitions}.  The notation convention we adopt for referring to a transition is the letter R or D, signifying a rise or decay type, respectively, followed by the year associated with the transition.  A rise (decay) type indicates a transition from the LHS (HSS) to the HSS (LHS).  The intervals of each transition were selected based on previously published dates and the emergence/disappearance of LFQPOs \citep{ShaposhnikovTitarchuk2009,  Motta2011, Stiele2011}.\footnote{The end date listed for transition ID GX339-D03 in Table 1 of \citet{ShaposhnikovTitarchuk2009} is mistyped and should read 06/05/03 (June 05 2003).}  We chose to extend the transition start/end times by $\sim$ 5 days beyond the intervals quoted in the literature to ensure that we capture the full transition behavior in our subsequent analysis.

Alarming discrepancies in the location of the inner accretion disc edge, particularly in the LHS, appear in the literature based on spectral fits to X-ray observations of GX 339--4.  The strongest case for disc truncation is presented in \citet{Tomsick2009}, which used fluorescent iron emission lines to find $R_{\rm in} > 175~R_{\rm g}$ (90\% confidence, assuming $i = 30^{\circ}$) when GX 339--4 was in the LHS with luminosity $L_{\rm X} \simeq 1.4 \times 10^{-3}~L_{\rm Edd}$ (assuming $D = 8 {\rm~kpc}$).  Claims of a truncated disc in the LHS also appear based on modelling of direct disc emission accounting for irradiation of the inner disc \citep{Cabanac2009}.  Even in the SIS, \citet{Yamada2009} suggest a recessed disc with $R_{\rm in} = 5-32~R_{\rm g}$ (68\% confidence, assuming $i = 30^{\circ}$).  These measurements suggest that the inner portion of the disc is absent in low luminosity states, perhaps replaced by a RIAF.  In stark contrast, there are numerous claims of the inner disc radius in GX 339--4 remaining consistent with the ISCO in the LHS.   Using physically motivated, sophisticated spectral fitting techniques, the disc inner radius was determined to lie within $R_{\rm in} = 2.0-3.0~R_{\rm g}$ in the SIS \citep{Miller2004b} and $R_{\rm in} = 3.0-5.0~R_{\rm g}$ in the LHS with luminosity $L_{\rm X} \simeq 0.05~L_{\rm Edd}$ (assuming $D = 8 {\rm~kpc}$) \citep{Miller2006}.  Both of these results were confirmed by a systematic reanalysis that employed detailed disc reflection modelling \citep{Reis2008}.  Using the same data set as in \citet{Miller2006}, \citet{WilkinsonUttley2009} also measured a disc extending down to $4~R_{\rm g}$ from their best-fit iron line model.  \citet{Tomsick2008} observed GX 339--4 in the LHS at luminosities $L_{\rm X} \simeq 0.023~L_{\rm Edd}$ and $L_{\rm X} \simeq 8 \times 10^{-3}~L_{\rm Edd}$, finding the disc to extend within $10~R_{\rm g}$ and being consistent with the disc remaining fixed at $\sim 4~R_{\rm g}$.  \citet{Reis2010} performed a systematic study of eight BHBs (including GX 339--4) in the LHS, robustly detecting a thermal component in all sources and placing a stringent upper limit of $10~R_{\rm g}$ on the truncation radius for six sources confirmed by both broad iron line and thermal disc modelling independently.

In order to measure black hole spin, one must associate $R_{\rm in}$ with the ISCO, located at $R_{\rm ISCO}$.  A measurement of the inner disc radius is the crucial diagnostic for measuring black hole spin, as $R_{\rm ISCO}$ is a monotonically decreasing function of the dimensionless spin parameter, $a_{\ast} \equiv J c / G M^{2}$, where $J$ is the black hole angular momentum \citep{Bardeen1972}.  Perhaps unsurprisingly, conflicting claims regarding the behaviour of $R_{\rm in}$ have led to discrepancies in spin measurements of GX 339--4.   Spin measurements employing the relativistically broadened iron line method \citep{Fabian1989, Laor1991} suggest that GX339--4 harbours a near-maximally spinning black hole with $a_{\ast} = 0.94 \pm 0.02$ \citep{Miller2009, Reis2008}, while the thermal disc continuum fitting technique \citep{Zhang1997} supports a low-to-moderately spinning black hole with $a_{\ast} < 0.4$ (\citeauthor{Yamada2009} \citeyear{Yamada2009}; see also \citeauthor{KolehmainenDone2010} \citeyear{KolehmainenDone2010}; \citeauthor{Kolehmainen2011} \citeyear{Kolehmainen2011}).

LFQPOs exhibit variability simultaneous with transitions between the LHS and HSS \citep{Rutledge1999b}.  Variable LFQPOs were observed in each GX 339--4 transition studied here, with the measured frequency ranges listed in Table \ref{tab:transitions} and typically spanning $0.2-8{\rm~Hz}$ \citep{Belloni2005, ShaposhnikovTitarchuk2009, Motta2011, Stiele2011,Nandi2012}.  The idea that LFQPOs may originate from oscillating modes of a quasi-spherical RIAF that is interior to, or partially overlapping with, a truncated thin disc was proposed by \citet{GianniosSpruit2004}.  Long time-scale variability of the disc blackbody component of GX 339--4 was associated with disc instabilities and interpreted as arising from intrinsic variability of the disc \citep{WilkinsonUttley2009}.  This suggests that the variable LFQPOs may be intricately linked to disc truncation; however, the truncation radius cannot exceed $\sim 20~R_{\rm g}$ if the observed variations are limited by the viscous timescale.

\subsection{{\it Rossi X-ray Timing Explorer}}
\label{sec:rxte}
The monitoring campaign of the {\it Rossi X-ray Timing Explorer} ({\it RXTE}) \citep{Bradt1993, Swank1999} observed GX 339--4 transitioning between states in 2002, 2003, 2004, 2007, 2010, and 2011, obtaining well-sampled (in time) observations spanning each transition before being decommissioned in early 2012.  We examined 283 {\it RXTE} observations of GX 339--4 covering seven time intervals, during which the source experienced state transitions (see Table \ref{tab:transitions}).  The {\it RXTE} standard products, comprised of the source and background spectra along with instrument response matrices, were obtained from the High Energy Astrophysics Science Archive Research Center data archive.\footnote{http://heasarc.gsfc.nasa.gov/cgi-bin/W3Browse/w3browse.pl}  The Proportional Counter Array (PCA) \citep{Jahoda1996} and High Energy X-ray Timing Experiment (HEXTE) \citep{Rothschild1998} standard source spectra were analysed jointly.  The PCA energy spectrum (Standard 2 mode) is built from summing all layers from various Proportional Counter Units (PCUs).  The PCU-1, PCU-2, and PCU-3 are used in all observations prior to 25 December 2006 (i.e., R02, D03, R04), while only the PCU-2 and PCU-3 are used in observations thereafter (i.e., R07, D07, R10, D11) due to a propane loss in the top layer of PCU-1.\footnote{http://heasarc.nasa.gov/docs/xte/recipes/stdprod\_guide.html}  For observations corresponding to transitions R02, D03, R04, R07, and D07, we used the HEXTE Cluster B spectrum, but could not use these data for R10 and D11 because the HEXTE Cluster B was permanently left in an off-source pointing position from 29 March 2010 onward.\footnote{http://heasarc.nasa.gov/docs/xte/xhp\_new.html}  The HEXTE Cluster A data for transitions R10 and D11 are affected by strong residuals, resulting from difficultly in determining the appropriate contribution of the background spectra.  Consequently, we excluded all HEXTE data while analysing transitions R10 and D11, using PCA data only for these observations.  This resulted in many observations with a strong soft component, which is indicative of a disc, being excised from our sample due to poor spectral modelling of the high-energy continuum (see \S \ref{sec:diskreq}).

\begin{figure}
\begin{center}
\includegraphics[width=84mm]{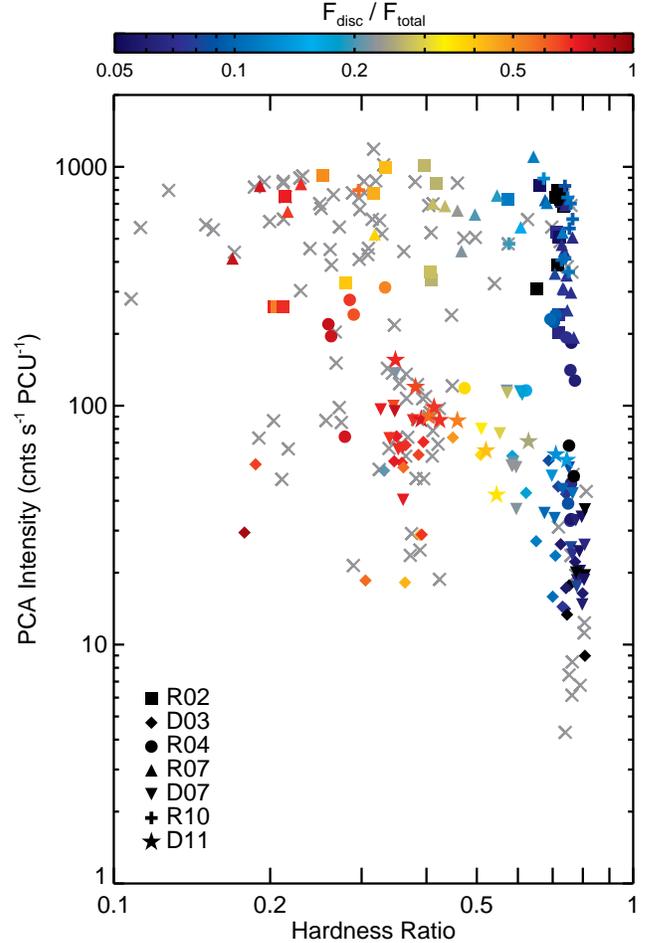}
\end{center}
\caption{Hardness-intensity diagram for all observations of GX 339--4 considered in this work generated from the results of the \texttt{pow/bkn} model (see \S \ref{sec:model}).  The disc fraction for each observation is given by the glyph colour and is an indicator of the spectral state.  Data points marked by a {\it grey cross} correspond to observations that were not deemed to statistically require a disc component.  The key in the {\it bottom left} identifies the glyph associated with each transition ID.  See \S \ref{sec:rxte} for a description of how the HID was generated.  A wide range of hardness is considered in this work.}
\label{fig:hid}
\end{figure}

While the time sampling capability of the {\it RXTE} satellite makes it well-suited for a study of BHB state transitions, which are a highly time-dependent phenomenon, {\it RXTE} has important limitations.  Determining the spectral properties of the disc component for a given observation is made difficult by the limited low-energy coverage of {\it RXTE}, which only extends down to $\sim$ 3 keV, while the disc spectrum peaks at $\sim$ 1 keV.  Especially in the HIS and LHS, the diminished disc component, if present, is difficult to characterise due to a combination of its weakness and the dominance of a Comptonised component.  Figure \ref{fig:hid} presents a hardness-intensity diagram (HID) of all 283 {\it RXTE} observations of GX 339--4 considered in this work.  Observations that do not statistically require a disc component (see \S \ref{sec:diskreq}) are indicated by {\it grey crosses}, while the various glyphs indicate the transition associated with the observation.  The plethora of IS and HSS observations (i.e., low-to-moderate hardnesses) deemed to not require a disc comes from the R10 and D11 data sets, where the absence of HEXTE data prevented acceptable fits to disc-dominated spectra.  Notably, Figure \ref{fig:hid} shows that the set of observations considered in the subsequent analysis span all spectral states, which indicates that we are studying the complete evolution of state transitions in GX 339--4.

% FIGURE 1

For a given observation, hardness and intensity are calculated from the PCA data as follows.  The distribution of counts across all channels is obtained from both the standard PCA source spectral file and background spectral file.  The background counts are then subtracted from the source counts channel-by-channel, yielding the counts in each channel associated with GX 339--4.  The hardness is then defined as the ratio of GX 339--4 counts in channels 15 - 24 (6.3 - 10.5 keV) to those in channels 8 - 14 (3.8 - 6.3 keV) \citep{Belloni2005}.  A count rate is obtained by summing the counts across all channels and dividing by the exposure time.  Good time intervals (GTIs) are identified from the GTI file included in the standard products.  The average number of PCUs that are turned on during the GTIs is determined from the filter file included in the standard products.  The intensity is then defined as the count rate divided by the average number of active PCUs during the observation GTIs.

To justify using the $\chi^{2}$ fitting statistic for deriving parameter constraints, both the PCA and HEXTE spectra were binned to require a minimum of 20 counts per energy bin using the \texttt{FTOOL} \texttt{grppha}.  A systematic error of 0.6\% was added to the PCA spectrum using \texttt{grppha}; however, the PCA spectra without systematic errors added are used when deriving relative measurements (see \S \ref{sec:relative}).  All spectral fits are performed with \texttt{XSPEC} version 12.7.1 \citep{Arnaud1996}.  The PCA and HEXTE spectra are fitted over the $2.8-25 {\rm~keV}$ and $20-100 {\rm~keV}$ energy ranges, respectively.   The response energy range over which spectral models are calculated is extended to 0.1 keV and 100 keV using the \texttt{energies extend} command in \texttt{XSPEC} with 50 logarithmic steps in each direction.   All observation times correspond to the midpoint of the PCA observation.  Quoted errors from spectral fits are $1\sigma$ uncertainties, unless specified otherwise.

%===========================================================================
% MULTICOLOUR DISC BLACKBODY MODELS
\section{Multi-colour Disc Blackbody Models}
The accretion disc in BHBs manifests itself as a soft, quasi-blackbody component with a typical peak inner disc temperature of $\sim 1~{\rm keV}$ \citep[e.g.,][]{FenderBelloni2004}.  While the accepted description for the origin of the soft thermal component in the HSS is a geometrically thin, optically thick accretion disc \citep{ShakuraSunyaev1973} extending down to the ISCO, no such consensus has been reached for the nature of the innermost flow in the HIS or LHS.  Appealing to spectral fits of the thermal continuum with multi-colour disc (MCD) blackbody models to isolate the disc component, one can measure the temporal evolution of the accretion disc over the course of a state transition.

\subsection{The Standard MCD Model}
The most commonly adopted disc model is the MCD blackbody model (\texttt{diskbb} in \texttt{XSPEC}) \citep{Mitsuda1984, Makishima1986}.  The attractiveness of the MCD model is owed to its legacy, applicability, and simplicity.  However, the familiarity and broad acceptance of \texttt{diskbb} often mask subtle features of the model, possibly leading to unintended misuse.  Here, we dissect the \texttt{diskbb} model in order to elucidate its underlying assumptions and expose its limitations.  Assuming a geometrically thin, optically thick accretion disc that locally emits a blackbody spectrum in concentric annuli, the colour temperature for a MCD model at a given radius is described as,
\begin{equation}
T_{\rm col}(r) = T_{\ast} \left( \frac{r}{R_{\rm in}} \right)^{-3/4} \left[1 - \beta \left( \frac{r}{R_{\rm in}}, \right)^{-1/2} \right]^{1/4}, \label{eq:Tcol}
\end{equation}
where $R_{\rm in}$ is the radius of the inner disc edge and $\beta$ describes the boundary condition at this location.  For discs in the HSS, $R_{\rm in}$ is commonly associated with $R_{\rm ISCO}$.  A zero-torque boundary condition at $R_{\rm in}$ corresponds to $\beta = 1$, while $\beta = 0$ allows for non-zero stresses at the inner disc and is the choice adopted in \texttt{diskbb} \citep{Zhang1997, Gierlinski1999, Zimmerman2005}.  The leading factor, which roughly describes the colour temperature at the inner disc, is given by,
\begin{equation}
T_{\ast} = f_{\rm col} \left( \frac{3 G M \dot{M}}{8 \pi \sigma R_{\rm in}^{3}} \right)^{1/4} \label{eq:Tstar},
\end{equation}
where $G$ is the gravitational constant, $\sigma$ is the Stefan-Boltzmann constant, $M$ is the compact object mass, $\dot{M}$ is the mass accretion rate, and $f_{\rm col}$ is the colour correction factor \citep{ShimuraTakahara1995b}, also referred to as the spectral hardening factor.  For a zero-torque condition at $R_{\rm in}$ (i.e., $\beta = 1$ in Equation \ref{eq:Tcol}), $T_{\ast}$ is related to the maximum temperature attained in the disk by $T_{\rm col, max} = 0.488 T_{\ast}$, which occurs at $r = (49/36) R_{\rm in}$.  Since the disc temperature does not reach $T_{\ast}$ at any radius, $T_{\rm col, max}$ is a more physically meaningful characterisation of the blackbody temperature associated with the spectrum.  For a thin disc that is torqued at the inner boundary, as is the case for \texttt{diskbb}, $T_{\rm col, max}$ coincides with $T_{\ast}$ because the temperature profile grows asymptotically with decreasing $r$ (see Figure 1 of \citeauthor{Zimmerman2005} \citeyear{Zimmerman2005}).  One of the two free parameters of the \texttt{diskbb} model is $T_{\ast}$, implicitly assuming $f_{\rm col} = 1$.  The `normalisation' is the other free parameter of \texttt{diskbb} and is given by,
\begin{equation}
K_{\rm bb} = 100 \frac{1}{f_{\rm col}^{4}} \left( \frac{R_{\rm in, km}}{D_{\rm kpc}} \right)^{2} {\rm cos}~i, \label{eq:Kbb}
\end{equation}
where, again, $f_{\rm col} = 1$, $D_{\rm kpc}$ is the distance to the source in kiloparsecs, $R_{\rm in, km}$ is the inner disc radius in kilometers, and $i$ is the inner disc inclination.  Inner disc radii are commonly derived from $K_{\rm bb}$ by measuring intrinsic system properties and imposing a {\it constant} $f_{\rm col}$ (not necessarily unity as assumed by \texttt{diskbb}), regardless of the source luminosity state \citep[e.g.,][]{Zdziarski2004}.  Remarkably, the simple two-parameter \texttt{diskbb} model fully describes the accretion disc spectrum under the appropriate assumptions of a non-zero torque inner disc boundary condition and a fixed colour correction factor for characterising the overall deviation from a blackbody spectrum.

\subsection{Zero-Torque Boundary Condition and Disc Vertical Structure}
If the inner disc radius can be identified with the ISCO, a zero-torque boundary condition becomes attractive from the physical consideration that the gas viscous inspiral timescale is long relative to the free-fall timescale for gas at the ISCO; therefore, the gas interior to $R_{\rm in}$ does not have time to radiate appreciably before being accreted \citep{ShakuraSunyaev1973, NovikovThorne1973, PageThorne1974, AbramowiczKato1989, FKR1992, AshfordiPaczynski2003}.  In principle, for a geometrically thin disc in the presence of a magnetic field, non-negligible torque resulting from magnetic stresses may be present at the ISCO \citep{Krolik1999, Gammie1999, AgolKrolik2000, AshfordiPaczynski2003}.  However, magnetohydrodynamic three-dimensional simulations of global thin accretion discs around black holes demonstrate that a weak magnetic field is unable to couple the gas within the plunging region to the disc, despite large fluctuating magnetic stresses inside the ISCO, in accordance with the zero-torque boundary condition \citep{Armitage2001, ReynoldsArmitage2001}.  Similar simulations exploring initially purely poloidal and toroidal field topologies instead observe significant stress at the ISCO and continuing deep within the plunging region, refuting the zero-torque assumption \citep{HawleyKrolik2001, HawleyKrolik2002}.  A difficulty with allowing for a non-zero torque in disc models is that there is no obvious choice for the magnitude of the torque, necessitating the introduction of a free parameter.  The zero-torque choice is both physically motivated and is not arbitrary; however, the presence of significant torque on the inner disc may be important in real systems depending on highly unknown factors such as the disc thickness, dissipation properties, and magnetisation.

The underlying physics responsible for altering the locally emitted spectrum of an optically thick disc from a pure blackbody are parameterised into the colour correction factor.   The result of this approximation is the colour-corrected, or `diluted',  blackbody,
\begin{equation}
I_{\nu} = \frac{1}{f_{\rm col}} B_{\nu} \left( f_{\rm col} T_{\rm eff} \right),
\end{equation}
where $I_{\nu}$ is the specific intensity, $B_{\nu}$ is the Planck function, $T_{\rm eff} = T_{\rm col} / f_{\rm col}$ is the effective temperature, and $f_{\rm col} = 1$ refers to a pure blackbody spectrum.  In reality, deviations in the emergent disc spectrum from a canonical blackbody arise from frequency-dependent opacities determined from the disc vertical structure, which is neglected in the standard thin disc treatment.  The degree of spectral modification will be governed by the combined effects of electron scattering (i.e., Comptonisation in the disc), free-free emission/absorption, and bound-free absorption, resulting in a departure from local thermodynamic equilibrium and a hardening of the spectrum \citep{FeltenRees1972, ShimuraTakahara1995b, Zavlin1996, Rutledge1999a, McClintock2004}.  A surrounding corona acting as a depository for some fraction of the disc accretion power provides yet another means for altering the emergent disc spectrum \citep{SvenssonZdziarski1994, Merloni2000}.  Furthermore, magnetic pressure support likely contributes significantly to hydrostatic balance in the disc photosphere and may act to vertically extend the disc atmosphere, producing a harder spectrum \citep{Blaes2006, BegelmanPringle2007}.  The phenomenological colour-correction prescription avoids all of these complications by neglecting any frequency dependence, only permitting translational (i.e., frequency-independent) hardening and attenuation of the Planck function.  Provided that this model yields a satisfactory fit to the observed disc spectrum, one presumes that $f_{\rm col}$, which is assumed to be constant for all radial locations in the disc, provides an adequate description of the disc vertical structure in an average sense.

The two disc models we appeal to in this work differ from \texttt{diskbb} in that they both adopt a zero-torque inner boundary condition and allow for the possibility of a variable colour correction factor.  The \texttt{ezdiskbb} model \citep{Zimmerman2005} is the analog of \texttt{diskbb}, but with the zero-torque boundary condition enforced.  The free parameters are the maximum disc colour temperature, $T_{\rm col, max}$ defined above, and the normalisation, given by
\begin{equation}
K_{\rm ez} = 100 \frac{1}{f_{\rm col}^{4}} \left( \frac{R_{\rm in, km}}{D_{\rm kpc}} \right)^{2} {\rm cos}~i, \label{eq:Kez}
\end{equation}
where $f_{\rm col}$ is not specified.  We emphasize that changes in $K_{\rm ez}$ do not necessarily imply changes in the disc inner radius, due to the parameter degeneracy between $R_{\rm in}$ and $f_{\rm col}$.  The \texttt{diskpn} model \citep{Gierlinski1999}, along with the zero-torque boundary condition, incorporates a pseudo-Newtonian potential \citep{PaczynskyWiita1980} to more accurately calculate the radial temperature distribution in the disc at the expense of introducing the inner disc radius as a free parameter.  The two remaining free parameters are $T_{\rm col, max}$ and the normalisation,
\begin{equation}
K_{\rm pn} = \frac{1}{f_{\rm col}^{4}} \left( \frac{M_{\rm M_{\odot}}}{D_{\rm kpc}} \right)^{2} {\rm cos}~i, \label{eq:Kpn}
\end{equation}
where $M_{\rm M_{\odot}}$ is the compact object mass in solar mass units.  In all of the MCD model normalisations, $f_{\rm col}$ enters to the fourth power.  Therefore, asking how much $f_{\rm col}$ could conceivably change during a state transition is not unreasonable, as only small changes in $f_{\rm col}$ are required to explain the observed spectral evolution, which may not necessitate disc truncation.

The disc models described up to this point neglect many important physical considerations.  We wish to emphasize that sophisticated, physically motivated models are available in \texttt{XSPEC} that account for irradiation of the disc \citep[\texttt{diskir}; ][]{Gierlinski2008}, general relativistic effects around a Schwarzschild \citep[\texttt{grad}; ][]{Ebisawa1991} and Kerr black hole \citep[\texttt{kerrbb}; ][]{Li2005}, and detailed modelling of the disc vertical structure around a Kerr black hole \citep[\texttt{bhspec}; ][]{Davis2005}.  Realistic disc models come with the expense of introducing additional parameters.  The limited quality of X-ray observations and the desire to limit the number of model parameters forces the use of over-simplified disc models, with the hope that they can adequately capture the most important aspects of the accretion physics.  The purpose of the subsequent analysis is to demonstrate that simplistic disc models produce good spectral fits but they cannot be used to make reliable claims regarding the behaviour of the inner disc radius.

%===========================================================================
% ANALYSIS AND RESULTS
\section{Analysis and Results}
\label{section:results}
In what follows, we exploit the timing and monitoring capabilities of {\it RXTE}, which provide multiple sequential spectra of moderate resolution tracking the transition between spectral states, to study the evolution of the accretion disc in GX 339--4.  The ultimate goal of our spectral analysis is to isolate the disc component in order to measure relative changes in disc properties during transitions over the rise (LHS $\rightarrow$ HSS) and decay (HSS $\rightarrow$ LHS) stages of an outburst.  We show that allowing for physically reasonable changes in $f_{\rm col}$ provides an alternative explanation to disc truncation for the observed disc spectral evolution during state transitions.  Table \ref{tab:bestfits_sample} provides a sample of best-fit parameters for the fits used in this work.  Table \ref{tab:bestfits_sample} in its entirety is available as an electronic supplement.

%===========================================================================
% Requirement of a Disc Component
\subsection{Requirement of a Disc Component}
\label{sec:diskreq}
We wish to verify the presence of a thermal disc component with adequate confidence using phenomenological models that have proven successful in characterising the general features of BHB X-ray spectra.  To determine whether or not a disc component is required in a given {\it RXTE} spectrum, we adopt a methodology based on $\chi^{2}$ statistics that uses phenomenological models to fit numerous observations in various spectral states, as motivated by previous works \citep{Dunn2008, Dunn2010, Dunn2011, ReynoldsMiller2011}.  The PCA and HEXTE data are fit jointly, allowing an energy-independent constant factor to vary for the HEXTE data to account for any normalisation offset relative to the PCA and linking all parameters between the PCA and HEXTE spectra.  The spectral models are modified by photoelectric absorption using the \texttt{phabs} model with a fixed equivalent neutral hydrogen column density of $N_{\rm H} = 5.7 \times 10^{21} {\rm~cm^{-2}}$ \citep{Miller2009}.  Both Galactic and intrinsic absorption are folded into the \texttt{phabs} model, leading to a potential ambiguity as to whether low-energy spectral evolution is due to changes in the intrinsic absorption or to evolution of the source.  \citet{MillerCackettReis2009} fitted high-resolution BHB spectra (including GX 339--4) over a wide range in luminosity and spectral states, demonstrating that the low-energy spectral evolution is a consequence of the evolution of the source itself, rather than a variable neutral photoelectric absorption.   Notably, \citet{Cabanac2009} present the case for possible variations in $N_{\rm H}$ during outbursts of GX 339--4; however, the restricted low-energy coverage of {\it RXTE} does not permit $N_{\rm H}$ to be a free parameter in our spectral fits.

As a first attempt, the spectra are fitted with a simple model consisting of an absorbed power-law and MCD blackbody (i.e., \texttt{phabs$\times$(powerlaw+diskbb)} in \texttt{XSPEC}) to model the continuum and accretion disc, respectively.  In many observations, this simple prescription does not yield an adequate fit due to the presence of a broad Fe K$\alpha$ emission line.  To check for the requirement of an iron line component, a Gaussian line profile (\texttt{gauss} in \texttt{XSPEC}) is added to the model at fixed centroid energy of $6.4 {\rm~keV}$ with the width, $\sigma$, and normalisation, $K_{\rm line}$, as free parameters.  An $F$-test is inappropriate for determining whether the inclusion of a spectral line is a statistically significant improvement \citep{Protassov2002}.  The alternative diagnostics outlined by \citet{Protassov2002} for testing for the presence of a line are not easily incorporated into the spectral fitting framework used here.  We are concerned with obtaining an acceptable fit in order to determine the presence of a disc component and do not wish to make any strong statements regarding Fe K$\alpha$ emission; therefore, we adopt the line detection procedure of \citet{Dunn2008}.  In order for a line to be considered significant, two criteria must be met: (1) An $F$-test must determine that the line is significant at $\ge 3\sigma$ ($F$-statistic probability $\mathcal P \le 0.0027$).  (2) Given the $1\sigma$ uncertainty in the line normalisation, $\sigma_{K_{\rm line}}$, obtained with the \texttt{XSPEC} \texttt{error} command, we require $K_{\rm line} \ge 3 \sigma_{K_{\rm line}}$.  This procedure for line determination is more conservative than appealing to an $F$-test alone, though we note that some detected line features may be erroneous.

We next test for the presence of a high energy break in the spectrum by replacing the power-law component with the broken power-law model, \texttt{bknpower}, and repeating the aforementioned fitting procedure.  If an $F$-test determines that the broken power-law model is a significant improvement over the power-law model at $\ge 3\sigma$ confidence, then we proceed with the broken power-law model imposing the parameter ranges for the spectral indices and break energy of $1 \le \Gamma \le 4$ and $E_{\rm b} \ge 10~{\rm keV}$, respectively.  In instances where a line component was deemed necessary in the power-law model but not the broken power-law model, or {\it vice versa}, we select the best fitting model determined by the lowest reduced $\chi^{2}$-statistic, $\chi^{2}_{\nu} \equiv \chi^{2}/\nu$, where $\nu$ is the number of degrees of freedom in the model.  For cases when the best-fit converges to the parameter limits of the \texttt{bknpower} model, the pipeline reverts to the \texttt{powerlaw} model for that observation.  If the best-fit \texttt{powerlaw} model also runs into a parameter bound, the observation is discarded.

At this point in the phenomenological fitting pipeline, for a fit to be considered acceptable for use in subsequent analysis, we require $\chi^{2}_{\nu} \le 2$.  Often, a poor fit to the high-energy spectrum is the culprit for failure to meet this criterion.  For instance, a poor fit may result from the inability of simplistic, unphysical models to properly fit a reflection component or a complicated spectrum arising from a dynamic corona.  Since we are using phenomenological models to characterize the soft disc component and are not interested in a detailed characterization of the high-energy spectrum, if the best-fit model to the combined PCA/HEXTE data gives $\chi^{2}_{\nu} > 2$, then we repeat the above fitting procedure on the PCA data alone in search of an acceptable fit.  In these cases where only the PCA data are being fitted, we use the \texttt{powerlaw} model only because any spectral energy break is expected to lie beyond the high-energy cutoff of the PCA, rendering the use of the \texttt{bknpower} model inappropriate.

The disc component (\texttt{diskbb}) is then removed from the best-fit model and the spectrum is re-fitted.  An $F$-test is performed to determine whether or not the inclusion of the disc improved the fit by $\ge 3\sigma$.  Observations that meet these criteria are kept for further analysis, while those that do not statistically require a disc are discarded.  Running this model fitting procedure on 283 total observations of GX 339--4 in different stages of transition between the low/hard and high/soft states resulted in 177 observations where a disc component was deemed statistically required from spectral fitting techniques (see Table \ref{tab:transitions}).

We caution that the absence of a quasi-blackbody component in a given {\it RXTE} spectrum does not imply the absence or truncation of the physical accretion disc.  The low energy cutoff of the {\it RXTE} PCA is restricted to $\sim 3 {\rm~keV}$; thus, only the Wien tail of the MCD blackbody spectrum will be in the observable band.  Although we may not detect a disc, especially in the HIS and LHS, this is not enough to claim the absence of a disc component, which may be dominant at energies below the sensitivity of {\it RXTE} \citep[e.g.,][]{Dunn2008} and has been observed with {\it XMM-Newton} in the LHS of GX 339--4 \citep[e.g.,][]{Miller2006, Reis2008}.

\begin{figure}
\begin{center}
\includegraphics[width=84mm]{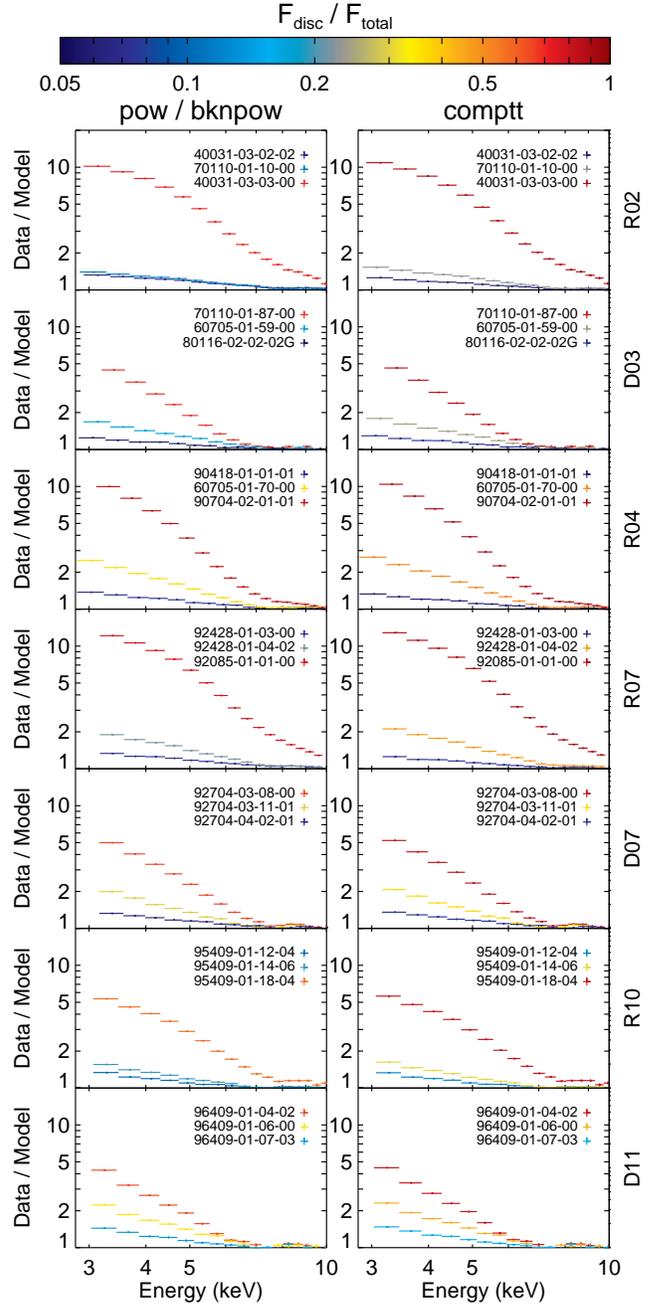}
\end{center}
\caption{Ratios of the {\it RXTE} data to the best-fit models with the \texttt{diskbb} component removed to highlight the need for a quasi-blackbody feature to model the soft excess.  For each transition, three data/model ratios are shown that representatively span the range in measured disk fractions.  Glyph colours denote the disc fraction for each fit.  For comparison, the best-fit models where Comptonisation is modelled by the \texttt{powerlaw} or \texttt{bknpower} ({\it left column}) and \texttt{comptt} ({\it right column}) \texttt{XSPEC} models are shown.  For each row, the observation IDs are given in the {\it upper right} corner of the plots and the transition ID on the {\it rightmost} ordinate.}
\label{fig:diskreq}
\end{figure}

Figure \ref{fig:diskreq} shows the data-to-model ratios with the disc component removed from the best-fit model for representative observations of each state transition.  Disk fractions are defined by,
\begin{equation}
DF \equiv  \frac{F_{\rm disc}^{0.1-10~{\rm keV}}}{F_{\rm total}^{0.1-100~{\rm keV}}}, \label{eq:DF}
\end{equation}
where $F_{\rm disc}$ and $F_{\rm total}$ are the unabsorbed fluxes of the disc component and overall model, respectively, taken over the energy ranges indicated by the superscripts.  Fluxes were obtained with the \texttt{cflux} command in \texttt{XSPEC}.  Disc components are robustly detected by the pipeline described above, even for moderately low disc fractions, which is essential for tracking the evolution of the disc throughout each transition in the subsequent analysis.

% FIGURE 2

%===========================================================================
\subsection{Sensitivity of Spectral Fits to the Continuum Model}
\label{sec:model}
The motivation for employing a simple power-law model with limited free parameters is to fit the portion of the spectrum that is thought to arise from Comptonisation of thermal disc photons by a low density, energetic electron corona of unknown geometry, which produces a power-law spectrum to first order.  The phenomenological power-law model extends to arbitrarily low energies, which does not mimic a physical Comptonisation process where a sharp cutoff is expected at an energy comparable to the characteristic energy of the input seed photon distribution, namely, the quasi-blackbody disc spectrum.  Therefore, the extension of the power-law component toward arbitrarily low energies results in `flux stealing' from the disc component, making a disc more difficult to detect than had we adopted a more physical Comptonisation model.

To investigate the sensitivity of fitting the disc component on the choice of Comptonisation model, we explore two different models to characterise the continuum.  The Comptonisation model \texttt{comptt} \citep{Titarchuk1994, HuaTitarchuk1995} provides an analytic description of the spectrum produced from a soft photon distribution propagating through a hot plasma cloud.  We re-ran the disc detection pipeline with \texttt{comptt} in place of the power-law component, allowing only the plasma temperature, $T_{\rm p}$, plasma optical depth, $\tau_{\rm p}$, and normalisation, $K_{\rm comptt}$, to be free parameters.  The remaining \texttt{comptt} parameters that we chose were a redshift $z = 0$, a seed photon temperature, $T_{0} = T_{\rm col, max}$ (i.e., 0.488 times the \texttt{diskbb} temperature), and a spherical plasma cloud geometry.  The additional free parameter introduced by \texttt{comptt} is justified by the physical, as opposed to phenomenological, nature of the model.  For cases where the best-fit converges to the enforced plasma temperature limit, $T_{\rm p} = 100~{\rm keV}$, the data are re-fitted with $T_{\rm p}$ fixed at 100 keV.  When the disc component is removed to check for the requirement of a disc, $T_{0}$ becomes a free parameter, as there is no longer an obvious, physically motivated choice for the seed photon distribution injected into the corona in the absence of a disc.  A total of 166/283 observations required a disc component when \texttt{comptt} was used, compared to the 177/283 using the standard phenomenological \texttt{powerlaw} and \texttt{bknpower} models.

% FIGURE 3
\begin{figure}
\begin{center}
\includegraphics[width=84mm]{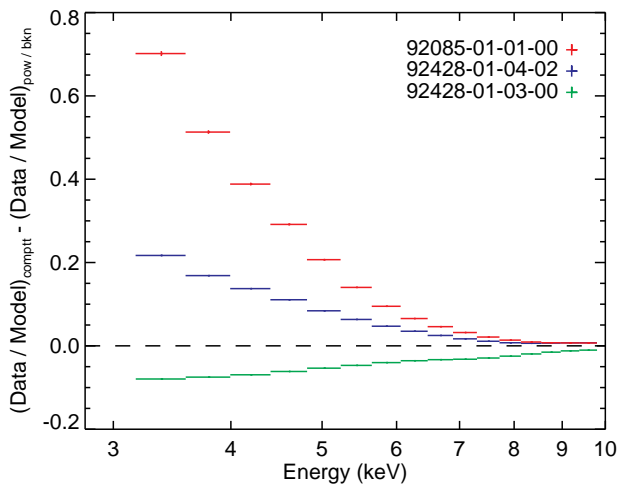}
\end{center}
\caption{Dependence of the strength and spectral shape of the disc component on the choice of Comptonisation model for transition R07.  The residuals plotted are computed by subtracting the data-to-model ratio for the \texttt{pow/bknpow} best-fit model from the best-fit \texttt{comptt} data-to-model ratio.  The data-to-model ratios used are those where the disc component was removed, as done in Figure \ref{fig:diskreq}.  Residuals falling on the {\it dashed} line indicate no difference in the measured disk component between the two Comptonisation models, while a positive excess implies flux stealing from the disc component by the \texttt{pow/bkn} model relative to the \texttt{comptt} model.  From {\it top} to {\it bottom}, the residuals were computed from observation IDs 92428-01-03-00 ({\it red glyphs}), 92428-01-04-02({\it blue glyphs}), and 92085-01-01-00 ({\it green glyphs}), which represent the HSS, IS, and LHS, respectively.  The flux stealing effect is most pronounced in the HSS and IS.}
\label{fig:fluxsteal}
\end{figure}

% FIGURE 4
\begin{figure}
\begin{center}
\includegraphics[width=84mm]{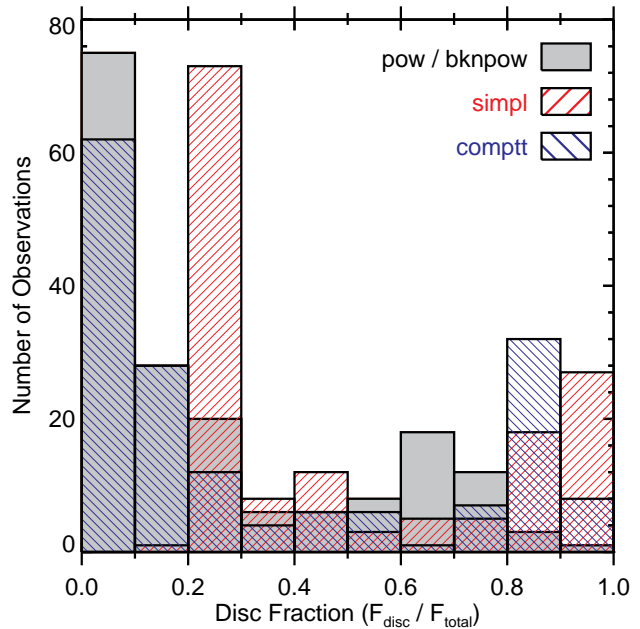}
\end{center}
\caption{Histogram of disc fractions for fits performed with Comptonisation models \texttt{pow/bknpow} ({\it grey filled}), \texttt{simpl} ({\it red forward slashes}), and \texttt{comptt} ({\it blue backward slashes}).  The excess of low disc fractions for the \texttt{pow/bknpow} model is a consequence of the power-law extending to arbitrarily low energies, effectively stealing flux from the disc component.}
\label{fig:diskfrac}
\end{figure}

An alternative empirical substitute for a power-law model is \texttt{simpl} \citep{Steiner2009}, which truncates the power-law spectrum at an energy comparable to the characteristic input seed photon energy; thus, avoiding the unphysical low-energy extension of the \texttt{powerlaw} and \texttt{bknpower} models.  Since \texttt{simpl} is a convolution model, we cannot apply the disc detection pipeline of \S \ref{sec:diskreq} to determine whether or not a disc component is required.  Instead, the previously determined 177 best-fit models are re-fitted with \texttt{simpl} as the component responsible for modelling Comptonisation, allowing for both up- and down-scattering of the input seed spectrum.  Again, as was done for the disc detection pipelines using \texttt{powerlaw}/\texttt{bknpower} and \texttt{comptt} as the continuum models, the observation is discarded if the best-fit converges on a parameter limit.  This returned 150/177 observations where the best-fit does not encounter a parameter limit.  While \texttt{simpl} cannot capture the high energy break commonly observed in BHB spectra, often resulting in the dismissal of the HEXTE data in order to obtain acceptable fits, this excision of data is justified because a detailed model of the high-energy spectrum is ancillary to a reliable model of the disc spectrum.

The choice of Comptonisation model influences the measured disc properties.  As discussed above, the three Comptonisation models we explore in this work are: (1) standard power-law or broken power-law (hereafter, \texttt{pow/bknpow}), (2) power-law with a low-energy cut-off (hereafter, \texttt{simpl}), and (3) physical Comptonisation model (hereafter, \texttt{comptt}).  For representative observations over all transitions fit with the \texttt{pow/bknpow} and \texttt{comptt} models, Figure \ref{fig:diskreq} shows data-to-model ratios with the disc component removed from the best-fit model, which is a probe of the spectral shape and strength of the disc component.  The disc is strong in the HSS and gradually weakens, but remains detectable, in the LHS down to $DF \simeq 0.05$.

Figure \ref{fig:fluxsteal} shows the residuals for the data-to-model ratios of the \texttt{comptt} fits compared to those for the best-fit \texttt{pow/bknpow} models for a representative transition.  The flux stealing phenomenon inherent to the low-energy extension of the \texttt{pow/bknpow} model is apparent from the positive excess in the residuals and affects the measured disc spectrum, particularly for high disc fractions that are characteristic of the HSS.  A histogram of measured disk fractions for all observations deemed to require a disk and for each choice of Comptonisation model is shown in Figure \ref{fig:diskfrac}.  Generally, replacing the power-law model with more physically motivated Comptonisation models produces higher disc fractions and yields disc model spectra that are unplagued by flux stealing.  The measured disk fraction for a given observation generally follows $DF_{\rm \texttt{simpl}} > DF_{\rm \texttt{comptt}} > DF_{\rm \texttt{pow/bknpow}}$.

Given this study on the sensitivity of the spectral fits to the choice of continuum model, we must decide how to proceed with the subsequent analysis in order to best represent the true character of the disc evolution during a state transition.  In the interest of comparison and because discrepancies in the measured disc properties arise between the three different, but seemingly adequate, Comptonisation models, we do not favour one model over another.  Instead, we elect to perform the subsequent analysis in three branches, identical in all respects except for the Comptonisation model (i.e., \texttt{pow/bknpow}, \texttt{simpl}, \texttt{comptt}) being used in the spectral fits.

%===========================================================================
\subsection{Disc Evolution Results from Spectral Fitting}
A parameter degeneracy exists between the inner disc radius and the colour correction factor within the normalisation of the standard \texttt{diskbb} model.  Therefore, decomposing the contributions to the disc evolution due to changes in the vertical disc structure, parametrized into $f_{\rm col}$, and a migrating inner disc region is nontrivial.  Combining the models \texttt{diskpn} and \texttt{ezdiskbb}, while appealing to relative measurements of the disc model normalisation evolution, we examine the degree of change in $f_{\rm col}$ and $R_{\rm in}$ required to explain the disc spectral evolution in state transitions of GX 339--4.

% FIGURE 5
\begin{figure}
\begin{center}
\includegraphics[width=84mm]{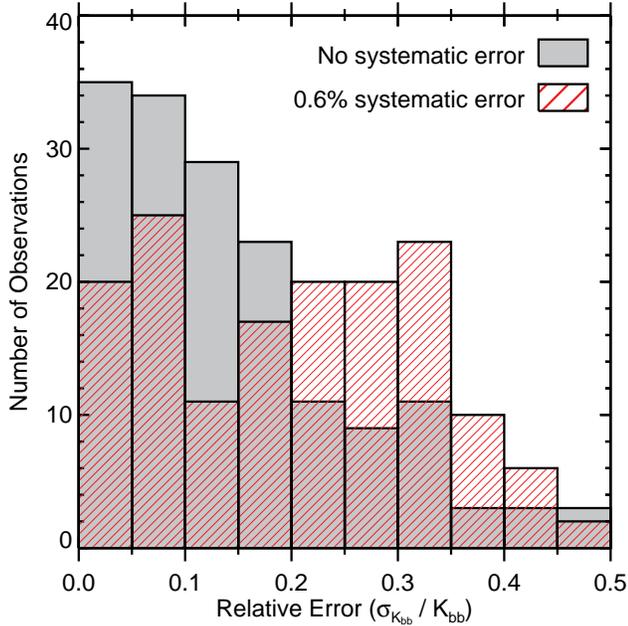}
\end{center}
\caption{Histogram of relative errors on the best-fit \texttt{diskbb} normalisation, $\sigma_{K_{\rm bb}} / K_{\rm bb}$, for all {\it RXTE} observations requiring a disc component.  Models fit with a systematic error of 0.6\% added to all PCA energy channels and no systematic error added to the PCA data are shown as {\it red forward slashes} and {\it grey} bars, respectively.  The relative error on the disc normalisation is reduced when systematic error is not added to the data.}
\label{fig:Kbb_err}
\end{figure}

%===========================================================================
\subsubsection{Relative Measurements}
\label{sec:relative}
Obtaining reliable {\it absolute} measurements of quantities locked up in the disc normalisation is limited by constraints on the observables, namely, the mass, distance, and disc inclination, which all appear in the normalisation.  However, {\it relative} changes in $f_{\rm col}$ and/or $R_{\rm in}$ are accessible from the evolution of the disc normalisation because the intrinsic parameters do not change between successive observations.

Imposing a small systematic error on {\it RXTE} PCA data is common practice, resulting in a substantial improvement in the quality of spectral fits.  Applying model fits to data where no systematic error is added will result in poor fits but with best-fit parameter errors dramatically reduced.  While the best-fit parameter values may be unreliable, they should be affected in a systematic sense and systematic errors associated with instrument calibration will be removed from the parameter errors.  Provided that relative, rather than absolute, measurements are of interest, the evolution of a parameter may be tracked from fits to spectra without systematic error incorporated into the PCA data.

% FIGURE 6
\begin{figure}
\begin{center}
\includegraphics[width=84mm]{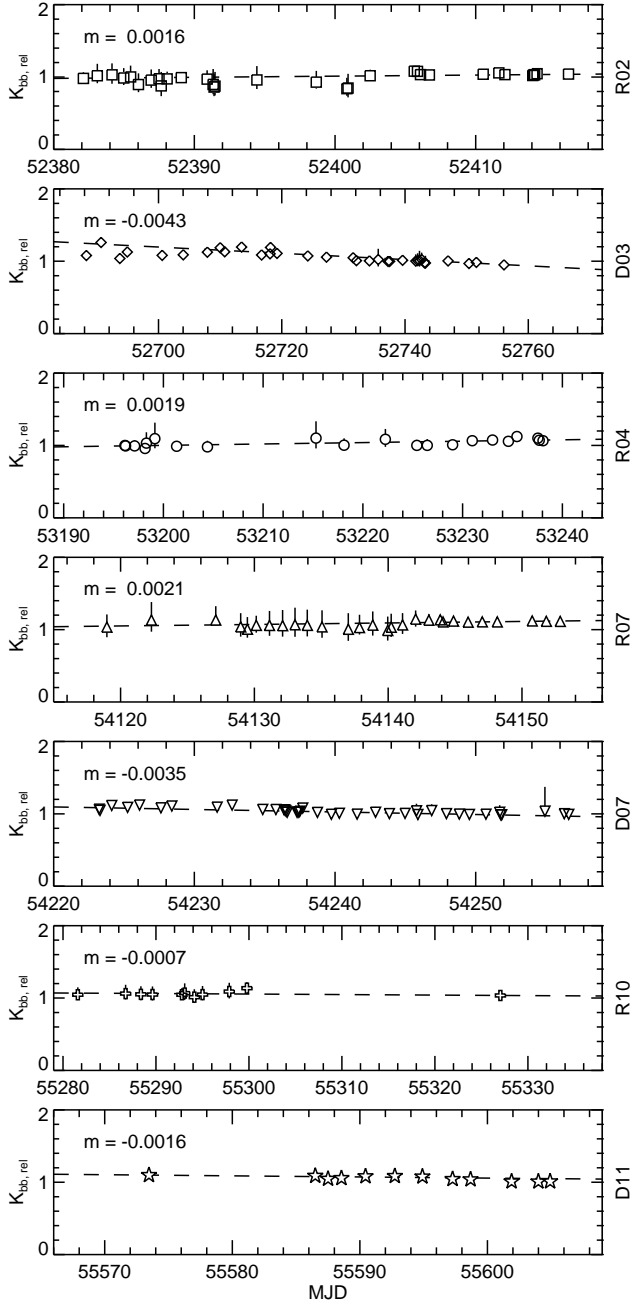}
\end{center}
\caption{Relative comparison of \texttt{diskbb} normalisations, $K_{\rm bb, rel}$ (see text for definition), for each transition.  The slope, $m$, from a linear fit to the data is displayed in the {\it upper left} corner for each plot.  A zero-slope means that relative changes in $K_{\rm bb}$ with time are unaffected by neglecting to add systematic errors to the PCA data, indicating that the systematics remain constant from observation to observation for the time intervals considered.  The flatness of $K_{\rm bb,rel}$ justifies the use of relative measurements for studying the disc evolution throughout a state transition.  From {\it top} to {\it bottom}, the panels show the time evolution of $K_{\rm bb, rel}$ for transitions R02, D03, R04, R07, D07, R10, and D11.  Glyphs without error bars are larger than their respective 1$\sigma$ error.}
\label{fig:Kbb_rel}
\end{figure}

To test the validity of this proposition, we re-fitted all spectra deemed to require a disc component with the best-fit models determined from the aforementioned fitting pipeline, but with no systematic error added to the PCA data.  Figure \ref{fig:Kbb_err} compares the distribution of the \texttt{diskbb} normalisation relative error, defined as $\sigma_{K_{\rm bb}} / K_{\rm bb}$, for fits to all GX 339--4 spectra when no systematic error and the standard 0.6\% systematic error were added to the PCA data \citep{Jahoda2006}.  As expected, not including systematic errors results in improved relative errors on the \texttt{diskbb} normalisation.  Adopting an analysis technique based on omitting systematic errors from the PCA data requires that the disc normalisation values are only affected in a systematic way as a result.  For each transition, Figure \ref{fig:Kbb_rel} shows $K_{\rm bb, rel} = K_{\rm bb, nosys} / K_{\rm bb}$, which is defined as the best-fit \texttt{diskbb} normalisation for fits with no systematic error added, $K_{\rm bb, nosys}$, relative to the normalisation for fits with 0.6\% systematic error added, $K_{\rm bb}$.  The constancy of this ratio throughout the spectral evolution of all transitions suggests that excluding systematic error for the PCA data to study the evolution of the disc normalisation in a relative sense is justified in practice.

Occasionally, an observation that was adequately fitted when a 0.6\% systematic error was added to the PCA data will converge on parameter limits or grossly unphysical values when the data without systematic error are fitted.  In these situations, the data without systematic error added are fitted by hand in order to achieve a fit with more realistic best-fit parameters.  All errors presented in this work are $1\sigma$ and are appropriately propagated when making relative measurements.\footnote{Two identical data sets, one with a small systematic error added ($X$) and one without ($Y$), are nearly perfectly correlated (i.e., the correlation coefficient, $\rho_{X,Y} = {\rm cov}(X,Y) / (\sigma_{X} \sigma_{Y}) \simeq 1$).  Since both $X$ and $Y$ only differ in the uncertainties associated with the data, their covariance is given by, ${\rm cov}(X,Y) = {\rm cov}(X,X) = \sigma_{X}^{2}$.  The resulting correlation coefficient becomes, $\rho_{X,Y} = \sigma_{X} / \sigma_{Y} \simeq 1$, assuming that the systematic errors added to $Y$ are small, which is true for our purposes.}

% FIGURE 7
\begin{figure}
\begin{center}
\includegraphics[width=84mm]{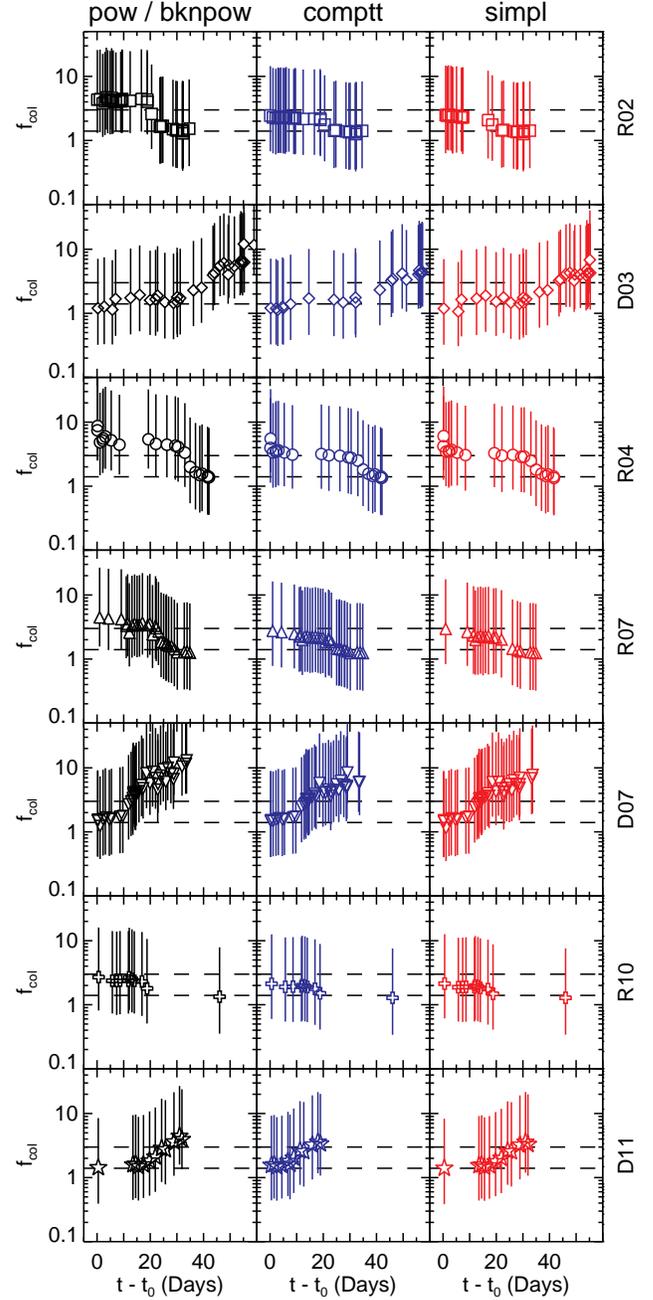}
\end{center}
\caption{Evolution of the colour correction factor over all state transitions derived from best-fit \texttt{diskpn} normalisations where the inner disc radius was held at 6 $R_{\rm g}$.  The {\it left}, {\it middle}, and {\it right} panels  correspond to fits using the \texttt{pow/bknpow}, \texttt{comptt}, and \texttt{simpl} continuum models, respectively.  The abscissa shows time elapsed since the start time, $t_{0}$, of the transition (see Table \ref{tab:transitions}).  Error bars account for the maximum conceivable ranges of the mass, distance, and inclination of GX 339--4 (see text).  The {\it dashed lines} enclose $1.4 < f_{\rm col} < 3$, which is the adopted allowable range in $f_{\rm col}$.  The transition ID corresponding to each row is indicated on the {\it rightmost} ordinate.  Within the large uncertainties, a non-truncated disc with a variable $f_{\rm col}$ can adequately explain the observed disc spectral evolution.}
\label{fig:fcol_all}
\end{figure}

%===========================================================================
\subsubsection{Disc Spectral Evolution: Variable $f_{\rm col}$}
\label{sec:fcol}
A variable colour correction factor has been considered from both a numerical \citep[e.g.,][]{ShimuraTakahara1995b, Merloni2000, Davis2005} and observational \citep[e.g.,][]{Dunn2011, ReynoldsMiller2011} approach, with $1.4 < f_{\rm col} < 3$ generally believed to be a conservatively broad allowable range, implying at most a factor of $\sim$ 2 change in $f_{\rm col}$.  However, this range in $f_{\rm col}$ does not consider regimes extending to the low disc luminosities associated with the HIS and LHS nor the effects of disc inhomogeneities and strong magnetisation.  Spectra that are less disc-dominated are harder; thus, requiring an increased $f_{\rm col}$.  The possibility of extending the upper $f_{\rm col}$ bound for discs in the HIS and LHS (i.e., non-disc-dominated regimes where $L_{\rm disc} \lesssim 0.01~L_{\rm Edd}$) and whether a colour-corrected blackbody is an appropriate description of the disc spectrum in this regime remain open questions.

Using the \texttt{diskpn} model with the inner disc radius fixed ensures that the evolution of the disc normalisation is due entirely to a variable colour correction factor, which can be computed directly from Equation \ref{eq:Kpn}, as demonstrated by \citet{ReynoldsMiller2011}.  We elect to explore the degree of change in $f_{\rm col}$ alone that is required to explain the disc spectral evolution for a fixed inner disc radius of $6~R_{\rm g}$.  This choice for $R_{\rm in}$ is motivated by considering a disc extending down to the ISCO for a Schwarzschild black hole and is the minimum value allowed by \texttt{diskpn}.

To measure the evolution of {\it absolute} $f_{\rm col}$ values required to explain the disc evolution in each transition for a fixed inner disc radius, we re-fit all of the spectra with 0.6\% systematic error added to the PCA data using the best-fit models, but replacing the \texttt{diskbb} component with \texttt{diskpn} and freezing $R_{\rm in}$ at 6 $R_{\rm g}$.  Figure \ref{fig:fcol_all} shows the evolution in $f_{\rm col}$ that is necessary to account for each transition, with the error bars incorporating the best-fit $K_{\rm pn}$ error and the range in intrinsic parameters of GX 339--4.  The values adopted for the GX 339--4 observables, along with their (lower, upper) limits, are $M = 7~(5.8, 12) {\rm~M_{\odot}}$, $D = 8~(6, 10) {\rm~kpc}$, and $i = 29^{\circ}~(10^{\circ}, 60^{\circ})$ (see \S \ref{sec:gx339} and references therein).  Figure \ref{fig:fcol_all} shows that, within the uncertainties, the disc spectral evolution can potentially be explained by a variable $1.4 \lesssim f_{\rm col} \lesssim 3$ for an inner disc radius that remains at 6 $R_{\rm g}$ throughout each state transition.

Studying the evolution of absolute $f_{\rm col}$ values for a fixed $R_{\rm in}$ is limited by the large uncertainties in the mass, distance, and inclination of GX 339--4.  Given that an inner disc that remains at 6 $R_{\rm g}$ combined with a modestly variable $f_{\rm col}$ may be a viable model for accretion disc evolution in the state transitions studied here, we use the relative measurement technique introduced in \S \ref{sec:relative} to better understand the degree of change in $f_{\rm col}$ that is required to explain the observations without appealing to disc truncation.

The PCA data are replaced with a duplicate copy where no systematic errors have been added.  Next, the \texttt{diskpn} fits where $R_{\rm in} = 6~R_{\rm g}$ are re-fitted and a best-fit is found for each observation.  This is done for each of the three Comptonisation models \texttt{pow/bknpow}, \texttt{simpl}, and \texttt{comptt}.  For each observation in a given transition, $f_{\rm col}$ is computed from $K_{\rm pn}$ using Equation \ref{eq:Kpn}, without specifying the intrinsic parameters (i.e., $M$, $D$, $i$).  The resulting $f_{\rm col}$ values are normalised to the $f_{\rm col}$ from a representative HSS observation, $f_{\rm col, HSS}$, and further normalised to the adopted $f_{\rm col} = 1.4$ value for the HSS \citep{Davis2005}, yielding the relative colour correction factor,
\begin{equation}
f_{\rm col, rel} \equiv 1.4 \left( \frac{f_{\rm col}}{f_{\rm col, HSS}} \right). \label{eq:fcol_rel}
\end{equation}

% FIGURE 8
\begin{figure}
\begin{center}
\includegraphics[width=84mm]{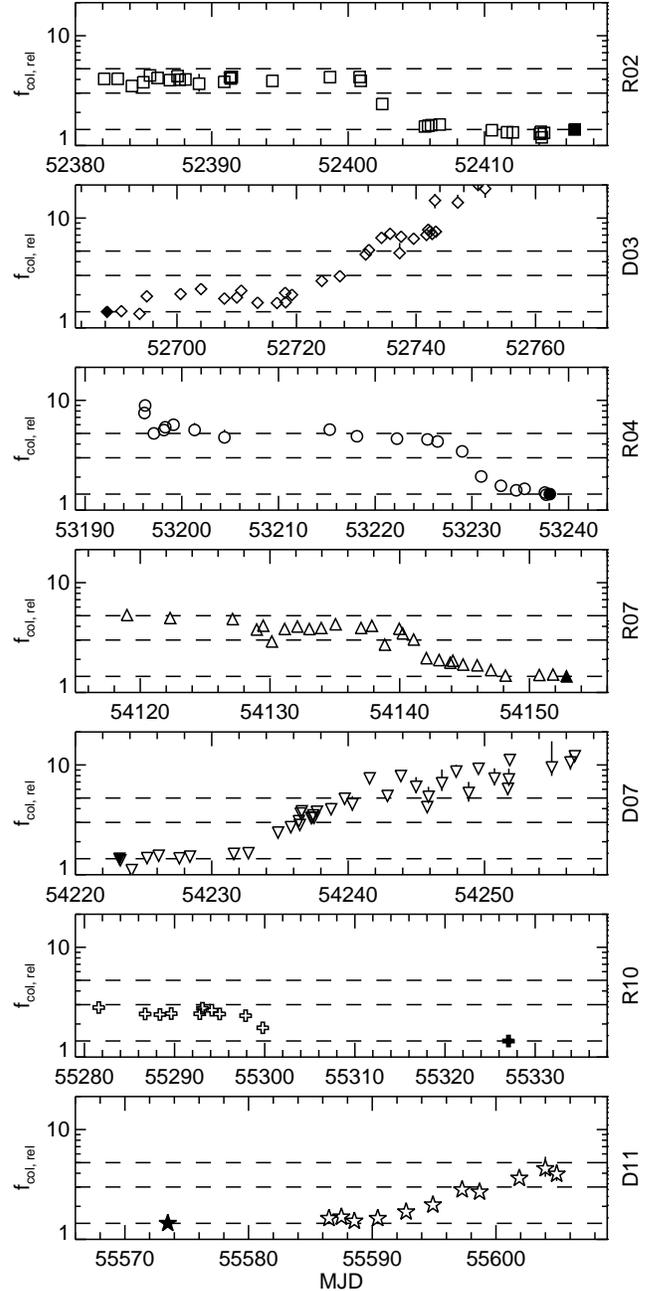}
\end{center}
\caption{Relative measurements of the evolution of $f_{\rm col}$ for each transition derived from fits using \texttt{pow/bknpow} to model Comptonisation.  The two {\it lower dashed lines} enclose $1.4 < f_{\rm col} < 3$, while the {\it upper dashed line} marks $f_{\rm col}= 5$.  The $f_{\rm col, rel}$ measurements for each transition were made relative to the filled glyphs and normalised to $f_{\rm col} = 1.4$, which is appropriate for the HSS.  The disc evolution can be explained for transitions R02, R07, R10, and D11 by a non-truncated disc with $1.4 \le f_{\rm col, rel}\le 5$.}
\label{fig:fcol_rel}
\end{figure}

\begin{figure}
\begin{center}
\includegraphics[width=84mm]{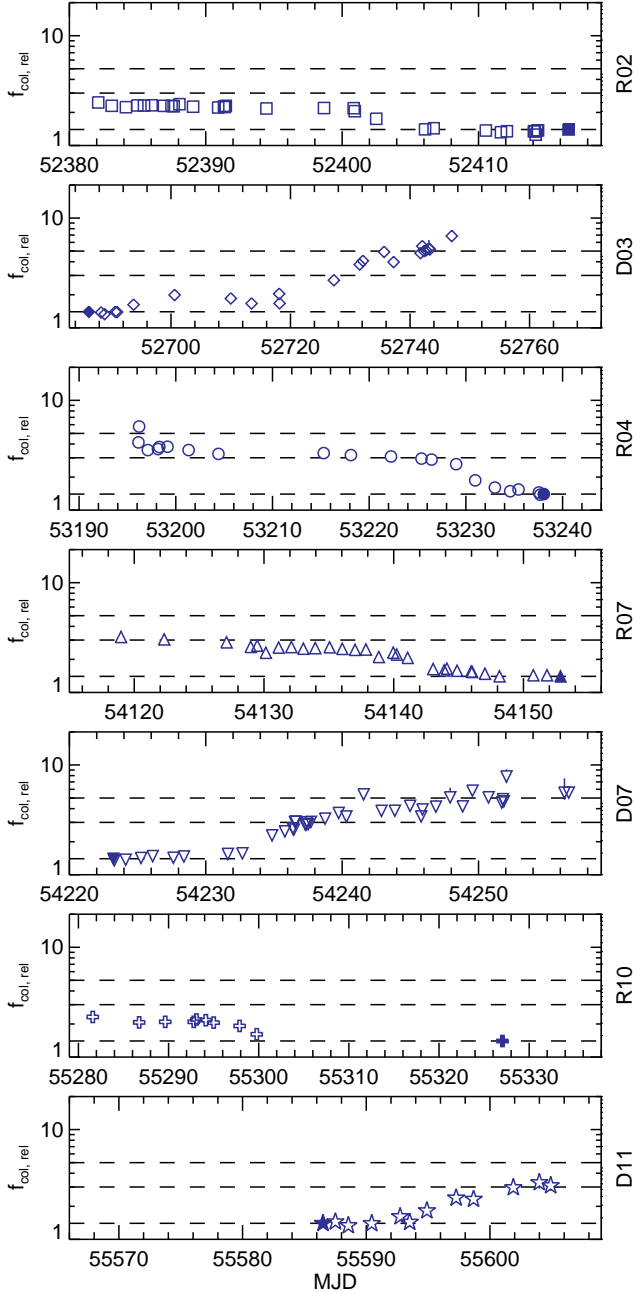}
\end{center}
\caption{Same as Figure \ref{fig:fcol_rel}, but with the Comptonisation model \texttt{comptt} replacing \texttt{pow/bknpow}.  The required evolution in $f_{\rm col, rel}$ is systematically reduced compared to results from the \texttt{pow/bknpow} continuum model.}
\label{fig:fcol_rel_comptt}
\end{figure}

\begin{figure}
\begin{center}
\includegraphics[width=84mm]{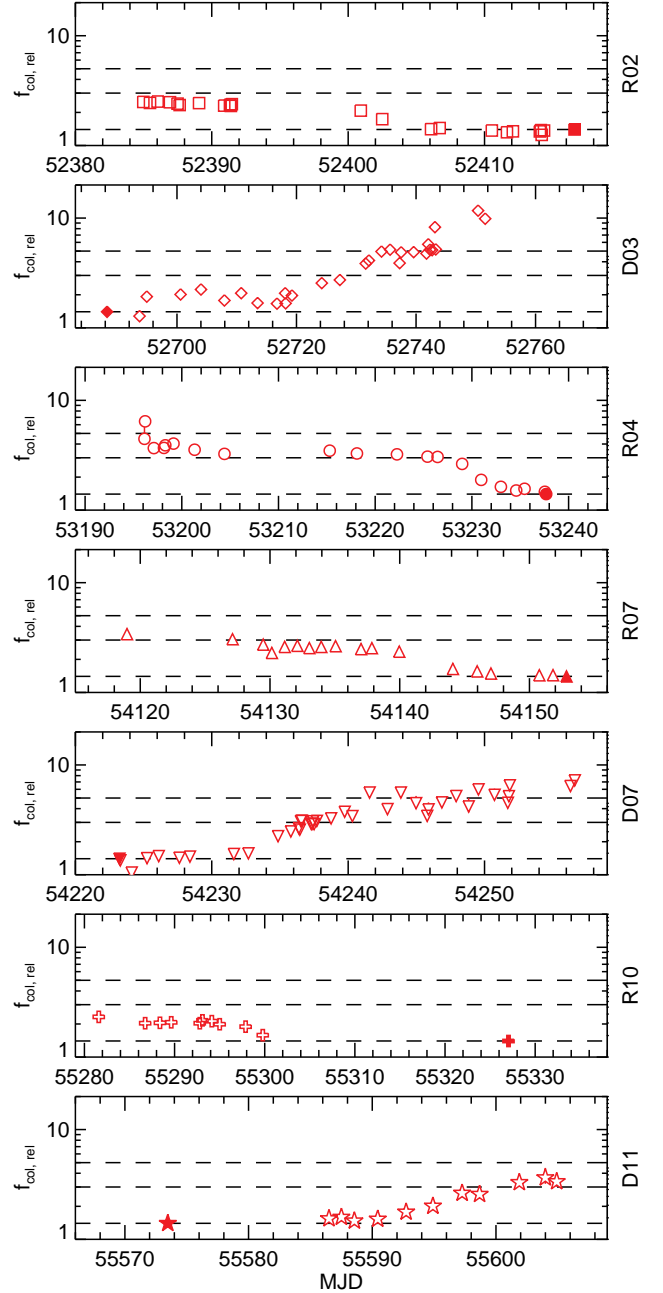}
\end{center}
\caption{Same as Figure \ref{fig:fcol_rel}, but with the Comptonisation model \texttt{simpl} replacing \texttt{pow/bknpow}.  The required evolution in $f_{\rm col, rel}$ is systematically reduced compared to results from the \texttt{pow/bknpow} continuum model.}
\label{fig:fcol_rel_simpl}
\end{figure}

Supposing that the inner disc remains fixed at $6~R_{\rm g}$ throughout each state transition, Figures \ref{fig:fcol_rel}-\ref{fig:fcol_rel_simpl} show the relative change in $f_{\rm col}$ alone required to explain the observed disc spectral evolution.  Notably, the \texttt{pow/bknpow} model requires more drastic change in $f_{\rm col}$ compared to \texttt{comptt} and \texttt{simpl}.  Under the presumption that $f_{\rm col, rel} = 3$ is a conservative upper bound, the notion of a variable $f_{\rm col}$ alone is perhaps possible for transitions R02, R07, R10, and D11; however, an evolving $f_{\rm col}$ is an inadequate description of the evolving disc spectrum for transitions D03, R04, and D07, regardless of the Comptonisation model employed.  For the later set of transitions, one must invoke disc truncation or an alternative disc model to satisfactorily explain the observed disc evolution.  \citet{ReynoldsMiller2011} speculate that if the colour-corrected blackbody is a viable model of the LHS, then $f_{\rm col}$ may be as large as $\sim 5$ based on spectral fits to a comprehensive set of BHB spectra.  Inspecting Figures \ref{fig:fcol_rel_comptt}-\ref{fig:fcol_rel_simpl}, where the \texttt{comptt} and \texttt{simpl} models were adopted, all seven transitions essentially conform to this extended $f_{\rm col}$ range; although, making definitive claims regarding the physically plausible $f_{\rm col}$ range and applicability of the MCD model in the LHS is beyond our scope.

% FIGURES 9, 10

%===========================================================================
\subsection{A Hot Inner Disc Can Masquerade as an Increased $f_{\rm col}$}
\label{sec:fakeit}
Observational evidence suggests that GX 339--4 harbors a near-maximally spinning black hole (see \S \ref{sec:gx339}).  Consequently, the inner disc extends deep into the black hole potential and relativistic effects become non-negligible, rendering Newtonian disc models inappropriate.  The disc models used in this work to derive $f_{\rm col}$ measurements impose an effective lower limit on the inner disc radius of $6~R_{\rm g}$, whereas the ISCO of a Kerr black hole with $a_{\ast} = 0.998$ lies at $1.24~R_{\rm g}$, corresponding to a substantially hotter inner disc.  Consider attempting to use a Newtonian prescription to model an accretion disc around a Kerr black hole, where $R_{\rm in} = R_{\rm ISCO}$.  The Newtonian disc model is restricted by $R_{\rm in} \ge 6~R_{\rm g}$ and will attempt to fit the true hotter disc, located at $R_{\rm ISCO} < 6~R_{\rm g}$, by increasing the colour correction factor as follows.  The colour temperature determined by the model depends on both $f_{\rm col}$ and $T_{\rm eff}$ according to $T_{\rm col} = f_{\rm col} T_{\rm eff}$.  The maximum effective temperature of the Kerr disc exceeds that of the Newtonian model with $R_{\rm in} = 6 R_{\rm g}$; therefore, the only way for the model to achieve the colour temperature demanded by the data is to increase $f_{\rm col}$ to an erroneously high value.

Using the \texttt{XSPEC} command \texttt{fakeit none}, we generate a suite of simulated disc spectra with the \texttt{kerrbb} model of a spinning black hole with an {\it RXTE} response matrix and integration time of 2000 seconds.  The cases considered are a Schwarzschild black hole ($a_{\ast} = 0$, $\eta = 0.057$) and maximal Kerr black hole ($a_{\ast} = 0.998$, $\eta = 0.30$) \citep{Thorne1974}, each having mass $7~M_{\odot}$ located at a distance of 8 kpc with a disc inclination of $30^{\circ}$.  The inner disc torque is set to zero and the effects of both self-irradiation and limb-darkening on the disc spectrum are switched on.  We pair mass accretion rates with colour correction factors that are reasonably representative of the spectral state corresponding to each $\dot{M}$ that we select.  For the $a_{\ast} = 0$ black hole, we simulate disc spectra for $\dot{M} = 0.05~\dot{M}_{\rm Edd}$ with $f_{\rm col} = 1.4$, attempting to mimic a HSS/SIS accretion disc; and $\dot{M} = 0.005~\dot{M}_{\rm Edd}$ with $f_{\rm col} = 3$, which is more representative of a HIS/LHS observation.  Similarly, for the $a_{\ast} = 0.998$ black hole, we choose $\dot{M} = 0.01~\dot{M}_{\rm Edd}$ with $f_{\rm col} = 1.4$ and $\dot{M} = 0.001~\dot{M}_{\rm Edd}$ with $f_{\rm col} = 3$.  Extending $\dot{M}$ beyond these ranges for a fixed $f_{\rm col}$ resulted in either simulated data with unconstrained uncertainties or unacceptable fits (i.e., $\chi^{2}_{\nu} \gg 1$) using the procedure described below.

Each spectrum is fit over the energy range 2.8-10 keV using the \texttt{diskpn} model with $R_{\rm in}$ fixed at $6~R_{\rm g}$.  The best-fit colour correction factor is computed from the normalisation using Equation \ref{eq:Kpn}.  Table \ref{tab:kerrbb} presents the results of this exercise.  In general, the psuedo-Newtonian model, \texttt{diskpn}, reliably recovers the true $f_{\rm col}$ value for a non-spinning black hole, but requires a factor of $\sim 1.9$ increase in $f_{\rm col}$ for a maximally spinning black hole.  This demonstrates that the inability of Newtonian disc models to properly model $R_{\rm in}$ and $T_{\rm eff}$ for Kerr black holes can result in unphysically large $f_{\rm col}$ measurements to account for the hotter inner disc.

One may wonder if relative measurements of $f_{\rm col}$ derived from Newtonian disc models, as done in this work, are affected by the inability of these models to properly represent the hotter discs around spinning black holes.  A colour correction factor evolving in a HSS $\rightarrow$ LHS transition from $f_{\rm col} = 1.4 \rightarrow 3$ corresponds to a factor of 2.1 increase in $f_{\rm col}$.  Inspection of Table \ref{tab:kerrbb} indicates that the \texttt{diskpn} model requires a factor of $\sim 2.1$ increase in $f_{\rm col}$ for an order of magnitude decrease in $\dot{M}$ for both the Schwarzschild and maximal Kerr black holes.  This result is fully consistent with the `true' $f_{\rm col}$ evolution and shows that relative measurements of $f_{\rm col}$ made with Newtonian disc models are reliable, even when the black hole under consideration is maximally spinning.

% TABLE 2
\begin{table}
\addtolength{\tabcolsep}{-2.5pt}
\centering
\begin{tabular}{l l l l l l l}
\hline
\hline
$a_{\ast}$ & $\dot{M}$ & $T_{\rm col, max}$ & $K_{\rm pn}$ & $f_{\rm col, true}$ & $f_{\rm col}$ & $\chi^{2}_{\nu}$ \\
 & ($\dot{M}_{\rm Edd}$) & (keV) & & & & \\
\hline
0 & 0.05 & 0.416(2) & 0.177(8) & 1.4 & 1.39(2) & 1.35 \\
0 & 0.005 & 0.493(5) & 0.0098(8) & 3.0 & 2.87(6) & 1.26 \\
0.998 & 0.01 & 0.455(5) & 0.015(2) & 1.4 & 2.58(9) & 0.87 \\
0.998 & 0.001 & 0.56(1) & 0.0006(1) & 3.0 & 5.8(2) & 0.71 \\
\hline
\end{tabular}
\caption{Results from fitting simulated spectra generated from the \texttt{kerrbb} model of a Kerr black hole with the pseudo-Newtonian model, \texttt{diskpn}, where the inner disc radius was set to $6~R_{\rm g}$.  The first two columns are the black hole spin and mass accretion rate in Eddington units used to generate the simulated disc spectra.  Proceeding from {\it left} to {\it right}, the columns are the best-fit maximum disc colour temperature, best-fit disc normalisation, colour correction factor used to generate the disc spectrum, colour correction factor recovered from the best-fit normalisation, and reduced $\chi^{2}$.  All models had 15 degrees of freedom.  Uncertainties on the last significant digit are given in parentheses and correspond to the 90\% level.}
\label{tab:kerrbb}
\end{table}

%===========================================================================
\subsection{Disc Temperature and Luminosity}
The luminosity of a geometrically thin, optically thick disc is theoretically expected to obey a relation of the form $L_{\rm disc} = \sigma A T_{\rm eff}^{4}$, where $A$ is the radiating surface area.  The disc luminosity depends on both $A$, which scales as $R_{\rm in}^{2}$, and effective temperature, which depends on $f_{\rm col}$.  The purpose of this section is to explore the $L_{\rm disc}-R_{\rm in}^{2}T_{\rm eff}^{4}$ relation, particularly extended to low luminosities (i.e., the HIS and LHS), and the effect of introducing the possibility of a variable colour correction factor.

% FIGURE 11
\begin{figure}
\begin{center}
\includegraphics[width=84mm]{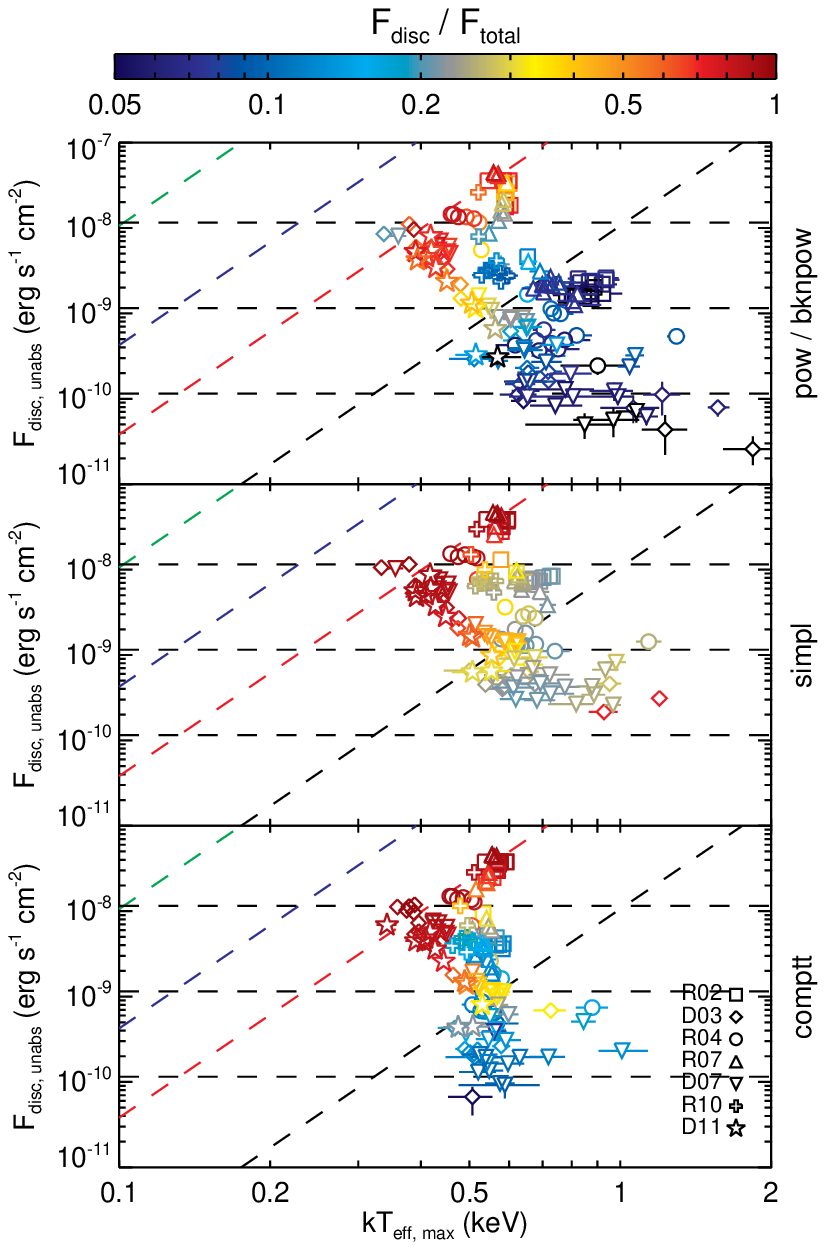}
\end{center}
\caption{Unabsorbed disc flux (0.1-10 keV) vs. maximum disc effective temperature ($T_{\rm eff, max} = T_{\rm col, max} / f_{\rm col}$) for all transitions, where $T_{\rm col, max}$ is measured using \texttt{ezdiskbb} to model the disc component and a constant $f_{\rm col} = 1.4$ is assumed.   Comptonisation is modelled using the \texttt{XSPEC} models \texttt{pow/bknpow} ({\it top panel}), \texttt{simpl} ({\it center panel}), and \texttt{comptt} ({\it bottom panel}).  From {\it top} to {\it bottom}, the {\it horizontal dashed lines} indicate $L_{\rm disc}/L_{\rm Edd} = 10^{-1}, 10^{-2}, 10^{-3}$, where $L_{\rm disc}$ was converted to $F_{\rm disc}$ using Equation \ref{eq:Fdisc} with $D = 8~{\rm kpc}$.  The Eddington luminosity is given by, $L_{\rm Edd} = 4 \pi G M m_{\rm p} c / \sigma_{\rm T}$, where $m_{\rm p}$ is the proton mass, $\sigma_{T}$ is the Thomson cross section, and we adopt $M = 7~{\rm M_{\odot}}$.  Uncertainties in $M$ and $D$ are not incorporated because the shape of the $F_{\rm disc}-T_{\rm eff, max}$ relation is insensitive to these intrinsic parameters.  The {\it diagonal dashed lines} represent the expected $F_{\rm disc} \propto T_{\rm eff,max}^{4}$ relation for an optically thick, geometrically thin, zero-torque disc (Equation \ref{eq:Ldisc}) assuming the aforementioned intrinsic parameters and constant $R_{\rm in} = 1~R_{\rm g}$ ({\it black line}), $6~R_{\rm g}$ ({\it red line}), $20~R_{\rm g}$ ({\it blue line}), and $100~R_{\rm g}$ ({\it green line}).  Glyph colours denote the disc fraction of each observation.  The key in the {\it bottom right} identifies the glyph associated with each transition ID.  When a constant $f_{\rm col}$ is assumed, departure from the $L_{\rm disc}-T_{\rm eff}^{4}$ is evident for low-luminosity discs characteristic of the HIS and LHS.}
\label{fig:FT4}
\end{figure}

For the \texttt{ezdiskbb} disc model of a Newtonian disc with a zero-torque inner boundary condition, \citet{Zimmerman2005} finds the disc luminosity is given by,
\begin{equation}
L_{\rm disc} = 73.9 \sigma \left( \frac{T_{\rm col, max}}{f_{\rm col}} \right)^{4} R_{\rm in}^{2}. \label{eq:Ldisc}
\end{equation}
Observationally, we measure an unabsorbed disc flux,
\begin{equation}
F_{\rm disc} = \frac{L_{\rm disc}}{4 \pi D^{2}}. \label{eq:Fdisc}
\end{equation}
In practice, $F_{\rm disc}$ is obtained by operating the \texttt{XSPEC} \texttt{cflux} command on the disc component with the normalisation frozen to its best-fit value.  Unabsorbed disc fluxes and unabsorbed total model fluxes, $F_{\rm total}$, are calculated over 0.1-10 keV and 0.1-100 keV, respectively.  Figure \ref{fig:FT4} shows the relation between unabsorbed disc flux and maximum effective temperature for each of the three Comptonisation models \texttt{pow/bknpow}, \texttt{comptt}, and \texttt{simpl}.  $T_{\rm col, max}$ is measured from fits to all observations with the disc component modelled by \texttt{ezdiskbb} and $T_{\rm eff, max} = T_{\rm col, max} / f_{\rm col}$, assuming a constant $f_{\rm col} = 1.4$.  The disc fractions, defined by Equation \ref{eq:DF}, are given by the glyph colours.  We note that this work includes observations with very low disc fractions compared to previous investigations of the luminosity-temperature relation for BHBs, which tend to be restricted to the HSS where $DF \gtrsim 0.8$ \citep[e.g.,][]{GierlinskiDone2004}.

If one assumes a constant disk geometry, the disk emitting area is unchanging and we expect to recover the scaling, $F_{\rm disc} \propto T_{\rm eff}^{4}$, denoted by the diagonal lines in Figure \ref{fig:FT4} for various choices of $R_{\rm in}$ and $f_{\rm col}$.  While the theoretical relation is achieved for disc luminosities, $L_{\rm disc} \gtrsim 0.05 L_{\rm Edd}$, clear departure is evident below this threshold where the BHB is categorized in the HIS or LHS.

The \texttt{diskbb} and \texttt{ezdiskbb} models are purely Newtonian and correction factors accounting for various general relativistic and viewing angle effects \citep{Cunningham1975, Zhang1997} have been applied in previous observational studies of the luminosity-temperature relation for BHBs \citep[e.g.,][]{GierlinskiDone2004, Dunn2011}.  Selecting the proper correction factors requires knowledge of the disc inclination and spin, for which there is considerable debate for GX 339--4 (see \S \ref{sec:gx339}).  Fortunately, the correction factors are independent of flux and temperature; therefore, we can safely neglect incorporating these modifications without altering the shape of the $F_{\rm disc}-T_{\rm eff}$ diagram, irrespective of the accuracy of the $F_{\rm disc}$ and $T_{\rm eff}$ values.

% FIGURE 12
\begin{figure}
\begin{center}
\includegraphics[width=84mm]{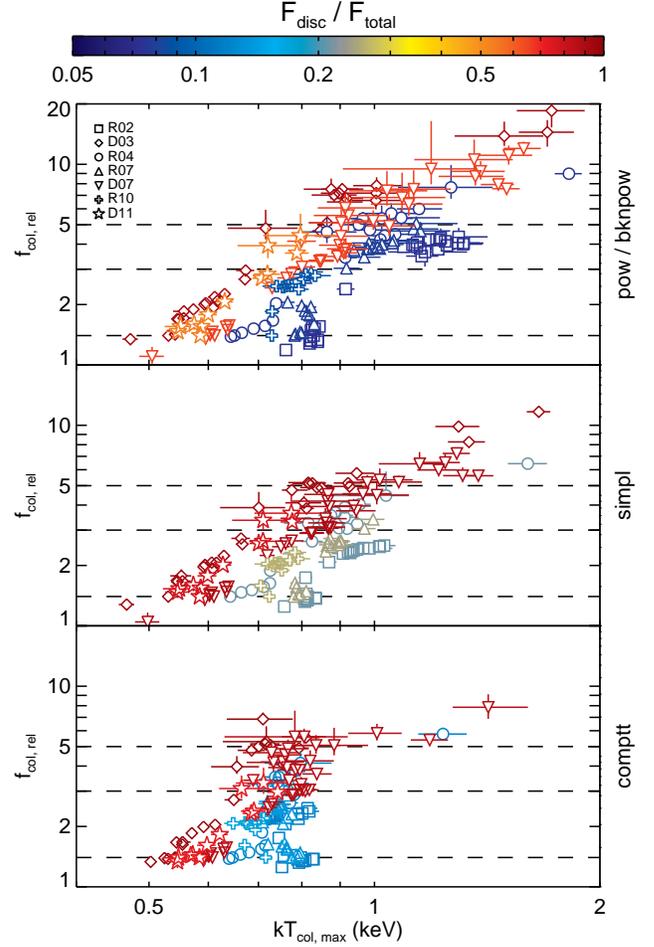}
\end{center}
\caption{Relative measurements of the colour correction factor shown against the maximum colour temperature of the disc from best-fits using the \texttt{XSPEC} Comptonisation models \texttt{pow/bknpow} ({\it top panel}), \texttt{simpl} ({\it center panel}), and \texttt{comptt} ({\it bottom panel}).  Results from all transitions are shown together, with the disc fraction given by the glyph colour.  The key in the {\it upper left} portion of the {\it top panel} identifies the glyph associated with each transition ID.  $T_{\rm col, max}$ is obtained from the best-fit \texttt{ezdiskbb} parameter and $f_{\rm col, rel}$ is computed using the same procedure as in Figure \ref{fig:fcol_rel} (see text).  Horizontal {\it dashed lines} mark $f_{\rm col, rel} = 1.4, 3, 5$.  For a disc inner radius fixed at $6 R_{\rm g}$, the majority of observations yield $1.4 \lesssim f_{\rm col, rel} \lesssim 5$ and show a roughly linear relation between $f_{\rm col}$ and $T_{\rm col}$.}
\label{fig:fcol_Tcol}
\end{figure}

% FIGURE 13
\begin{figure}
\begin{center}
\includegraphics[width=84mm]{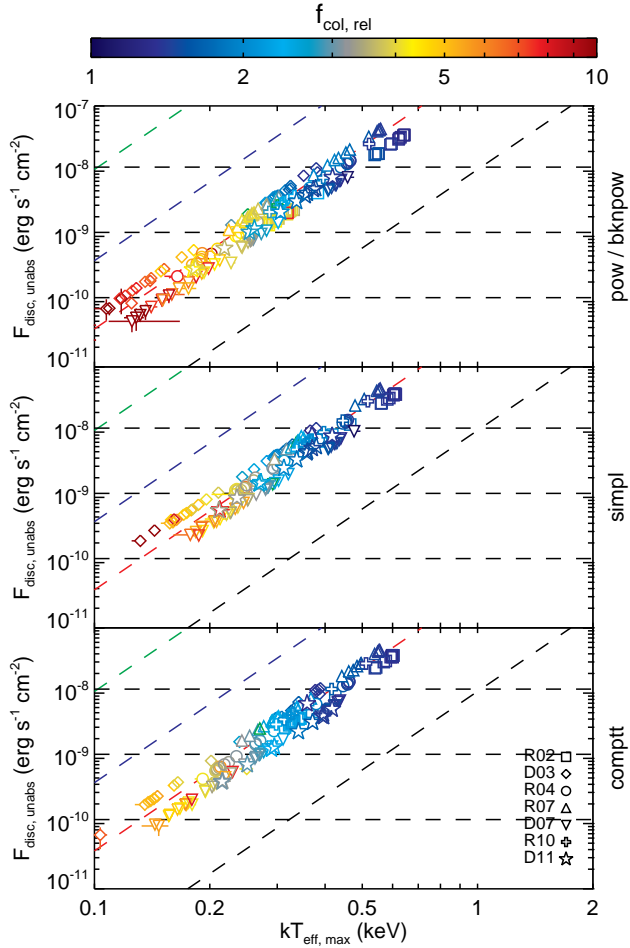}
\end{center}
\caption{Unabsorbed disc flux (0.1-10 keV) vs. maximum disc effective temperature for all transitions, where $T_{\rm col, max}$ is measured using \texttt{ezdiskbb}.  Relaxing the assumption of a constant $f_{\rm col}$, all observations were re-fitted with \texttt{ezdiskbb} replaced by \texttt{diskpn} with $R_{\rm in} = 6~R_{\rm g}$.  Relative values of $f_{\rm col}$ were computed from the best-fit \texttt{diskpn} normalisation, using the same procedure as in Figure \ref{fig:fcol_rel}.  The abscissa shows $T_{\rm eff, max} = T_{\rm col, max} / f_{\rm col, rel}$.  Glyph colours correspond to the $f_{\rm col, rel}$ of each observation.  The various lines are the same as those in Figure \ref{fig:FT4}.  The $L_{\rm disc} \propto T_{\rm eff}^{4}$ relation is recovered for disc luminosities, $L_{\rm disc} \gtrsim 10^{-3}~L_{\rm Edd}$, when $f_{\rm col}$ is considered to be variable.}
\label{fig:FT4_fcol}
\end{figure}

% FIGURE 14
\begin{figure}
\begin{center}
\includegraphics[width=84mm]{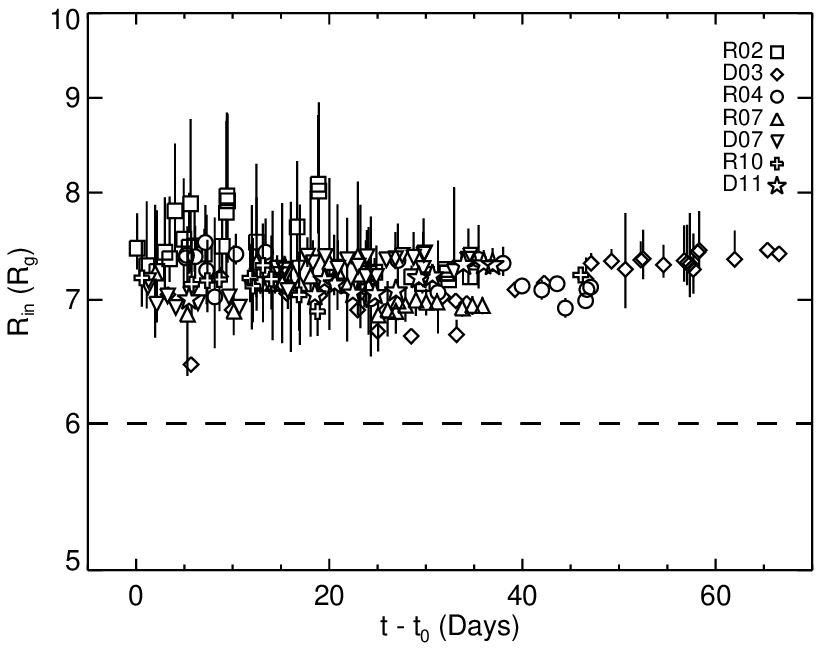}
\end{center}
\caption{Time evolution, as measured from the start of the transition listed in Table \ref{tab:transitions}, of the inner disc radius computed from the \texttt{ezdiskbb} and \texttt{diskpn} normalisations (Equation \ref{eq:KezKpn}) when $R_{\rm in}$ was fixed in the \texttt{diskpn} model at $6~R_{\rm g}$.  The {\it dashed line} marks $R_{\rm in} = 6~R_{\rm g}$.  The key in the {\it upper right} corner denotes the glyphs corresponding to each transition.  The constancy of $R_{\rm in}$ implies that relative measurements made from combining results from fits made with \texttt{ezdiskbb} and \texttt{diskpn} are unaffected by the minor differences between those disc models.}
\label{fig:fcol_Rin}
\end{figure}

The effective temperature plotted in Figure \ref{fig:FT4} is calculated under the assumption of a constant colour correction factor, which will subsequently be relaxed.  Deriving relative measurements for the colour correction factor according to Equation \ref{eq:fcol_rel} and as described in \S \ref{sec:fcol}, Figure \ref{fig:fcol_Tcol} shows the relationship between $f_{\rm col}$ and the maximum colour temperature, which is a parameter in the disc models.  Supposing that the inner disc remains fixed at $R_{\rm in} = 6~R_{\rm g}$, which is the implicit assumption behind all of our $f_{\rm col, rel}$ measurements, the colour correction factor increases with colour temperature approximately linearly.  The majority of observations yield colour corrections that fall within the range $1.4 \lesssim f_{\rm col, rel} \lesssim 5$, in agreement with \citet{ReynoldsMiller2011}.  Notably, the models \texttt{simpl} and \texttt{comptt}, which are rooted in physical considerations of Comptonisation, constrict the range of $f_{\rm col, rel}$ as compared to the phenomenological and widely used \texttt{powerlaw} and \texttt{bknpower} models.  Particularly for best-fit colour temperatures, $T_{\rm col, max} \gtrsim 1~{\rm keV}$, the \texttt{powerlaw} and \texttt{bknpower} continuum models may return erroneously low values for the disc normalisation.

We attempt to explore whether the luminosity-temperature relation can be extended to lower luminosities characteristic of the HIS and LHS by allowing for a variable $f_{\rm col}$.  Figure \ref{fig:FT4_fcol} shows the $F_{\rm disc}-T_{\rm eff}$ diagram with the only difference to Figure \ref{fig:FT4} being that the effective temperature is now calculated by, $T_{\rm eff, max} = T_{\rm col, max} / f_{\rm col, rel}$.  In other words, Figure \ref{fig:FT4} is recomputed using the $f_{\rm col, rel}$ values of Figure \ref{fig:fcol_Tcol} in place of the constant $f_{\rm col} = 1.4$.  Accepting the dynamic range in $f_{\rm col}$ suggested by Figure \ref{fig:fcol_Tcol}, $F_{\rm disc} \propto T_{\rm eff}^{4}$ is recovered for disc luminosities, $L_{\rm disc} \gtrsim 10^{-3}~L_{\rm Edd}$.  Furthermore, the relation falls neatly on the expected track corresponding to the emitting surface area remaining constant with the inner disc located at $R_{\rm in} = 6~R_{\rm g}$.

Caution must be exercised when combining different disc models.  The relative measurements $f_{\rm col, rel}$ used to recompute $T_{\rm eff, max}$ in Figure \ref{fig:FT4_fcol} called upon \texttt{diskpn} and \texttt{ezdiskbb} separately to model the disc component in order to take advantage of the strengths of each model for our purposes.  One may reasonably ask if these two disc models, which both adopt the zero-torque inner boundary condition, are consistent.  The significant difference between the models is that \texttt{diskpn} approximates the general relativistic effects on the accretion disc with a pseudo-Newtonian potential, while \texttt{ezdiskbb} is strictly Newtonian; therefore, discrepancies between the two may materialize in the spectral fits.  Combining the normalisations from each disc model, given by Equations \ref{eq:Kez} and \ref{eq:Kpn}, the inner disc radius is obtained,
\begin{equation}
\frac{R_{\rm in}}{R_{\rm g}} = 0.0677 \left( \frac{K_{\rm ez}}{K_{\rm pn}} \right)^{1/2}. \label{eq:KezKpn}
\end{equation}
Since $R_{\rm in}$ is a separate parameter in \texttt{diskpn} instead of being incorporated into the normalisation, we expect to exactly recover this value for $R_{\rm in}$ from Equation \ref{eq:KezKpn} if \texttt{diskpn} and \texttt{ezdiskbb} are identical.  Figure \ref{fig:fcol_Rin} shows $R_{\rm in}$ computed from Equation \ref{eq:KezKpn} for fits to all observations where $R_{\rm in} = 6~R_{\rm g}$ in the \texttt{diskpn} fits.  Values of $R_{\rm in}$ falling on $6~R_{\rm g}$ would indicate that the inner radius enforced in the \texttt{diskpn} fits was exactly recovered in the independent \texttt{ezdiskbb} fits, while deviations from $6~R_{\rm g}$ arise from differences in the \texttt{ezdiskbb} and \texttt{diskpn} models.  Although the inner disc radius value of $6~R_{\rm g}$ set in \texttt{diskpn} is not exactly recovered, $R_{\rm in}$ remains constant throughout each transition at $\sim 7.5~R_{\rm g}$, merely a systematic offset that does not affect the interpretation of Figure \ref{fig:FT4_fcol}.

%===========================================================================
\subsection{Low Frequency Quasi-Periodic Oscillations}
LFQPOs are observed in the temporal power spectra of BHBs and remain a long-standing puzzle of active observational and theoretical interest.  Explanations for the origin of LFQPOs appeal to an exhaustive array of production models including trapped waves \citep{Kato1990}, discoseismic modes \citep{NowakWagoner1991, NowakWagoner1992, NowakWagoner1993, ReynoldsMiller2009}, Lense-Thirring precession \citep{Ipser1996, Stella1999, Ingram2009}, accretion-ejection instability \citep{TaggerPellat1999, Varniere2002, VarniereTagger2002}, and a truncated disc LHS geometry \citep{GianniosSpruit2004}.  Recent simulations of global, magnetised accretion discs suggest that LFQPOs have an origin in the corona rather than the disc directly \citep{ONeill2011}.  While the origin of LFQPOs is not firmly established, the details of the disc geometry may provide clues to their nature.

% FIGURE 15
\begin{figure}
\begin{center}
\includegraphics[width=84mm]{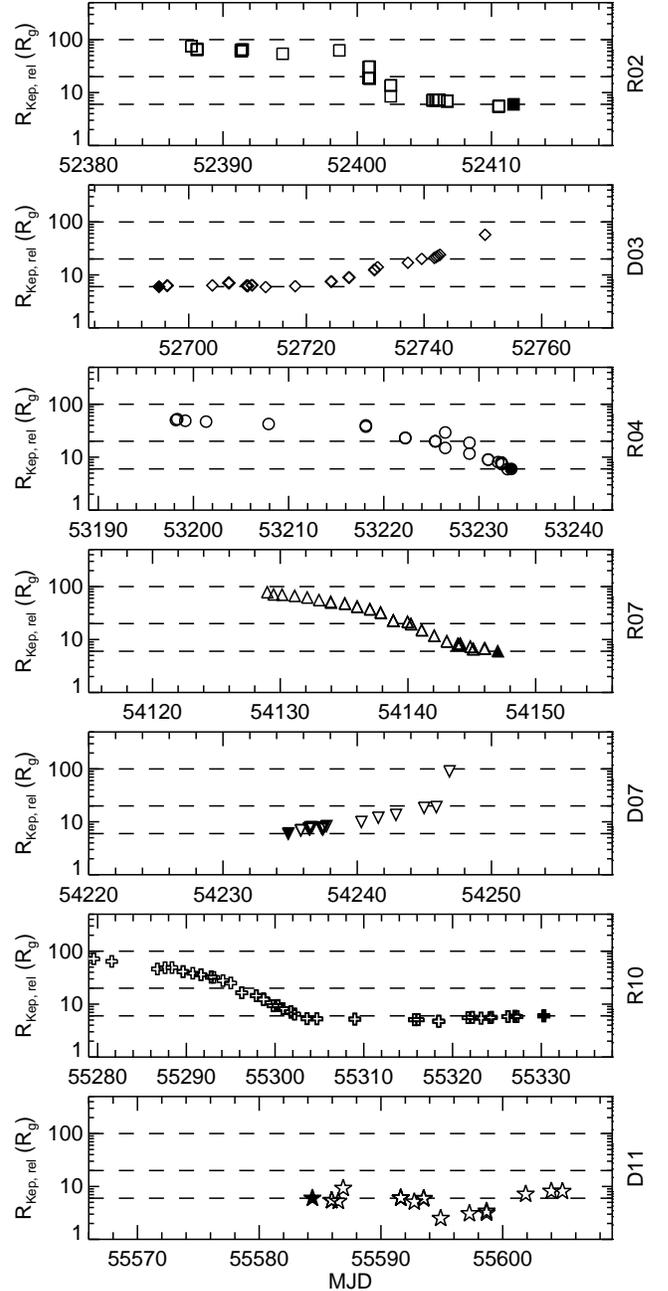}
\end{center}
\caption{Relative measurements of the evolution of $R_{\rm Kep, rel}$ for each transition (see text for definition) derived from LFQPO frequency evolution.  The {\it dashed lines} mark $R_{\rm in} = 6, 20, 100~R_{\rm g}$.  The $R_{\rm Kep, rel}$ measurements for each transition were made relative to the filled glyphs and normalized to $R_{\rm in} = 6~R_{\rm g}$, which is appropriate for a Schwarzschild black hole in the HSS.}
\label{fig:qpo_rel}
\end{figure}

X-ray spectra alone cannot distinguish between a truncated disc and changing disc structure/coronal activity as the underlying accretion geometry.  Given a plausible production mechanism, LFQPOs have the potential to be diagnostics of the inner accretion flow.  In the LFQPO excitation model envisioned by \citet{GianniosSpruit2004}, quasi-spherical oscillations of the RIAF interior to a truncated disc in the LHS produce modes with frequencies near the Keplerian frequency of the inner disc,
\begin{equation}
\frac{R_{\rm Kep}}{R_{\rm g}} = \left( \frac{G M}{c^{3}} 2 \pi \nu_{\rm QPO} \right)^{-2/3}, \label{eq:RKep}
\end{equation}
where $\nu_{\rm QPO}$ is the frequency of the LFQPO.  In this framework, \citet{GianniosSpruit2004} require that the RIAF, and hence the recessed inner disc, extend out to $R_{\rm in} \sim 100~R_{\rm g}$.

LFQPOs observed in GX 339--4 and their frequency migration over the course of the state transitions studied here has received considerable attention in the literature.  The frequency of LFQPOs is observed to decrease (increase) over a decay (rise) transition with a typical range $0.1~{\rm Hz} \lesssim \nu_{\rm QPO} \lesssim 10~{\rm Hz}$, implying that accretion rate and QPO frequency are intricately linked.  In an attempt to test the truncated disc LFQPO production model, which predicts that $R_{\rm Kep} \simeq R_{\rm in} \gtrsim 100~R_{\rm g}$ in the LHS, we collect measured $\nu_{\rm QPO}$ values for all of the GX 339--4 state transitions considered \citep{Belloni2005, ShaposhnikovTitarchuk2009, Stiele2011, Motta2011, Nandi2012}.  Given the uncertainty in black hole mass for GX 339--4, we seek relative measurements of $R_{\rm Kep}$ normalized to the ISCO for a non-spinning black hole in the HSS,
\begin{equation}
R_{\rm Kep, rel} \equiv 6 R_{\rm g} \left( \frac{R_{\rm Kep}}{R_{\rm Kep, HSS}} \right), \label{eq:RKep_rel}
\end{equation}
where $R_{\rm Kep, HSS}$ is the Keplerian radius computed for a given transition from the $\nu_{\rm QPO}$ of the observation nearest to the HSS in time using Equation \ref{eq:RKep}, without specifying $M$.  Supposing that $R_{\rm Kep}$ is associated with $R_{\rm in}$, Figure \ref{fig:qpo_rel} shows the evolution in disc inner radius implied by the observed migrating LFQPOs for each transition.  With the exception of transition D11, the Keplerian frequencies implied by the LFQPOs reach out to $\sim 50-100~R_{\rm g}$ in the LHS, which is broadly consistent with the degree of recession required by the RIAF/truncated disc LFQPO excitation model.  If \mbox{LFQPOs} are associated with the inner disc in a similar manner to that described by \citet{GianniosSpruit2004}, then a truncated disc geometry may be a plausible description for state transitions.  However, the spectral fits in this work suggest that $R_{\rm in}$ does not truncate to such extreme disc locations, which would demand reconciling.

%===========================================================================
% DISCUSSION
\section{Discussion}
\label{section:discussion}
We have shown that a moderately increasing $f_{\rm col}$ as the disc fraction decreases can adequately and consistently describe the spectral evolution of the accretion disc in GX 339--4 over multiple state transitions.  The purpose of this work is {\it not} to debunk a truncated disc geometry as a viable model for the LHS of BHBs.  We recognize that the observations in this work interpreted with simplistic disc models are unable to distinguish between changes in $R_{\rm in}$ and/or $f_{\rm col}$.  Rather, we demonstrate that reasonable variations in $f_{\rm col}$ provide an alternative explanation to disc truncation for accretion disc evolution.  Motivated by this result and supposing that $f_{\rm col}$ is variable, we consider the impacts that this would have on the interpretation of BHB observations, ultraluminous X-ray sources, and models of the accretion flow across spectral states.  We then discuss the physically expected range of attainable values for $f_{\rm col}$ and conclude the discussion with a cautionary commentary on the widely used disc models.

\subsection{Implications of a Variable Colour Correction}
Imposing an unchanging inner accretion disc radius located at the ISCO of a Schwarzschild black hole, we measured the relative changes in $f_{\rm col}$ necessary to adequately fit {\it RXTE} spectra of GX 339--4 over seven state transitions.  Arbitrarily selecting a normalisation of $f_{\rm col} = 1.4$ for the HSS and denoting measurements of $f_{\rm col}$ relative to this value by $f_{\rm col, rel}$ (see Equation \ref{eq:fcol_rel}), we tracked the evolution of $f_{\rm col, rel}$ for each transition (see Figures \ref{fig:fcol_rel}-\ref{fig:fcol_rel_simpl}).  Transitions R02, R07, R10, and D11 require colour corrections conforming to the range $1.4 \lesssim f_{\rm col, rel} \lesssim 3$, while transition R04 demands the range be extended to $1.4 \lesssim f_{\rm col, rel} \lesssim 5$.  These ranges in $f_{\rm col}$ evolution, which can account for the majority of the transitions, correspond to a factor of $\sim 2.0-3.5$ increase in $f_{\rm col}$ for a HSS $\rightarrow$ LHS transition.  Transitions D03 and D07 require more extreme evolution in $f_{\rm col, rel}$, with values as high as $f_{\rm col, rel} \sim 10$.  Our main finding is to demonstrate that an evolving $f_{\rm col}$ provides an alternative and adequate description of the changing disc spectrum during a state transition, which is perhaps a more attractive scenario to imagine than the sudden evacuation or replenishment of the inner accretion flow.

\subsubsection{Recovering $L-T_{\rm eff}^{4}$ at Low Luminosities}
A blackbody source of constant emitting area is expected to obey the $L-T_{\rm eff}^{4}$ scaling, which relates the source luminosity to its effective temperature.  Accretion flows corresponding to the HSS of BHBs, where the thermal accretion disc dominates the total flux and the inner disc location does not deviate from the ISCO, nicely follow this expected relation \citep[e.g.,][]{GierlinskiDone2004}; however, deviations arise in the HIS and LHS \citep[e.g.,][]{Dunn2011}.  This work encompasses the disc luminosity range, $10^{-3} \lesssim L_{\rm disc}/L_{\rm Edd} \lesssim 0.5$, finding that a $f_{\rm col}$ that evolves with spectral state can reproduce the theoretical $L_{\rm disc}-T_{\rm eff}^{4}$ across all spectral states.  Studying the disc-dominated states (i.e., $DF \gtrsim 0.8$) of many BHBs over a typical luminosity range, $0.01 \lesssim L_{\rm disc} / L_{\rm Edd} \lesssim 0.5$, \citet{GierlinskiDone2004} found that the $L_{\rm disc}-T_{\rm eff}^{4}$ relation holds for a {\it constant} $f_{\rm col}$.  We observe similar and consistent behaviour for GX 339--4 in the HSS.  As the disc fraction drops and the source enters the HIS or LHS, significant departures from $L_{\rm disc}-T_{\rm eff}^{4}$ are apparent in Figure \ref{fig:FT4}, in agreement with \citet{Dunn2011}.  These `spurs' on the luminosity-temperature diagram can be interpreted as either a decrease in the inner disc radius, and hence the emitting area, or the result of neglecting to account for a decrease in $T_{\rm eff} = T_{\rm col} / f_{\rm col}$ caused by an increased $f_{\rm col}$ in the LHS.  Of course, some combination of these two effects is possible as well.  \citet{Dunn2011} measured the incremental degree of change in $f_{\rm col}$ from a canonical HSS value that was required to return the spurs to the expected $L_{\rm disc} \propto T_{\rm eff}^{4}$ law for a constant emitting area (see their Figure A.1).  For GX 339--4, they found that fairly modest changes in $f_{\rm col}$, consistent with the relative changes we determined, could restore the theoretical relation; however, they argue that the most severe departures from $L_{\rm disc}-T_{\rm eff}^{4}$, which occur for the lowest disc fractions, are not likely the result of a variable $f_{\rm col}$.  Here, we suggest that the physically realizable range of $f_{\rm col}$ is perhaps not as narrow as commonly assumed (see \S \ref{sec:fcol_range}) and that a thin disc geometry with a fixed inner radius and evolving $f_{\rm col}$ is a viable phenomenological model for BHB state transitions.

We expand on the work of \citet{Dunn2011} by breaking the observations of a BHB into individual transitions, exploiting relative measurements, and exploring empirical and physical Comptonisation models.  Comparing Figures \ref{fig:fcol_rel}-\ref{fig:fcol_rel_simpl}, we find that employing physically motivated Comptonisation models that avoid the `flux stealing' phenomenon from the disc component require less severe $f_{\rm col}$ evolution for any given transition.   Inspection of the panels of Figure \ref{fig:FT4_fcol} shows that the $L_{\rm disc}-T_{\rm eff}^{4}$ relation is recovered when a changing $f_{\rm col}$ is incorporated, regardless of the choice of Comptonisation model.  Relative measurements of $f_{\rm col}$ broken down by transition in Figures \ref{fig:fcol_rel}-\ref{fig:fcol_rel_simpl} hint at an underlying difference in the details of the accretion flow depending on whether the source is in the rise or decay stage of outburst.  Qualitatively, the decay transitions demand more extreme and perhaps physically unattainable $f_{\rm col}$ evolution, while the rise transitions fit more comfortably into the idea of a moderately variable $f_{\rm col}$.  Given that decoupling the true behaviour of the inner disc from that of the colour correction factor is impossible with simplistic disc models and {\it RXTE} data, we leave the possibility of different accretion scenarios between rise and decay transitions as pure speculation.

\subsubsection{The Ultraluminous State}
\label{sec:ulx}
Adopting a variable colour correction factor parameterises the various physical processes in BHB accretion into a single factor.  This work suggests that moderate changes in $f_{\rm col}$ may conceivably account for the wide range of spectral states observed when a BHB undergoes a state transition.  Recently, a new spectral class, dubbed the ultraluminous state (ULS), has been postulated for the class of ultraluminous X-ray sources (ULXs) \citep{Gladstone2009}.  Here, we explore whether the ULS, which has never been observed in a confirmed BHB, is consistent with the interpretation of spectral states arising from physical processes that are observationally manifested as an evolving colour correction factor.

ULXs are X-ray sources that exceed the Eddington luminosity of stellar mass black holes ($L_{\rm X} \gtrsim 10^{39}~{\rm erg~s^{-1}}$), but are not spatially coincident with galactic nuclei; therefore, ULXs are distinct from AGNs.  Two possible explanations for ULXs are: (1) intermediate mass black holes (IMBHs; $M \sim 10^{2} - 10^{4}~M_{\odot}$) accreting at sub-Eddington rates, but the production rate of binaries composed of an IMBH and donor star is likely too low to account for the number of observed ULXs \citep{Madhusudhan2006}; and/or (2) stellar mass black holes accreting at super-Eddington rates, which requires a mechanism for achieving super-Eddington accretion, such as beaming/viewing effects of outflows \citep[e.g.,][]{King2001, BegelmanKingPringle2006} or disc inhomogeneities \citep{Begelman2001, Begelman2002, Begelman2006}.

The spectral characteristics of ULXs were studied by \citet{Gladstone2009}, leading them to propose that ULXs are super-Eddington accreting stellar mass black holes with ubiquitous spectral features.  These commonalities suggest that ULXs are indeed a spectral class of BHBs, dubbed the ultraluminous state (ULS); however, this accretion state does not appear to arise in the well-studied Galactic BHBs.  The ULS is characterised by both a broadened, hardened thermal disc component and a break in the power-law component at lower energies ($\gtrsim 3~{\rm keV}$) than typically observed in less luminous BHB states.  The \citet{Gladstone2009} ULX sample was adequately fitted with a physically motivated disc plus Comptonising corona model, where the best-fit models called for optically thick coronae ($\tau_{\rm p} \sim 5-30$) and cool disc colour temperatures ($T_{\rm col, max} \sim 0.2~{\rm keV}$).  Physically, the broadened/hardened disc and low-energy spectral break may be attributed to a complicated inner geometry where an optically thick corona obscures and alters the energetics of the inner disc.  Normally, disc temperatures this low would imply a black hole of mass $\sim 10^{3}~M_{\odot}$ \citep[e.g.,][]{Kaaret2003}; however, the measured disc temperature may be lowered if a significant fraction of the accretion energy is dissipated in the corona \citep[e.g.,][]{SvenssonZdziarski1994}.  Based on the requirement of an optically thick corona, as opposed to the $\tau_{\rm p} \sim 1$ coronae typically observed in BHBs, sub-Eddingtion accretion onto an IMBH was ruled out.

% FIGURE 16
\begin{figure}
\begin{center}
\includegraphics[width=84mm]{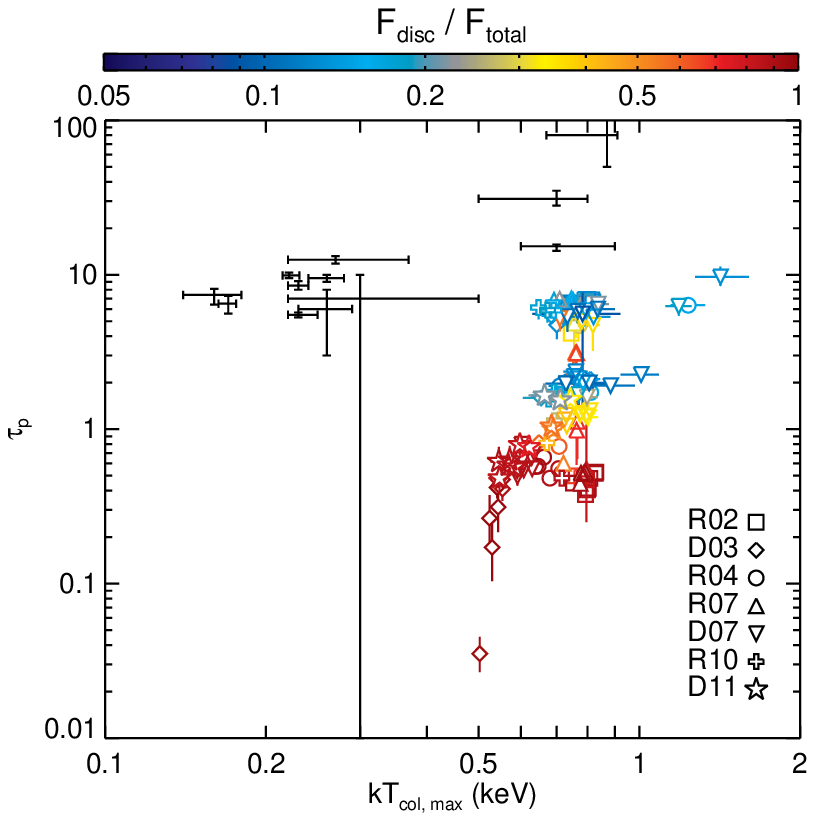}
\end{center}
\caption{Coronal optical depth vs. maximum disc colour temperature for all transitions, where $\tau_{\rm p}$ and $T_{\rm col, max}$ are measured using \texttt{comptt} and \texttt{diskpn} with $R_{\rm in} = 6~R_{\rm g}$, respectively.  The disc fraction is given by the glyph colour.  The results from a sample of 12 ULXs (Table 8 of \citet{Gladstone2009}) are overlaid in {\it black} {\it crosshairs}.  The HIS/LHS GX 339--4 {\it RXTE} data analysed in this work permit moderately optically thick coronae in the same range as found for ULXs.}
\label{fig:taup_Tcolmax}
\end{figure}

Figure \ref{fig:taup_Tcolmax} shows the coronal optical depth versus the maximum disc colour temperature for GX 339--4 determined from the best-fits to the disc plus Comptonising corona \texttt{XSPEC} model,  \texttt{phabs$\times$(comptt+diskpn)}, with $R_{\rm in} = 6~R_{\rm g}$.  Replacing the disc component with either \texttt{diskbb} or \texttt{ezdiskbb} produces essentially indistinguishable results.  For comparison, the results from Table 8 of \citet{Gladstone2009} are shown for a sample of 12 ULXs determined from fits to high-quality {\it XMM-Newton} data, where \texttt{diskpn} and \texttt{comptt} were also adopted as the disc and Comptonising models, respectively.\footnote{\citet{Gladstone2009} do not specify the $R_{\rm in}$ value used in the \texttt{diskpn} model.  Our results suggest that the coronal properties and disc temperature are largely unchanged in the spectral fits, regardless of what one chooses for $R_{\rm in}$.}  When GX 339--4 is in the HIS/LHS (i.e., low disc fractions), the typical best-fit coronal optical depth falls in the range, $\tau_{\rm p} \sim 2-10$.  The coronal optical depth of the ULX sample is described by $\tau_{\rm p} \sim 5-15$ ($\tau_{\rm p} \sim 10-100$) for discs with temperatures $T_{\rm col, max} \lesssim 0.5~{\rm keV}$ ($T_{\rm col, max} \gtrsim 0.5~{\rm keV}$).  Figure \ref{fig:taup_Tcolmax} challenges the claim that spectral fits to sub-Eddington BHBs do not permit moderately optically thick coronae.

Although the data are consistent with an optically thick corona model, here we point out a strong parameter degeneracy between the colour correction factor and the coronal optical depth.  Of the three models we considered to represent Comptonisation, all of which provided equally good fits to the data, only \texttt{comptt} incorporates the coronal temperature and optical depth as physical parameters.  Inspection of Figure \ref{fig:fcol_Tcol} reveals that the \texttt{pow/bkn} and \texttt{simpl} models require higher colour correction factors than \texttt{comptt}.  Particularly for the HIS/LHS observations of GX 339--4, introducing a physical corona drives $f_{\rm col}$ to lower values with preference for an optically thick corona (see Figure \ref{fig:fcol_Tcol}).  Therefore, the data cannot distinguish between an increased colour correction factor and an optically thick corona.

\begin{figure}
\begin{center}
\includegraphics[width=84mm]{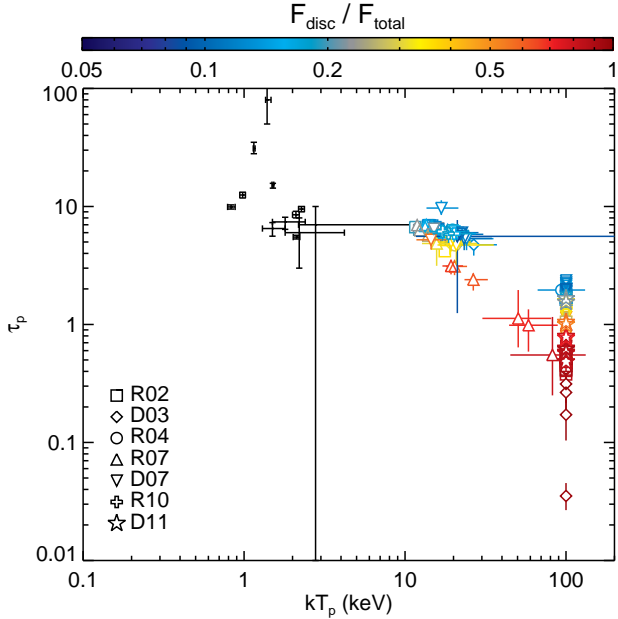}
\end{center}
\caption{Coronal optical depth vs. coronal temperature, both measured using \texttt{comptt}.  The disk model was \texttt{diskpn} with $R_{\rm in} = 6~R_{\rm g}$.  The disc fraction is given by the glyph colour.  The results from a sample of 12 ULXs (Table 8 of \citet{Gladstone2009}) are overlaid in {\it black} {\it crosshairs}.}
\label{fig:taup_Tp}
\end{figure}

In the IMBH model, ULXs are massive scaled up versions of BHBs in the LHS (i.e., IMBHs accreting at sub-Eddington rates).  The IMBH model for ULXs was discarded by \citet{Gladstone2009} because the measured coronal optical depths were larger than commonly observed in BHBs.  In the case of GX 339--4 in the HIS/LHS, using the same disc plus Comptonisation model, we see that the best-fit coronal optical depths lie precisely in the same moderately optically thick range as the ULX sample of \citet{Gladstone2009}.  Therefore, the data {\it do} permit optically thick coronae in the HIS/LHS of GX 339--4, but we stress that the measurement of coronal optical depth from basic disc + corona modelling cannot be made definitive due to the degeneracy between $\tau_{\rm p}$ and $f_{\rm col}$.  This possibility alleviates the hesitation for supposing that ULXs are analogous to scaled up BHBs on the grounds of coronal optical depth.  Allowing for a variable colour correction factor, the broadened/hardened ULX disc spectra and low measured ULX disc temperatures naturally fit into the IMBH description; therefore, introducing arguments for reducing the disk temperature by dissipating accretion energy into the corona become unnecessary.  An important discrepancy in coronal properties between the HIS/LHS GX 339--4 data and the ULX sample is the temperature of the corona.  While the data permit $\tau_{\rm p} \sim 5-10$ for both the HIS/LHS and ULXs (see Figure \ref{fig:taup_Tcolmax}), the associated coronal temperature for ULXs ($T_{\rm p} \sim 1-3~{\rm keV}$) is an order of magnitude below that of GX 339--4 in the HIS/LHS ($T_{\rm p} \sim 10-30~{\rm keV}$) (see Figure \ref{fig:taup_Tp}).  We stress that the purpose of this section is to highlight an interesting possibility that the ULS may not be an intrinsic feature of BHBs; however, this is highly speculative and further investigations are required to shed more light on the nature of ULXs.

% FIGURE 17

\subsubsection{The Role of $f_{\rm col}$ in State Transitions}
A triggering mechanism for state transitions is not yet well understood.  \citet{Schnittman2012} demonstrated that all of the spectral states of BHBs could be recovered by varying only the mass accretion rate, $\dot{M}$, (i.e., effectively changing the location of the disc photosphere, or the disc/corona boundary) in the magnetohydrodynamic general relativistic global disc simulation of \citet{Noble2010}.  The disc was assumed to produce a diluted blackbody spectrum at the photosphere with a constant $f_{\rm col} = 1.8$, which could be further altered (e.g., broadened/hardened) by bremsstrahlung, synchrotron, and inverse Compton processes in the corona.  The emergent spectrum was computed with a post-processing radiative transfer technique for choices of $\dot{M}$ representative of the observed BHB states.  In this simulation, the inner accretion disc remained extended down toward the ISCO.  Only details of the coronal physics for a given location of the disc photosphere (i.e., for a given value of $\dot{M}$) provided alterations to a diluted blackbody spectrum and all BHB spectral states were reproduced within this framework.  This result emphasises the importance of coronal physics in modifying the underlying disc spectrum, suggesting that perhaps $f_{\rm col}$ is a variable function of $\dot{M}$ and can be dependent on processes external to the disc.

A changing mass accretion rate plays a key role in state transitions, but is likely not the whole picture, nor is it clear what mechanism drives changes in $\dot{M}$.  For an inner disc remaining at the ISCO during a state transition, a large change in $\dot{M}$ is implied.  In the HSS, the inner portions of the disc are expected to be radiation pressure-dominated, where instabilities may lead to an inhomogeneous disc structure and a low $f_{\rm col} \sim 1$ \citep{Begelman2006}.  As $\dot{M}$ decreases, the ratio of radiation-to-gas pressures drops and below some $L_{\rm disc}/L_{\rm Edd}$ the density inhomogeneities would disappear and $f_{\rm col}$ would increase.  If the disc is strongly magnetised, values of $f_{\rm col} \gtrsim 5$ are conceivable \citep{BegelmanPringle2007}.  For a disc that instead transitions from a standard thin disc to a truncated disc with an interior RIAF, a relatively constant $\dot{M}$ is implied at this stage with a decreased radiative efficiency.  Both scenarios of an inner disc remaining extended to the ISCO and a truncated disc + RIAF predict different $\dot{M}$ behaviour, which is difficult to distinguish observationally.

The phenomenological picture of a variable colour correction factor being capable of describing the disc spectral evolution observed in BHB outbursts advocates for the development and consideration of models that appeal to an evolving vertical disc structure and/or intricate disc-corona-jet connection.  Recent observational work supports the idea of a coupled disc + corona playing a critical role in the LHS.  Adopting sophisticated, self-consistent modelling techniques, \citet{Reis2012} studied the BHB XTE J1650-500 during its 2001/2002 outburst and found definite correlations between the reflected X-ray flux and power-law flux associated with the corona.  These results fit into the interpretation of a light bending + collapsing corona model for state transitions \citep[e.g.,][]{MiniuttiFabian2004} and do not lend support to the truncated disc picture.  An organized campaign of contemporaneous and frequent X-ray/radio monitoring of Cygnus X-1 in the LHS reveal that the reflected X-ray flux and the radio flux are positively correlated \citep{Miller2012}, which is consistent with the coronal ejection model of \citet{Beloborodov1999}.  As the promising prospect of disc + corona coupling in BHB outbursts receives more attention, we hope that the relative measurements of $f_{\rm col}$ presented here serve as a useful reference for physical models that seek to generate predictive diagnostics of spectral hardening in accretion systems.

\subsection{Physical Range of the Colour Correction Factor}
\label{sec:fcol_range}
The colour correction factor phenomenologically parameterises our ignorance of spectral alterations arising from the highly unknown and complicated combinations of disc vertical structure, accretion power dissipation properties, magnetisation, and inhomogeneities.  Lumping the effects of these physical processes into the frequency-independent $f_{\rm col}$ parameter is attractive from the standpoint of limiting the parameters in MCD blackbody models; however, whether this treatment is an appropriate description of BHB discs in all spectral stages remains disputed.  Here, we discuss the results from theoretical endeavors to constrain the physically realisable range of the colour correction factor.

\subsubsection{Radiative Transfer in One-Dimensional Disc Atmospheres}
Investigations into the physically reasonable $f_{\rm col}$ range and the regimes for which the colour-corrected blackbody model is justified have focused on one-dimensional atmospheric disc models.  These atmospheres are assumed to have no azimuthal structure and are constructed at many disc radii from self-similar standard thin disc equations.  In brief, frequency-dependent opacities are computed at each radial position, allowing the modification of the locally emitted blackbody spectrum to be determined for multiple concentric annuli.  The emergent spectrum, which is modified by the vertical structure, is computed for each annulus.  The integrated disc spectrum is obtained by summing over all annuli and is fitted with the colour-corrected blackbody model, yielding a best-fit $f_{\rm col}$.

Following the technique of self-consistently computing the radiative transfer and vertical disc structure \citep{Ross1992}, \citet{ShimuraTakahara1995b} considered the effects of electron scattering and free-free emission/absorption on the emergent disc spectrum.   Regardless of the black hole mass and radial position in the disc, the local spectra (i.e., the emergent spectrum from an annulus) were adequately fit with $f_{\rm col} \simeq 1.8 - 2.0$ and $f_{\rm col} \simeq 1.7$ for $\dot{M} \simeq \dot{M}_{\rm Edd}$ and $\dot{M} \simeq 0.1~\dot{M}_{\rm Edd}$, respectively, suggesting a relatively constant $f_{\rm col}$ over a large range of mass accretion rates expected in the HSS.  For $\dot{M} \lesssim 0.01~\dot{M}_{\rm Edd}$, which is characteristic of the LHS, the local spectrum was not well-described by the colour-corrected blackbody model due to ineffective (i.e., unsaturated) Comptonisation within the disc.

\citet{Merloni2000} extended the study of \citet{ShimuraTakahara1995b} by considering the effects of Doppler blurring, gravitational redshift, and allowing the dissipated accretion energy to be partitioned between the corona and disc \citep{SvenssonZdziarski1994}.  Assuming all of the accretion energy is dissipated within the disc, \citet{Merloni2000} found $f_{\rm col} \simeq 1.8$ for $\dot{M} = 0.1 - 0.3~\dot{M}_{\rm Edd}$, fully consistent with \citet{ShimuraTakahara1995b}.  Exploring the lower accretion rate, $\dot{M} = 0.05~\dot{M}_{\rm Edd}$ and allowing a significant fraction of the energy to be dissipated in the corona, \citet{Merloni2000} instead found a variable $f_{\rm col}$ ranging from $\sim 1.9 - 2.7$.  Therefore, if coronal activity is important, $f_{\rm col}$ cannot be considered constant, with the suggested physical range being $1.7 \lesssim f_{\rm col} \lesssim 3$.

\citet{Davis2005} presents the most sophisticated study to date of the effects of disc vertical structure on the emergent spectrum, incorporating bound-free opacities and a fully relativistic radiative transfer treatment.  For disc luminosities, $L_{\rm disc} = 0.01 - 0.3~L_{\rm Edd}$, which spans the luminosity regimes of the LHS and HSS, \citet{Davis2005} found a weakly evolving $f_{\rm col} \simeq 1.4 - 1.6$, concluding that $f_{\rm col}$ remains fairly constant over broad ranges in mass accretion rate.  The results of \citet{Davis2005} are at odds with those of \citet{Merloni2000}; however, a direct comparison is complicated by the different methodologies used in each work.

\subsubsection{Magnetisation}
Although the details of magnetic fields (e.g., topology, strength, variability) in accreting black hole systems are poorly understood, magnetisation is central to many aspects of accretion physics.  Perhaps most notable is the role of the magnetorotational instability (MRI) \citep{BalbusHawley1991}, which is widely accepted as the mechanism for angular momentum transport; hence, allowing accretion to persist.  The turbulence generated within the disc by the MRI amplifies the toroidal component of magnetic field, perhaps reaching a state where magnetic pressure dominates over the combined gas and radiation pressures.  \citet{BegelmanPringle2007} present analytic arguments for the maximal toroidal magnetic field strength achievable by the MRI and compute the corresponding disc structure, finding that magnetically dominated discs are thicker and have harder spectra compared to the standard Shakura-Sunyaev thin disc.  Under the conjecture that the toroidal field attains the limiting value set by the MRI, \citet{BegelmanPringle2007} derive a lower limit for the colour correction factor,
\begin{equation}
\resizebox{1.0\hsize}{!}{$f_{\rm col} \ge 6.9 \left( \frac{\rho_{\ast}}{\rho} \right)^{-2/9} \alpha^{5/27} \left( \frac{M}{M_{\odot}} \right)^{-1/108} \left( \frac{\dot{M}}{\dot{M}_{\rm Edd}} \right)^{5/108} \left( \frac{r}{R_{\rm g}} \right)^{25/108}$}, \nonumber
\end{equation}
where $\rho$ is the mid-plane density, $\rho_{\ast}$ is the density corresponding to the thermalization layer (i.e., effective photosphere) of the disc, and $\rho_{\ast} / \rho < 1$.  Shearing box simulations place the disc photosphere a few scale heights above the mid-plane, corresponding to density contrasts, $\rho_{\ast} / \rho \simeq 10^{-2} - 10^{-4}$ \citep{Hirose2006, Hirose2009}.  Therefore, if the disc around a stellar mass black hole is magnetically dominated, the colour correction factor can far surpass the values predicted by thin disc treatments that neglect the effects of a magnetic field on the disc structure.

Conceivably, the details of vertical dissipation and coronal properties could alter the emergent disc spectrum considerably.  \citet{Beloborodov1999} envisaged a picture where buoyant magnetic loops subject to Parker instability rise into the corona and the magnetic energy available within the loops is subsequently dissipated by reconnection.  Recently, \citet{Uzdensky2012} analytically computed the vertical structure of a gas pressure-dominated disc threaded by a poloidal magnetic field where turbulent (i.e., dissipative) heating arising from the MRI is balanced by radiative cooling.  Based on this disc model, \citet{Uzdensky2012} argue that the {\it minimum} fraction of available accretion energy that ends up being released in the corona is, $f_{\rm min} \sim \beta_{\rm 0, max}^{-1/5} \sim 0.4$, where $\beta_{\rm 0, max} \sim 100$ is a reasonable estimate of the gas-to-magnetic pressure ratio at the disc mid-plane.  Under conditions where a substantial fraction of accretion power is dissipated in the corona, \citet{Merloni2000} demonstrate that substantial spectral hardening is expected.

Numerical simulations show that magnetically supported, gas pressure-dominated discs have harder emergent spectra than their unmagnetised counterparts \citep{Blaes2006}.  \citet{Blaes2006} calculated local accretion disc spectra from stratified shearing box simulations and found the disc spectrum hardened due to magnetic pressure support dominating in the upper atmosphere.  This magnetic pressure acts to vertically extend the disc atmosphere and decrease the density at the effective photosphere, leading to saturated Comptonisation and spectral hardening.  In deriving the local disc spectrum, \citet{Blaes2006} considered two vertical dissipation profiles: (1) a constant dissipation per unit mass and (2) a broken power-law fit to the numerical dissipation profile.  Both treatments produced identical emergent spectra, implying that the local emergent disc spectrum is insensitive to the vertical dissipation profile and that magnetic pressure support is solely responsible for the observed spectral hardening.  Motivated by the result that magnetic pressure support drives the spectral hardening and assuming the dissipation profile has no radial dependence, \citet{Blaes2006} constructed global disc spectra and found only modest hardening, with $f_{\rm col}$ increasing from 1.48 to 1.74 upon introducing magnetic pressure support.  \citet{Blaes2006} caution that the numerical dissipation occurs at the grid scale and may be sensitive to grid resolution.  Recent work indicates that numerical dissipation in ideal magnetohydrodynamic grid-based simulations does behave like physical dissipation, but acts to effectively diminish the resolution (Salvesen et al. 2013, in preparation).  Adopting dissipation profiles motivated by numerical simulations is an important step toward merging accretion disc theory with simulations.

\subsubsection{Inhomogeneous Discs}
Strong disc inhomogeneities are expected to arise in moderate-to-strongly magnetised, radiation pressure-dominated regions of accretion discs \citep{Begelman2002, Begelman2006} due to photon bubble instability \citep{Arons1992, Gammie1998, Begelman2001}.  In this picture, strong density contrasts arise on scales smaller than the disc scale height, resulting in a porous disc where radiation preferentially `leaks' through the underdense regions.  Due to the typical densities in photon bubble-dominated discs being higher than in standard thin discs of the same luminosity, thermalization of radiation is more effective in inhomogeneous discs.  Consequently, radiation that is thermalized near the disc mid-plane and escapes through optically thin chutes can result in a colour correction factor near unity for inhomogeneous discs \citep{Begelman2006}.  \citet{DexterQuataert2012} computed inhomogeneous disc spectra based on a toy model \citep{DexterAgol2011} and found that the disc spectrum broadens and hardens as the amplitude of the local temperature fluctuations increases.  Inhomogeneous discs can substantially alter the standard disc spectrum and further investigation is required to characterise their properties.

\subsection{Limitations of Simplistic Accretion Disc Models}
\label{sec:limitations}
While significant strides have been made in exploring the imprints of the disc structure on the emergent spectrum in the HSS, the applicability of the colour-corrected blackbody model to discs in the LHS remains elusive.  From a theoretical perspective, thin discs with low mass accretion rates will be fully gas pressure-dominated with absorption overtaking electron scattering as the dominant opacity source, nullifying the approximation of the emergent spectrum as a colour-corrected blackbody \citep[e.g.,][]{Ebisuzaki1984}.  Nevertheless, standard MCD models consistently produce exceptional fits to LHS accretion disc spectra, giving credence to the notion that the MCD model successfully describes real astrophysical discs in the LHS.  Analogously, standard thin discs are expected to suffer from thermal \citep{ShakuraSunyaev1976} and viscous \citep{LightmanEardley1974} instabilities, yet the MCD model provides an acceptable description of the observed disc spectrum, which should not be the case if the disc were plagued by catastrophic instabilities.  While it may turn out that $f_{\rm col}$ has a very narrow allowable range and/or it is an inappropriate parameterisation to use for low disc luminosities, there is currently no compelling evidence that this should be so based on fits to observed disc spectra alone.

Direct measurements of $f_{\rm col}$ should not be made with Newtonian disc models when the black hole is known to possess a nonzero spin and an inner disc extending to the ISCO.  Newtonian disc models have the strength of being able to describe a standard black hole accretion disc spectrum with very few parameters; however, the important physical effects of light bending, Doppler blurring, and gravitational redshift inherent to Kerr black holes are neglected. The pseudo-Newtonian disc model, \texttt{diskpn}, provides accurate fits to fully relativistic disc spectra from a non-spinning black hole, but requires uncomfortably large values of $f_{\rm col}$ when fitting a disc spectrum from an extreme Kerr black hole (see \S \ref{sec:fakeit}).  When attempting to fit the hotter disc spectra intrinsic to spinning black holes, Newtonian disc models are forced to artificially increase the colour correction factor to achieve the colour temperature demanded by the data.  Fortunately, relative measurements of $f_{\rm col}$ do not appear to be appreciably affected in the way that absolute measurements are, at least for the low disc inclination considered in this work.  This implies that the relative measurements of $f_{\rm col}$ evolution during state transitions of GX 339--4 are sound, regardless of the intrinsic spin possessed by GX 339--4.  The possibility that the inner disc recedes radially outward or progresses inward between successive observations cannot be ruled out and would manifest itself in the relative measurements of $f_{\rm col}$, rendering them moot.

The main outcome of this work is that an evolving colour correction factor in a standard thin disc with inner radius fixed at the ISCO is a viable, albeit not the only, explanation for the changing disc spectrum of the BHB GX 339--4 over many state transitions.  Given the uncertainty in accretion flow geometry as BHBs transition between spectral states, we emphasise that inner disc radii cannot be reliably computed from spectral fits made with simplified disc models.  Instead, self-consistent, physically motivated spectral modelling applied to high-quality CCD X-ray spectra is necessary to distinguish between opposing accretion scenarios.  Future insights into the complicated disc-corona-jet connections in BHBs will likely progress by concentrated observing campaigns and the development of physical models that can provide testable diagnostics.

%===========================================================================
% SUMMARY AND CONCLUSIONS
\section{Summary \& Conclusions}
\label{section:conclusions}
The purpose of this work is to show that physically reasonable changes in the colour correction factor provide a valid, alternative explanation to disc truncation for explaining the accretion disc spectral evolution observed in black hole binary state transitions.  To this end, we systematically performed spectral fits to {\it RXTE} observations of the low-mass X-ray binary GX 339--4 over seven state transitions and collected the observations that statistically required a disc component.  The disc detection pipeline measured discs with luminosities as low as $L_{\rm disc} \simeq 10^{-3}~L_{\rm Edd}$ and disc fractions down to $DF \simeq 0.05$.  In a suite of spectral fits to the observations deemed to require a disc, we imposed an unchanging inner disc radius located at the ISCO for a Schwarzschild black hole in order to ensure that a changing colour correction factor, rather than a radially receding/advancing inner disc, was responsible for any spectral evolution of the disc component.  We give our conclusions as a bulleted list below.

\begin{enumerate}[label={\bfseries $\bullet$},leftmargin=*]
\item For nearly all state transitions of GX 339--4 investigated here, relative changes in $f_{\rm col}$ in the range $1.4 \lesssim f_{\rm col, rel} \lesssim 5$, which corresponds to a factor of $\sim$ 3.5 change, can fully account for disc evolution in a state transition without invoking a truncated disc geometry.

\item Accounting for an $f_{\rm col}$ evolution in the range quoted above over the course of state transitions in GX 339--4, the $F_{\rm disc} \propto T_{\rm eff}^{4}$ relation is recovered for a constant emitting area (i.e., non-truncated disc) down to the low luminosities characteristic of the HIS and LHS.

\item The validity of the results mentioned above are insensitive to the Comptonisation model adopted for fitting the high-energy continuum.  Adopting a continuum model that avoids the low-energy flux stealing character of a simple power-law reduces the degree of $f_{\rm col}$ variability required to explain the observed disc evolution in state transitions.

\item The practice of measuring the inner disc radius from the best-fit disc normalisation is unreliable when both $R_{\rm in}$ and $f_{\rm col}$ are worked into the disc normalisation.  Instead, we suggest that disc models where these two parameters are decoupled (e.g., \texttt{diskpn}) be used to break the degeneracy.

\item We stress that this work {\it does not} demonstrate that the inner accretion disc remains at the radius of marginal stability during a state transition.  Instead, we argue that variations of $f_{\rm col}$ within a theoretically reasonable range can account for most, if not all, of the disc spectral evolution observed over the course of a state transition.

\item Relative measurements, unlike absolute measurements, of $f_{\rm col}$ are not appreciably affected by our ignorance of black hole spin.

\item When a physically motivated Comptonisation model is used in conjunction with a disc component to fit the {\it RXTE} data of GX 339--4 in the HIS and LHS, we observe a parameter degeneracy between the coronal optical depth and the colour correction factor.  This calls into question the association of ULXs with super-Eddington stellar mass black hole binaries based on measurements of coronal optical depth without allowing for a variable $f_{\rm col}$.
\end{enumerate}

Simple disc models of the MCD family are attractive and useful from the standpoint that they consistently achieve superb fits with limited free parameters to accretion disc spectra, where the intrinsic system parameters are often poorly constrained.  This work illustrates that allowing for the possibility of reasonable variations in the colour correction factor provides an adequate description of disc spectra over a wide range in luminosity states.  We advocate that $f_{\rm col}$ may be dynamic and should not be dismissed as a mere constant in studies of accretion disc spectra spanning a broad range in luminosity, as is often done in practice.  Numerous factors may contribute to altering the emergent disc spectrum including the vertical disc structure, details of accretion energy dissipation, coronal dynamics, magnetisation, and disc inhomogeneities.  Given that the available X-ray data cannot isolate these effects, common practice is to adopt a blanket term, the colour correction factor, which is supposed to account for the overall deviations from a multi-colour blackbody disc spectrum.  We encourage an increased awareness for the implications of presuming a constant $f_{\rm col}$.  Important physics are wrapped up into $f_{\rm col}$ that should be considered before making definitive claims based on the assumption of an invariant $f_{\rm col}$.

The standard theoretical emergent disk spectrum hinges on the assumptions that the accretion energy is dissipated within the disk (i.e., where the majority of the mass resides) and that the density structure is smooth and homogeneous.  The effects of magnetisation, which surely play a crucial role in accretion physics and disc structure, are also ignored in the disc models applied to observations.  Radiation pressure-dominated, magnetised discs, as expected in the HSS, are subject to clumping instabilities, resulting in a porous disc structure and lowered colour correction factor \citep{Begelman2006}.  Considerations of the corona as an energy dissipation reservoir \citep{SvenssonZdziarski1994, Merloni2000, MerloniFabian2002} and of strongly magnetised discs \citep{BegelmanPringle2007} indicate that $f_{\rm col}$ may extend well beyond the canonical value of $\sim 1.7$ and into a realm consistent with the values required here and in similar observational studies \citep{Dunn2011, ReynoldsMiller2011}.  With the advent of sophisticated codes capable of treating magnetohydrodynamics with radiative transport and general relativity, state-of-the-art numerical simulations provide a promising avenue for studying accretion disc structure with the aforementioned assumptions relaxed.  Striving toward the development of sophisticated models that capture the relevant physics of realistic accreting black hole systems, provide predictive diagnostics, and are accessible to observers is a crucial step toward disentangling the details of the accretion flow and distinguishing between different accretion scenarios.

%===========================================================================
% ACKNOWLEDGEMENTS
\section*{Acknowledgments}
The authors thank the anonymous referee for her/his constructive comments and suggestions, which improved this paper.  GS thanks Jordan Mirocha for thoughtful discussions and the National Science Foundation for support through the Graduate Research Fellowship Program.  RCR thanks the Michigan Society of Fellows and NASA for support through the Einstein Fellowship Program, grant number PF1-120087.  This work used the \texttt{JANUS} supercomputer, which is supported by the National Science Foundation (award number CNS-0821794) and the University of Colorado Boulder.  The \texttt{JANUS} supercomputer is a joint effort of the University of Colorado Boulder, the University of Colorado Denver, and the National Center for Atmospheric Research.

\bibliographystyle{mn2e}
\bibliography{ms}
\label{lastpage}

% TABLE 3
\onecolumn
\begin{tiny}
\begin{landscape}
\begin{longtable}{c c c c c c c c c c c c c c c c c}

\hline \hline

\multicolumn{1}{c}{Trans.} & \multicolumn{1}{c}{Observation} & \multicolumn{1}{c}{MJD} & \multicolumn{1}{c}{Best--fit} & \multicolumn{1}{c}{$\Gamma_{1}$} & \multicolumn{1}{c}{$E_{\rm b}$} & \multicolumn{1}{c}{$\Gamma_{2}$} & \multicolumn{1}{c}{$K_{\rm pow}$} & \multicolumn{1}{c}{$\sigma$} & \multicolumn{1}{c}{$K_{\rm line}$} & \multicolumn{1}{c}{$kT_{\rm in}$} & \multicolumn{1}{c}{$K_{\rm disc}$} & \multicolumn{1}{c}{$C$} & \multicolumn{1}{c}{$\nu$} & \multicolumn{1}{c}{$\chi^{2}_{\nu}$} & \multicolumn{1}{c}{log($F_{\rm disc}$)} & \multicolumn{1}{c}{$DF$} \\

\multicolumn{1}{c}{ID} & \multicolumn{1}{c}{ID} & \multicolumn{1}{c}{} & \multicolumn{1}{c}{Model} & \multicolumn{1}{c}{} & \multicolumn{1}{c}{(keV)} & \multicolumn{1}{c}{} & \multicolumn{1}{c}{} & \multicolumn{1}{c}{(eV)} & \multicolumn{1}{c}{} & \multicolumn{1}{c}{(keV)} & \multicolumn{1}{c}{} & \multicolumn{1}{c}{} & \multicolumn{1}{c}{} & \multicolumn{1}{c}{} & \multicolumn{1}{c}{erg/s/cm$^{2}$} & \multicolumn{1}{c}{(\%)} \\
\hline
\endfirsthead

\hline \hline
\multicolumn{1}{c}{Trans.} & \multicolumn{1}{c}{Observation} & \multicolumn{1}{c}{MJD} & \multicolumn{1}{c}{Best--fit} & \multicolumn{1}{c}{$\Gamma_{1}$} & \multicolumn{1}{c}{$E_{\rm b}$} & \multicolumn{1}{c}{$\Gamma_{2}$} & \multicolumn{1}{c}{$K_{\rm pow}$} & \multicolumn{1}{c}{$\sigma$} & \multicolumn{1}{c}{$K_{\rm line}$} & \multicolumn{1}{c}{$kT_{\rm in}$} & \multicolumn{1}{c}{$K_{\rm disc}$} & \multicolumn{1}{c}{$C$} & \multicolumn{1}{c}{$\nu$} & \multicolumn{1}{c}{$\chi^{2}_{\nu}$} & \multicolumn{1}{c}{log($F_{\rm disc}$)} & \multicolumn{1}{c}{$DF$} \\

\multicolumn{1}{c}{ID} & \multicolumn{1}{c}{ID} & \multicolumn{1}{c}{} & \multicolumn{1}{c}{Model} & \multicolumn{1}{c}{} & \multicolumn{1}{c}{(keV)} & \multicolumn{1}{c}{} & \multicolumn{1}{c}{} & \multicolumn{1}{c}{(eV)} & \multicolumn{1}{c}{} & \multicolumn{1}{c}{(keV)} & \multicolumn{1}{c}{} & \multicolumn{1}{c}{} & \multicolumn{1}{c}{} & \multicolumn{1}{c}{} & \multicolumn{1}{c}{erg/s/cm$^{2}$} & \multicolumn{1}{c}{(\%)} \\
\hline
\endhead

\hline
\endfoot

\endlastfoot
R02 & 40031-03-01-00 & 52382.114 & \texttt{bkn$\times$(bb+ga)} & 1.47(2) & 33(1) & 2.43(6) & 1.13(5) & 1.26(7) & 0.040(4) & 1.38(8) & 31(8) & 0.76(1) & 70 & 0.91 & -8.62(3) & 7.2(4) \\
R02 & 40031-03-02-00 & 52383.103 & \texttt{bkn$\times$(bb+ga)} & 1.52(1) & 32(1) & 2.34(6) & 1.38(4) & 1.32(6) & 0.044(4) & 1.35(9) & 30(9) & 0.719(8) & 70 & 1.08 & -8.67(2) & 5.8(3) \\
R02 & 40031-03-02-01 & 52384.147 & \texttt{bkn$\times$(bb+ga)} & 1.526(8) & 34.0(9) & 2.42(4) & 1.36(3) & 1.31(5) & 0.046(3) & 1.21(6) & 54(12) & 0.749(6) & 70 & 0.96 & -8.60(2) & 6.9(2) \\
R02 & 40031-03-02-02 & 52384.979 & \texttt{bkn$\times$(bb+ga)} & 1.517(9) & 33.6(6) & 2.44(3) & 1.31(3) & 1.29(5) & 0.044(3) & 1.28(7) & 41(10) & 0.765(5) & 70 & 1.30 & -8.63(2) & 6.6(3) \\
R02 & 70110-01-05-00 & 52385.469 & \texttt{bkn$\times$(bb+ga)} & 1.58(2) & 33(1) & 2.40(6) & 1.34(6) & 0.9(1) & 0.024(4) & 1.3(1) & 24(10) & 0.91(1) & 70 & 0.84 & -8.78(4) & 5.4(4) \\
R02 & 40031-03-02-03 & 52386.026 & \texttt{pow$\times$(bb+ga)} & 1.565(7) & - & - & 1.51(3) & 1.23(5) & 0.042(3) & 1.30(7) & 32(8) & - & 43 & 1.62 & -8.71(2) & 4.7(2) \\
R02 & 40031-03-02-04 & 52386.934 & \texttt{bkn$\times$(bb+ga)} & 1.540(8) & 32.0(7) & 2.37(3) & 1.43(3) & 1.30(5) & 0.045(3) & 1.31(7) & 36(8) & 0.775(6) & 70 & 1.26 & -8.65(2) & 6.2(2) \\
R02 & 70110-01-06-00 & 52387.490 & \texttt{bkn$\times$(bb+ga)} & 1.60(1) & 34(2) & 2.43(8) & 1.42(6) & 1.04(9) & 0.029(3) & 1.3(1) & 25(10) & 0.91(2) & 70 & 0.72 & -8.84(5) & 4.7(4) \\
R02 & 70109-01-04-00 & 52387.646 & \texttt{bkn$\times$(bb+ga)} & 1.593(6) & 33.1(3) & 2.44(1) & 1.44(2) & 1.05(5) & 0.033(2) & 1.18(7) & 38(10) & 0.908(4) & 70 & 1.90 & -8.81(2) & 4.9(2) \\
R02 & 40031-03-02-05 & 52388.065 & \texttt{bkn$\times$(bb+ga)} & 1.51(1) & 32.1(9) & 2.39(4) & 1.33(5) & 1.26(6) & 0.045(4) & 1.37(8) & 33(8) & 0.739(8) & 70 & 1.16 & -8.60(2) & 6.9(4) \\
R02 & 40031-03-02-06 & 52389.071 & \texttt{bkn$\times$(bb+ga)} & 1.53(3) & 35(3) & 2.4(1) & 1.4(1) & 1.2(1) & 0.043(8) & 1.2(2) & 46(29) & 0.79(2) & 70 & 0.79 & -8.65(7) & 6.3(8) \\
R02 & 70110-01-07-00 & 52390.910 & \texttt{bkn$\times$(bb+ga)} & 1.59(1) & 32(1) & 2.51(8) & 1.45(6) & 1.11(8) & 0.038(4) & 1.2(1) & 40(15) & 0.88(2) & 70 & 1.07 & -8.74(4) & 5.7(5) \\
R02 & 70109-01-05-01G & 52391.336 & \texttt{bkn$\times$(bb+ga)} & 1.608(8) & 33.9(7) & 2.46(3) & 1.50(3) & 1.05(6) & 0.032(3) & 1.24(9) & 29(10) & 0.914(6) & 70 & 0.98 & -8.83(2) & 4.7(2) \\
R02 & 70109-01-05-00 & 52391.402 & \texttt{bkn$\times$(bb+ga)} & 1.616(7) & 32.5(8) & 2.41(3) & 1.55(3) & 0.99(6) & 0.031(2) & 1.20(8) & 35(10) & 0.901(6) & 71 & 1.31 & -8.82(2) & 4.7(2) \\
R02 & 70109-01-05-02 & 52391.469 & \texttt{bkn$\times$(bb+ga)} & 1.610(7) & 32.7(7) & 2.44(3) & 1.52(3) & 1.04(6) & 0.033(3) & 1.25(8) & 30(9) & 0.901(6) & 70 & 1.07 & -8.82(2) & 4.8(2) \\
R02 & 70110-01-08-00 & 52394.445 & \texttt{bkn$\times$(bb+ga)} & 1.623(9) & 30.6(9) & 2.37(3) & 1.60(4) & 1.05(6) & 0.036(3) & 1.20(8) & 38(12) & 0.901(8) & 70 & 1.19 & -8.79(3) & 5.0(3) \\
R02 & 70110-01-09-00 & 52398.668 & \texttt{bkn$\times$(bb+ga)} & 1.68(1) & 33(1) & 2.61(6) & 1.81(5) & 1.10(6) & 0.037(3) & 1.28(9) & 29(9) & 0.93(1) & 70 & 0.88 & -8.79(3) & 5.0(3) \\
R02 & 70109-01-06-00 & 52400.842 & \texttt{bkn$\times$(bb+ga)} & 1.785(8) & 30(1) & 2.56(4) & 2.23(5) & 1.05(5) & 0.036(3) & 1.27(7) & 31(8) & 0.925(9) & 70 & 1.05 & -8.76(3) & 5.3(3) \\
R02 & 70108-03-01-00 & 52400.924 & \texttt{pow$\times$(bb+ga)} & 1.836(6) & - & - & 2.50(4) & 1.01(4) & 0.036(2) & 1.15(7) & 42(13) & - & 43 & 1.80 & -8.80(3) & 4.3(3) \\
R02 & 70110-01-10-00 & 52402.499 & \texttt{bkn$\times$(bb+ga)} & 2.07(1) & 28(3) & 2.6(1) & 3.5(1) & 1.10(4) & 0.043(2) & 0.96(3) & 243(36) & 0.96(2) & 70 & 1.16 & -8.35(2) & 11.5(5) \\
R02 & 70109-04-01-00 & 52405.632 & \texttt{bkn$\times$(bb+ga)} & 2.41(1) & 32(3) & 2.85(8) & 5.4(1) & 1.11(3) & 0.040(2) & 0.867(8) & 1507(81) & 0.99(1) & 70 & 1.76 & -7.742(8) & 26.3(2) \\
R02 & 70109-04-01-01 & 52405.865 & \texttt{bkn$\times$(bb+ga)} & 2.386(7) & 30(2) & 2.79(5) & 5.1(1) & 1.12(3) & 0.041(2) & 0.877(8) & 1431(73) & 0.972(7) & 70 & 1.78 & -7.745(7) & 27.8(1) \\
R02 & 70109-04-01-02 & 52406.077 & \texttt{bkn$\times$(bb+ga)} & 2.38(1) & 28(7) & 2.8(3) & 5.1(2) & 1.05(4) & 0.042(2) & 0.871(8) & 1411(76) & 0.97(6) & 70 & 1.11 & -7.763(9) & 26.4(3) \\
R02 & 70110-01-11-00 & 52406.707 & \texttt{pow$\times$(bb+ga)} & 2.42(1) & - & - & 5.2(2) & 1.08(4) & 0.036(2) & 0.890(8) & 1342(70) & 0.99(2) & 73 & 1.11 & -7.747(8) & 26.53(6) \\
R02 & 70110-01-12-00 & 52410.535 & \texttt{pow$\times$(bb+ga)} & 2.43(1) & - & - & 3.8(1) & 1.12(4) & 0.030(2) & 0.867(6) & 2147(80) & 0.99(2) & 73 & 1.03 & -7.589(5) & 41.5(1) \\
R02 & 70109-01-07-00 & 52411.645 & \texttt{pow$\times$(bb+ga)} & 2.58(1) & - & - & 4.3(1) & 0.99(3) & 0.027(1) & 0.866(5) & 2507(84) & 1.02(2) & 72 & 1.48 & -7.523(5) & 39.9(4) \\
R02 & 70110-01-13-00 & 52412.080 & \texttt{pow$\times$(bb+ga)} & 2.53(2) & - & - & 3.0(2) & 0.98(4) & 0.024(1) & 0.867(5) & 2591(78) & 0.99(4) & 73 & 0.83 & -7.507(4) & 50.4(5) \\
R02 & 40031-03-03-00 & 52414.057 & \texttt{pow$\times$(bb+ga)} & 2.42(6) & - & - & 1.5(2) & 1.04(9) & 0.017(2) & 0.864(5) & 2957(103) & 0.95(8) & 72 & 0.66 & -7.457(6) & 70.0(3) \\
R02 & 40031-03-03-01 & 52414.126 & \texttt{pow$\times$bb} & 2.63(9) & - & - & 2.6(6) & - & - & 0.885(6) & 2491(77) & 1.3(2) & 74 & 1.37 & -7.488(6) & 53.6(5) \\
R02 & 40031-03-03-02 & 52414.193 & \texttt{pow$\times$(bb+ga)} & 2.46(7) & - & - & 1.6(3) & 0.89(9) & 0.016(2) & 0.803(5) & 3967(154) & - & 43 & 1.22 & -7.457(7) & 69.4(5) \\
R02 & 40031-03-03-04 & 52414.363 & \texttt{pow$\times$(bb+ga)} & 2.36(2) & - & - & 1.49(8) & 1.10(5) & 0.015(1) & 0.885(4) & 2669(72) & 0.87(3) & 72 & 1.13 & -7.457(4) & 71.1(2) \\
R02 & 70110-01-14-00 & 52416.601 & \texttt{pow$\times$(bb+ga)} & 2.43(2) & - & - & 3.3(2) & 1.08(5) & 0.028(2) & 0.874(6) & 2017(73) & - & 43 & 1.67 & -7.601(6) & 44.4(3) \\
\hline
D03 & 70109-01-36-02 & 52688.301 & \texttt{pow$\times$(bb+ga)} & 3.0(3) & - & - & 0.3(3) & 0.4(2) & 0.0009(3) & 0.557(8) & 5467(443) & - & 43 & 1.05 & -7.96(1) & 66(1) \\
D03 & 70111-01-01-000 & 52690.691 & \texttt{pow$\times$(bb+ga)} & 2.5(1) & - & - & 0.08(2) & 0.76(5) & 0.0014(1) & 0.567(3) & 4401(144) & 1.2(3) & 72 & 1.63 & -8.019(6) & 92.31(9) \\
D03 & 70110-01-86-00 & 52693.734 & \texttt{pow$\times$bb} & 3.20(6) & - & - & 1.5(2) & - & - & 0.50(1) & 6782(1224) & 2.9(6) & 74 & 1.72 & -8.07(3) & 21(3) \\
D03 & 70109-01-37-00 & 52694.944 & \texttt{pow$\times$(bb+ga)} & 2.28(4) & - & - & 0.21(2) & 0.82(4) & 0.0033(2) & 0.640(6) & 1429(87) & 1.3(1) & 72 & 1.55 & -8.30(1) & 71.7(8) \\
D03 & 70110-01-87-00 & 52700.602 & \texttt{pow$\times$(bb+ga)} & 2.32(5) & - & - & 0.19(3) & 0.72(6) & 0.0024(2) & 0.624(8) & 1230(100) & 1.1(2) & 72 & 1.12 & -8.40(1) & 68.3(9) \\
D03 & 70110-01-88-00 & 52704.021 & \texttt{pow$\times$(bb+ga)} & 2.10(6) & - & - & 0.14(2) & 0.96(6) & 0.0040(3) & 0.662(9) & 799(72) & 0.7(2) & 72 & 1.28 & -8.49(1) & 70(1) \\
D03 & 70110-01-89-00 & 52707.932 & \texttt{pow$\times$(bb+ga)} & 2.7(1) & - & - & 0.5(1) & 0.4(1) & 0.0020(4) & 0.58(2) & 1712(334) & - & 43 & 1.30 & -8.38(2) & 43.1(4) \\
D03 & 70109-02-01-01 & 52710.012 & \texttt{pow$\times$(bb+ga)} & 2.29(6) & - & - & 0.21(3) & 0.81(5) & 0.0032(3) & 0.601(8) & 1492(126) & - & 43 & 1.97 & -8.39(1) & 67.4(8) \\
D03 & 60705-01-56-00 & 52710.747 & \texttt{pow$\times$(bb+ga)} & 2.48(6) & - & - & 0.43(7) & 0.91(5) & 0.0038(4) & 0.64(1) & 878(97) & 1.3(1) & 71 & 1.73 & -8.50(1) & 42.7(2) \\
D03 & 70110-01-91-00 & 52713.486 & \texttt{pow$\times$(bb+ga)} & 2.43(9) & - & - & 0.20(5) & 0.75(7) & 0.0021(3) & 0.572(9) & 2295(217) & 1.4(3) & 72 & 1.36 & -8.29(1) & 73.1(5) \\
D03 & 60705-01-57-00 & 52716.732 & \texttt{pow$\times$(bb+ga)} & 2.8(1) & - & - & 0.4(1) & 0.88(8) & 0.0021(3) & 0.57(1) & 2581(260) & 2.2(4) & 71 & 1.02 & -8.23(1) & 57.6(6) \\
D03 & 50117-01-04-01 & 52718.102 & \texttt{pow$\times$(bb+ga)} & 2.23(5) & - & - & 0.17(2) & 0.93(4) & 0.0035(2) & 0.647(7) & 1096(75) & 1.0(1) & 71 & 1.72 & -8.39(1) & 71(1) \\
D03 & 50117-01-04-02 & 52718.180 & \texttt{pow$\times$(bb+ga)} & 2.5(1) & - & - & 0.4(1) & 0.90(8) & 0.0035(5) & 0.58(2) & 2212(377) & - & 42 & 1.53 & -8.27(2) & 57.7(6) \\
D03 & 70110-01-93-00 & 52719.234 & \texttt{pow$\times$(bb+ga)} & 2.27(7) & - & - & 0.22(4) & 0.78(7) & 0.0035(4) & 0.62(1) & 1270(119) & 1.1(2) & 72 & 1.58 & -8.40(1) & 66.0(6) \\
D03 & 70110-01-94-00 & 52724.232 & \texttt{pow$\times$(bb+ga)} & 2.14(5) & - & - & 0.23(3) & 0.99(5) & 0.0053(4) & 0.71(1) & 408(49) & 0.9(1) & 72 & 1.19 & -8.67(2) & 48(1) \\
D03 & 70110-01-95-00 & 52727.257 & \texttt{pow$\times$(bb+ga)} & 2.07(5) & - & - & 0.24(3) & 0.96(7) & 0.0049(5) & 0.71(2) & 276(42) & 1.2(1) & 72 & 1.25 & -8.83(2) & 37(1) \\
D03 & 60705-01-59-00 & 52731.583 & \texttt{pow$\times$(bb+ga)} & 1.82(2) & - & - & 0.19(1) & 1.22(5) & 0.0058(3) & 0.89(2) & 44(6) & 0.68(3) & 71 & 1.43 & -9.22(2) & 17.4(8) \\
D03 & 70110-01-96-00 & 52732.119 & \texttt{bkn$\times$(bb+ga)} & 1.72(5) & 34(9) & 2.6(7) & 0.14(2) & 1.0(1) & 0.0034(6) & 0.91(5) & 32(10) & 1.09(8) & 70 & 0.72 & -9.33(5) & 18(1) \\
D03 & 70110-01-97-00 & 52734.228 & \texttt{pow$\times$(bb+ga)} & 1.70(5) & - & - & 0.13(2) & 1.0(2) & 0.0026(7) & 1.06(9) & 11(5) & 0.93(7) & 72 & 0.85 & -9.51(7) & 12(2) \\
D03 & 80116-02-01-00 & 52735.660 & \texttt{pow$\times$(bb+ga)} & 1.65(1) & - & - & 0.109(4) & 0.86(9) & 0.0019(2) & 1.06(7) & 8(3) & 0.92(3) & 72 & 1.18 & -9.67(3) & 8.6(6) \\
D03 & 70110-01-98-00 & 52737.253 & \texttt{pow$\times$(bb+ga)} & 1.66(4) & - & - & 0.11(1) & 1.0(2) & 0.0024(6) & 0.75(8) & 42(34) & 0.86(6) & 72 & 0.88 & -9.54(8) & 11(2) \\
D03 & 80116-02-01-01G & 52737.502 & \texttt{pow$\times$(bb+ga)} & 1.63(2) & - & - & 0.100(5) & 0.9(1) & 0.0020(2) & 0.97(5) & 11(3) & 0.91(3) & 72 & 0.84 & -9.71(4) & 8.2(7) \\
D03 & 80116-02-01-02 & 52739.606 & \texttt{bkn$\times$(bb+ga)} & 1.57(3) & 35(9) & 2.0(3) & 0.079(6) & 1.0(1) & 0.0021(3) & 0.96(5) & 12(4) & 0.97(4) & 70 & 1.09 & -9.65(4) & 10.6(8) \\
D03 & 80116-02-02-00 & 52741.749 & \texttt{bkn$\times$(bb+ga)} & 1.60(2) & 34(4) & 2.1(2) & 0.082(5) & 0.8(1) & 0.0013(2) & 0.93(7) & 9(4) & 0.96(3) & 70 & 0.89 & -9.84(5) & 7.6(7) \\
D03 & 80116-02-02-01 & 52742.051 & \texttt{pow$\times$(bb+ga)} & 1.51(2) & - & - & 0.064(4) & 1.0(1) & 0.0014(2) & 1.06(9) & 6(2) & - & 43 & 0.66 & -9.80(4) & 7.5(7) \\
D03 & 70128-02-03-00 & 52742.346 & \texttt{pow$\times$(bb+ga)} & 1.60(2) & - & - & 0.078(3) & 0.7(1) & 0.0010(1) & 0.96(6) & 7(2) & - & 43 & 0.91 & -9.92(4) & 6.1(6) \\
D03 & 70128-02-03-01 & 52742.699 & \texttt{bkn$\times$(bb+ga)} & 1.57(3) & 52(17) & 2.2(8) & 0.069(6) & 1.2(1) & 0.0015(2) & 0.95(6) & 8(3) & 0.92(2) & 70 & 0.99 & -10.04(6) & 4.9(5) \\
D03 & 70110-01-02-10 & 52743.163 & \texttt{pow$\times$bb} & 1.59(9) & - & - & 0.07(2) & - & - & 1.8(2) & 0.5(3) & - & 45 & 1.09 & -8.0(2) & 6(2) \\
D03 & 80116-02-02-02G & 52743.305 & \texttt{pow$\times$(bb+ga)} & 1.60(2) & - & - & 0.075(4) & 0.8(1) & 0.0009(2) & 0.92(8) & 7(4) & 0.94(3) & 72 & 0.87 & -9.96(6) & 5.8(8) \\
D03 & 60705-01-61-00 & 52746.983 & \texttt{bkn$\times$(bb+ga)} & 1.42(8) & 17(2) & 1.8(1) & 0.041(8) & 1.0(2) & 0.0008(3) & 1.6(2) & 0.6(4) & 0.94(7) & 69 & 0.96 & -10.1(2) & 6(2) \\
D03 & 60705-01-62-00 & 52750.384 & \texttt{pow$\times$bb} & 1.37(6) & - & - & 0.026(4) & - & - & 2.3(1) & 0.14(4) & - & 44 & 1.05 & -10.1(1) & 6(1) \\
D03 & 70110-01-04-10 & 52751.569 & \texttt{pow$\times$bb} & 1.53(9) & - & - & 0.029(7) & - & - & 1.8(2) & 0.2(1) & - & 45 & 0.73 & -10.4(3) & 5(2) \\
D03 & 60705-01-63-00 & 52756.060 & \texttt{pow$\times$bb} & 1.51(5) & - & - & 0.022(3) & - & - & 2.7(7) & 0.02(3) & 0.79(8) & 73 & 1.19 & -10.6(2) & 3(1) \\
\hline
R04 & 90418-01-01-04 & 53196.146 & \texttt{pow$\times$(bb+ga)} & 1.52(4) & - & - & 0.19(2) & 0.0(2) & 0.0017(3) & 1.3(2) & 4(3) & 0.91(5) & 72 & 1.03 & -9.62(3) & 4.2(3) \\
R04 & 90418-01-01-00 & 53196.212 & \texttt{pow$\times$bb} & 1.38(7) & - & - & 0.12(3) & - & - & 1.90(8) & 1.9(3) & - & 45 & 1.21 & -9.29(8) & 9(1) \\
R04 & 90418-01-01-01 & 53197.113 & \texttt{bkn$\times$(bb+ga)} & 1.49(2) & 44(7) & 2.1(2) & 0.182(9) & 1.0(1) & 0.0045(6) & 1.00(5) & 20(5) & 0.90(2) & 70 & 1.51 & -9.36(3) & 7.4(5) \\
R04 & 90418-01-01-02 & 53198.162 & \texttt{pow$\times$(bb+ga)} & 1.54(1) & - & - & 0.209(7) & 0.8(1) & 0.0033(4) & 1.02(6) & 16(5) & 0.91(2) & 72 & 1.53 & -9.45(3) & 5.7(4) \\
R04 & 60705-01-65-00 & 53198.295 & \texttt{bkn$\times$(bb+ga)} & 1.48(2) & 32(7) & 1.8(1) & 0.187(8) & 0.9(1) & 0.0039(7) & 1.1(1) & 11(5) & 0.90(3) & 70 & 0.71 & -9.43(4) & 6.1(5) \\
R04 & 90418-01-01-03 & 53199.146 & \texttt{bkn$\times$(bb+ga)} & 1.48(1) & 36(3) & 1.92(9) & 0.192(6) & 0.8(1) & 0.0034(5) & 1.21(9) & 9(3) & 0.91(1) & 70 & 0.55 & -9.40(2) & 6.5(2) \\
R04 & 60705-01-65-01 & 53201.330 & \texttt{bkn$\times$(bb+ga)} & 1.52(3) & 39(5) & 2.2(2) & 0.22(2) & 1.0(2) & 0.004(1) & 1.1(1) & 15(8) & 0.94(3) & 70 & 0.65 & -9.36(6) & 6.8(8) \\
R04 & 60705-01-66-00 & 53204.415 & \texttt{pow$\times$(bb+ga)} & 1.57(2) & - & - & 0.28(1) & 0.9(1) & 0.0047(8) & 0.91(9) & 28(16) & 0.92(3) & 72 & 1.05 & -9.50(9) & 4.2(7) \\
R04 & 60705-01-67-01 & 53215.299 & \texttt{bkn$\times$(bb+ga)} & 1.55(1) & 39(6) & 1.9(1) & 0.31(1) & 0.8(1) & 0.005(1) & 1.1(1) & 13(7) & 0.93(2) & 70 & 0.89 & -9.32(4) & 5.7(5) \\
R04 & 60705-01-68-00 & 53218.113 & \texttt{bkn$\times$(bb+ga)} & 1.54(1) & 32(5) & 1.9(1) & 0.32(1) & 0.99(9) & 0.0072(8) & 1.04(6) & 25(7) & 0.98(2) & 70 & 0.91 & -9.20(3) & 7.1(4) \\
R04 & 60705-01-68-01 & 53222.252 & \texttt{bkn$\times$(bb+ga)} & 1.58(1) & 38(2) & 2.4(1) & 0.41(1) & 1.03(7) & 0.010(1) & 1.12(7) & 28(8) & 0.93(2) & 70 & 1.07 & -9.02(2) & 9.6(4) \\
R04 & 60705-01-69-00 & 53225.412 & \texttt{bkn$\times$(bb+ga)} & 1.63(1) & 33(3) & 2.07(9) & 0.48(2) & 1.05(6) & 0.012(1) & 1.09(5) & 33(7) & 0.91(2) & 70 & 1.11 & -9.01(2) & 9.1(4) \\
R04 & 90704-01-01-00 & 53226.452 & \texttt{bkn$\times$(bb+ga)} & 1.64(1) & 34(2) & 2.22(7) & 0.50(1) & 1.08(5) & 0.0128(8) & 1.07(4) & 39(6) & 0.90(1) & 70 & 1.44 & -8.96(2) & 10.2(4) \\
R04 & 60705-01-69-01 & 53228.997 & \texttt{bkn$\times$(bb+ga)} & 1.78(2) & 28(3) & 2.2(1) & 0.62(4) & 1.18(7) & 0.016(1) & 0.96(4) & 89(17) & 0.96(3) & 70 & 1.18 & -8.79(3) & 15.0(8) \\
R04 & 60705-01-70-00 & 53230.964 & \texttt{pow$\times$(bb+ga)} & 2.12(3) & - & - & 0.93(8) & 1.03(6) & 0.018(1) & 0.78(1) & 694(62) & 0.98(4) & 72 & 1.70 & -8.27(1) & 36.6(8) \\
R04 & 90110-02-01-03 & 53233.000 & \texttt{pow$\times$(bb+ga)} & 2.37(2) & - & - & 1.10(6) & 0.87(4) & 0.0123(6) & 0.770(6) & 1509(73) & 1.19(6) & 72 & 1.27 & -7.948(8) & 52.0(2) \\
R04 & 60705-01-70-01 & 53234.587 & \texttt{pow$\times$(bb+ga)} & 2.42(4) & - & - & 0.81(9) & 0.88(4) & 0.0111(6) & 0.726(4) & 2243(81) & 1.1(1) & 72 & 1.76 & -7.880(6) & 62.66(8) \\
R04 & 90704-01-03-00 & 53235.445 & \texttt{pow$\times$(bb+ga)} & 2.33(2) & - & - & 0.63(4) & 1.00(3) & 0.0110(4) & 0.752(4) & 1863(71) & 1.04(5) & 72 & 1.81 & -7.900(6) & 67.7(3) \\
R04 & 90704-02-01-01 & 53237.547 & \texttt{pow$\times$(bb+ga)} & 2.21(3) & - & - & 0.30(2) & 1.01(4) & 0.0072(3) & 0.699(4) & 2629(93) & 1.06(8) & 72 & 1.67 & -7.878(6) & 82.0(4) \\
R04 & 90704-02-01-03 & 53237.682 & \texttt{pow$\times$(bb+ga)} & 2.31(5) & - & - & 0.43(6) & 0.87(5) & 0.0079(5) & 0.676(4) & 3229(126) & 0.94(9) & 72 & 1.49 & -7.847(7) & 77.82(5) \\
R04 & 60705-01-71-00 & 53238.058 & \texttt{pow$\times$(bb+ga)} & 2.30(6) & - & - & 0.39(6) & 0.91(5) & 0.0076(5) & 0.683(4) & 3078(108) & - & 43 & 1.82 & -7.848(6) & 79.26(2) \\
\hline
R07 & 92052-07-06-01 & 54118.975 & \texttt{bkn$\times$(bb+ga)} & 1.47(1) & 37(2) & 2.12(6) & 0.52(2) & 0.9(2) & 0.011(3) & 1.2(1) & 28(12) & 0.93(1) & 70 & 1.32 & -8.91(3) & 7.3(4) \\
R07 & 92428-01-02-00 & 54122.304 & \texttt{bkn$\times$(bb+ga)} & 1.49(1) & 35.5(9) & 2.21(3) & 0.64(2) & 0.8(1) & 0.013(2) & 1.18(9) & 33(12) & 0.918(7) & 70 & 1.69 & -8.87(2) & 7.2(3) \\
R07 & 92428-01-03-00 & 54127.115 & \texttt{bkn$\times$(bb+ga)} & 1.53(2) & 37(2) & 2.41(9) & 0.84(4) & 0.7(1) & 0.015(3) & 1.2(1) & 36(18) & 0.93(1) & 70 & 0.82 & -8.79(3) & 7.3(4) \\
R07 & 92035-01-01-01 & 54128.976 & \texttt{bkn$\times$(bb+ga)} & 1.556(8) & 34.2(9) & 2.27(4) & 0.97(2) & 0.95(6) & 0.023(2) & 0.99(6) & 93(29) & 0.936(7) & 70 & 1.20 & -8.72(3) & 7.7(6) \\
R07 & 92035-01-01-00 & 54129.476 & \texttt{bkn$\times$(bb+ga)} & 1.558(9) & 37(2) & 2.33(7) & 0.99(2) & 1.02(6) & 0.024(2) & 1.04(6) & 70(21) & 0.92(1) & 70 & 1.00 & -8.76(3) & 6.9(4) \\
R07 & 92035-01-01-03 & 54130.140 & \texttt{bkn$\times$(bb+ga)} & 1.531(8) & 32(1) & 2.26(4) & 0.96(2) & 0.97(5) & 0.029(2) & 0.87(3) & 249(56) & 0.921(9) & 70 & 1.45 & -8.52(3) & 11.4(7) \\
R07 & 92035-01-01-02 & 54131.127 & \texttt{bkn$\times$(bb+ga)} & 1.572(8) & 33.3(9) & 2.35(3) & 1.14(2) & 1.03(6) & 0.027(2) & 1.03(6) & 87(27) & 0.923(7) & 70 & 1.89 & -8.68(3) & 7.6(5) \\
R07 & 92035-01-01-04 & 54132.109 & \texttt{bkn$\times$(bb+ga)} & 1.577(8) & 32.3(7) & 2.35(3) & 1.19(3) & 0.94(7) & 0.026(2) & 1.08(7) & 71(24) & 0.937(7) & 70 & 1.43 & -8.69(3) & 7.4(5) \\
R07 & 92035-01-02-00 & 54133.035 & \texttt{bkn$\times$(bb+ga)} & 1.600(7) & 33.7(7) & 2.41(3) & 1.30(3) & 0.90(6) & 0.027(2) & 1.03(7) & 87(30) & 0.924(6) & 70 & 1.55 & -8.68(3) & 7.2(5) \\
R07 & 92035-01-02-01 & 54133.953 & \texttt{bkn$\times$(bb+ga)} & 1.594(8) & 33.3(7) & 2.42(3) & 1.29(3) & 0.98(6) & 0.029(3) & 1.06(7) & 80(27) & 0.916(7) & 70 & 1.41 & -8.66(3) & 7.6(5) \\
R07 & 92035-01-02-02 & 54135.054 & \texttt{bkn$\times$(bb+ga)} & 1.612(8) & 32.9(8) & 2.41(3) & 1.36(3) & 0.94(7) & 0.027(3) & 1.12(8) & 60(21) & 0.900(6) & 70 & 1.41 & -8.70(3) & 6.8(4) \\
R07 & 92035-01-02-04 & 54137.018 & \texttt{bkn$\times$(bb+ga)} & 1.640(7) & 32.6(7) & 2.46(3) & 1.60(3) & 1.03(5) & 0.036(2) & 1.05(6) & 84(25) & 0.905(6) & 70 & 1.19 & -8.66(3) & 6.8(4) \\
R07 & 92035-01-02-08 & 54137.857 & \texttt{bkn$\times$(bb+ga)} & 1.68(1) & 34(1) & 2.56(7) & 1.76(6) & 0.96(7) & 0.033(3) & 1.11(9) & 69(29) & 0.92(1) & 70 & 1.48 & -8.65(3) & 8.2(6) \\
R07 & 92035-01-02-07 & 54138.846 & \texttt{bkn$\times$(bb+ga)} & 1.702(6) & 30.4(8) & 2.49(4) & 1.82(3) & 0.96(4) & 0.039(2) & 0.85(3) & 325(80) & 0.935(7) & 70 & 1.68 & -8.44(3) & 11.0(8) \\
R07 & 92035-01-02-06 & 54139.963 & \texttt{bkn$\times$(bb+ga)} & 1.827(7) & 30.8(7) & 2.64(4) & 2.39(5) & 1.06(4) & 0.040(2) & 1.07(5) & 92(24) & 0.912(7) & 70 & 1.28 & -8.59(2) & 9.7(5) \\
R07 & 92035-01-03-00 & 54140.225 & \texttt{bkn$\times$(bb+ga)} & 1.871(8) & 30.0(9) & 2.59(4) & 2.59(6) & 0.99(4) & 0.038(2) & 1.02(5) & 130(36) & 0.935(8) & 70 & 1.49 & -8.52(3) & 11.6(7) \\
R07 & 92035-01-03-01 & 54141.075 & \texttt{bkn$\times$(bb+ga)} & 1.979(8) & 31(1) & 2.74(5) & 3.14(7) & 0.97(4) & 0.039(2) & 0.97(4) & 211(49) & 0.937(8) & 70 & 1.38 & -8.41(3) & 15.4(8) \\
R07 & 92035-01-03-02 & 54142.056 & \texttt{bkn$\times$(bb+ga)} & 2.088(7) & 31(1) & 2.82(6) & 3.62(7) & 0.98(3) & 0.046(2) & 0.81(2) & 942(129) & 0.944(9) & 70 & 1.75 & -8.07(2) & 19.2(8) \\
R07 & 92035-01-03-03 & 54143.039 & \texttt{bkn$\times$(bb+ga)} & 2.26(1) & 26(5) & 2.7(1) & 5.1(1) & 1.00(4) & 0.046(2) & 0.84(2) & 1114(119) & 0.98(3) & 70 & 1.82 & -7.92(2) & 20.3(5) \\
R07 & 92428-01-04-00 & 54143.890 & \texttt{bkn$\times$(bb+ga)} & 2.33(1) & 34(2) & 2.9(1) & 5.4(1) & 1.02(4) & 0.044(2) & 0.85(1) & 1391(119) & 0.97(1) & 70 & 1.71 & -7.81(1) & 23.5(4) \\
R07 & 92428-01-04-02 & 54144.098 & \texttt{bkn$\times$(bb+ga)} & 2.32(1) & 33(3) & 3.1(2) & 5.5(2) & 1.01(5) & 0.044(3) & 0.87(1) & 1201(116) & 0.99(2) & 70 & 1.54 & -7.84(1) & 22.2(4) \\
R07 & 92428-01-04-03 & 54144.883 & \texttt{bkn$\times$(bb+ga)} & 2.36(2) & 22(8) & 2.8(2) & 5.5(3) & 1.10(5) & 0.047(3) & 0.86(1) & 1645(129) & 1.04(5) & 70 & 1.72 & -7.73(1) & 27.0(3) \\
R07 & 92428-01-04-04 & 54145.974 & \texttt{bkn$\times$(bb+ga)} & 2.41(1) & 41(6) & 3.3(5) & 5.8(2) & 1.01(5) & 0.039(2) & 0.86(1) & 1770(125) & 0.94(2) & 70 & 1.51 & -7.68(1) & 27.3(4) \\
R07 & 92035-01-04-00 & 54147.031 & \texttt{bkn$\times$(bb+ga)} & 2.51(1) & 29(3) & 2.9(1) & 5.4(2) & 1.03(4) & 0.036(2) & 0.876(7) & 2628(123) & 1.00(2) & 70 & 1.48 & -7.484(7) & 38.0(3) \\
R07 & 92035-01-04-01 & 54148.153 & \texttt{pow$\times$(bb+ga)} & 2.47(3) & - & - & 2.1(1) & 0.84(5) & 0.019(1) & 0.826(5) & 4196(154) & 1.09(4) & 72 & 1.90 & -7.381(6) & 66.5(5) \\
R07 & 92085-01-01-00 & 54150.761 & \texttt{pow$\times$(bb+ga)} & 2.36(2) & - & - & 1.29(6) & 0.94(4) & 0.018(1) & 0.842(4) & 3936(115) & 0.99(3) & 72 & 1.49 & -7.377(4) & 77.3(2) \\
R07 & 92085-01-01-04 & 54151.816 & \texttt{pow$\times$(bb+ga)} & 2.38(2) & - & - & 1.86(9) & 1.03(4) & 0.024(1) & 0.838(4) & 3806(123) & 0.97(3) & 72 & 1.74 & -7.400(5) & 69.2(3) \\
R07 & 92085-01-01-05 & 54152.855 & \texttt{pow$\times$(bb+ga)} & 2.37(2) & - & - & 1.04(7) & 0.91(4) & 0.016(1) & 0.823(4) & 4454(137) & 1.01(4) & 72 & 1.33 & -7.362(5) & 81.5(2) \\
\hline
D07 & 92704-03-06-02 & 54223.227 & \texttt{pow$\times$(bb+ga)} & 2.34(7) & - & - & 0.30(5) & 0.71(6) & 0.0048(4) & 0.642(7) & 1939(136) & 1.3(2) & 72 & 1.27 & -8.16(1) & 71.2(5) \\
D07 & 92704-03-06-03 & 54223.293 & \texttt{pow$\times$(bb+ga)} & 2.40(9) & - & - & 0.35(9) & 0.73(8) & 0.0044(6) & 0.62(1) & 2273(248) & 1.4(3) & 72 & 1.14 & -8.14(2) & 68.3(6) \\
D07 & 92704-03-07-00 & 54224.124 & \texttt{pow$\times$bb} & 3.06(7) & - & - & 1.6(3) & - & - & 0.53(2) & 4845(1518) & 1.9(4) & 74 & 1.74 & -8.10(5) & 22(2) \\
D07 & 92704-03-07-01 & 54225.253 & \texttt{pow$\times$(bb+ga)} & 2.34(7) & - & - & 0.28(5) & 0.75(7) & 0.0038(4) & 0.633(9) & 1802(157) & - & 43 & 1.64 & -8.22(1) & 69.4(7) \\
D07 & 92704-03-08-00 & 54226.106 & \texttt{pow$\times$(bb+ga)} & 2.33(4) & - & - & 0.34(4) & 0.82(4) & 0.0054(3) & 0.668(6) & 1431(81) & 1.0(1) & 72 & 1.82 & -8.220(9) & 65.4(5) \\
D07 & 92704-03-08-01 & 54227.624 & \texttt{pow$\times$(bb+ga)} & 2.08(7) & - & - & 0.13(3) & 0.86(6) & 0.0042(4) & 0.633(7) & 1843(137) & 1.1(2) & 72 & 1.85 & -8.21(1) & 82(1) \\
D07 & 92704-03-09-00 & 54228.399 & \texttt{pow$\times$(bb+ga)} & 2.24(6) & - & - & 0.22(4) & 0.76(6) & 0.0038(3) & 0.637(8) & 1575(132) & 1.0(1) & 72 & 1.92 & -8.26(1) & 73(1) \\
D07 & 92704-03-10-00 & 54231.616 & \texttt{pow$\times$(bb+ga)} & 2.22(6) & - & - & 0.24(4) & 0.83(5) & 0.0052(4) & 0.665(8) & 1256(97) & 1.0(1) & 72 & 1.60 & -8.29(1) & 70(1) \\
D07 & 92704-03-10-11 & 54232.682 & \texttt{pow$\times$(bb+ga)} & 2.16(4) & - & - & 0.22(3) & 0.93(4) & 0.0060(3) & 0.672(7) & 1153(72) & 1.1(1) & 72 & 1.94 & -8.30(1) & 69.3(9) \\
D07 & 92704-03-11-00 & 54234.858 & \texttt{pow$\times$(bb+ga)} & 1.98(3) & - & - & 0.26(2) & 0.82(6) & 0.0056(4) & 0.77(2) & 214(32) & 0.90(7) & 72 & 1.41 & -8.80(2) & 35(2) \\
D07 & 92704-03-11-01 & 54235.797 & \texttt{pow$\times$(bb+ga)} & 1.85(3) & - & - & 0.20(2) & 0.96(7) & 0.0058(5) & 0.80(2) & 135(23) & 0.91(6) & 72 & 1.67 & -8.92(3) & 30(2) \\
D07 & 92704-04-01-00 & 54236.383 & \texttt{pow$\times$(bb+ga)} & 1.90(3) & - & - & 0.25(2) & 0.76(7) & 0.0049(5) & 0.84(4) & 81(21) & 0.94(5) & 72 & 1.12 & -9.06(4) & 22(2) \\
D07 & 92704-04-01-01 & 54236.455 & \texttt{pow$\times$(bb+ga)} & 1.80(3) & - & - & 0.19(1) & 0.90(6) & 0.0055(4) & 0.82(2) & 113(18) & 0.94(5) & 72 & 1.30 & -8.97(2) & 27(1) \\
D07 & 92704-04-01-02 & 54236.523 & \texttt{pow$\times$(bb+ga)} & 1.80(3) & - & - & 0.20(2) & 0.86(8) & 0.0046(5) & 0.96(4) & 43(10) & 0.89(5) & 72 & 1.12 & -9.11(3) & 21(1) \\
D07 & 92704-03-12-00 & 54236.597 & \texttt{pow$\times$(bb+ga)} & 1.83(5) & - & - & 0.21(3) & 0.9(2) & 0.0039(9) & 0.97(7) & 36(13) & 0.92(7) & 72 & 1.41 & -9.28(4) & 14(1) \\
D07 & 92704-04-01-03 & 54237.298 & \texttt{pow$\times$(bb+ga)} & 1.78(4) & - & - & 0.18(2) & 1.0(1) & 0.0052(7) & 0.89(4) & 64(15) & - & 43 & 0.81 & -9.06(3) & 23(2) \\
D07 & 92704-04-01-04 & 54237.365 & \texttt{pow$\times$(bb+ga)} & 1.78(3) & - & - & 0.18(2) & 0.96(7) & 0.0054(5) & 0.90(4) & 62(14) & - & 43 & 1.14 & -9.07(3) & 22(2) \\
D07 & 92704-04-01-05 & 54237.433 & \texttt{pow$\times$(bb+ga)} & 1.74(3) & - & - & 0.16(1) & 1.00(7) & 0.0053(5) & 0.94(3) & 51(10) & - & 43 & 1.21 & -9.07(3) & 23(1) \\
D07 & 92704-03-12-01 & 54237.501 & \texttt{pow$\times$(bb+ga)} & 1.78(3) & - & - & 0.18(1) & 0.93(7) & 0.0050(5) & 0.89(3) & 62(13) & 0.89(4) & 72 & 1.67 & -9.08(3) & 22(1) \\
D07 & 92704-04-01-06 & 54237.704 & \texttt{pow$\times$(bb+ga)} & 1.80(2) & - & - & 0.20(1) & 0.87(9) & 0.0039(4) & 0.96(4) & 36(8) & - & 43 & 1.65 & -9.18(3) & 18(1) \\
D07 & 92704-03-13-00 & 54238.750 & \texttt{pow$\times$(bb+ga)} & 1.66(4) & - & - & 0.15(2) & 1.1(1) & 0.0039(7) & 0.95(6) & 31(12) & 0.95(5) & 72 & 1.19 & -9.42(5) & 11(1) \\
D07 & 92704-03-13-01 & 54239.743 & \texttt{bkn$\times$(bb+ga)} & 1.57(3) & 39(5) & 2.8(6) & 0.12(1) & 0.8(2) & 0.0025(6) & 1.10(9) & 13(5) & 0.93(4) & 70 & 0.98 & -9.38(4) & 13.4(9) \\
D07 & 92704-03-13-02 & 54240.308 & \texttt{pow$\times$(bb+ga)} & 1.63(3) & - & - & 0.14(1) & 0.6(1) & 0.0024(4) & 0.95(8) & 21(11) & 0.88(4) & 72 & 0.79 & -9.44(5) & 10(1) \\
D07 & 92704-03-14-00 & 54241.570 & \texttt{pow$\times$bb} & 1.49(5) & - & - & 0.09(1) & - & - & 1.58(7) & 2.4(4) & 0.91(5) & 74 & 1.13 & -9.51(6) & 10(1) \\
D07 & 92704-03-14-01 & 54242.884 & \texttt{pow$\times$(bb+ga)} & 1.52(2) & - & - & 0.094(5) & 0.7(1) & 0.0018(3) & 1.05(8) & 10(4) & 0.82(4) & 72 & 0.83 & -9.58(4) & 8.5(7) \\
D07 & 92704-03-14-02 & 54243.864 & \texttt{pow$\times$bb} & 1.48(5) & - & - & 0.08(1) & - & - & 1.53(8) & 2.0(3) & - & 45 & 1.07 & -9.64(6) & 8(1) \\
D07 & 92704-03-14-03 & 54244.979 & \texttt{pow$\times$(bb+ga)} & 1.52(4) & - & - & 0.083(4) & 0.0(2) & 0.0008(1) & 1.2(1) & 5(2) & - & 43 & 0.89 & -9.72(5) & 7.2(8) \\
D07 & 92704-04-02-00 & 54245.794 & \texttt{pow$\times$(bb+ga)} & 1.48(2) & - & - & 0.070(3) & 0.8(1) & 0.0015(2) & 0.84(4) & 25(7) & - & 43 & 1.09 & -9.57(4) & 10.6(9) \\
D07 & 92704-04-02-02 & 54245.892 & \texttt{pow$\times$(bb+ga)} & 1.51(3) & - & - & 0.078(6) & 0.8(3) & 0.0012(4) & 1.0(1) & 11(7) & 0.89(4) & 72 & 0.95 & -9.80(5) & 6.2(6) \\
D07 & 92704-03-15-00 & 54246.873 & \texttt{pow$\times$(bb+ga)} & 1.52(4) & - & - & 0.074(8) & 0.0(4) & 0.0005(1) & 1.1(2) & 3(3) & - & 43 & 0.86 & -9.90(7) & 5.5(7) \\
D07 & 92704-03-16-00 & 54247.923 & \texttt{bkn$\times$bb} & 1.49(6) & 29(11) & 1.8(3) & 0.06(1) & - & - & 1.4(2) & 1.4(5) & 1.01(9) & 72 & 1.18 & -9.9(1) & 6(1) \\
D07 & 92704-03-17-00 & 54248.836 & \texttt{bkn$\times$(bb+ga)} & 1.46(4) & 29(12) & 1.8(3) & 0.055(6) & 0.9(2) & 0.0011(3) & 1.0(1) & 8(6) & 0.91(8) & 70 & 0.71 & -9.81(7) & 8(1) \\
D07 & 92704-03-18-00 & 54249.556 & \texttt{pow$\times$bb} & 1.48(5) & - & - & 0.053(7) & - & - & 1.5(1) & 1.1(4) & 0.91(6) & 74 & 1.23 & -8.0(1) & 5(1) \\
D07 & 92704-03-19-00 & 54250.731 & \texttt{pow$\times$(bb+ga)} & 1.45(4) & - & - & 0.044(5) & 0.2(2) & 0.0004(1) & 1.2(1) & 2(2) & 0.72(5) & 72 & 0.83 & -9.99(6) & 6.0(7) \\
D07 & 92704-04-02-01 & 54251.708 & \texttt{pow$\times$(bb+ga)} & 1.47(3) & - & - & 0.042(3) & 1.1(2) & 0.0009(2) & 0.96(9) & 6(3) & 0.87(4) & 72 & 0.82 & -9.99(7) & 6.6(9) \\
D07 & 92704-04-02-03 & 54251.780 & \texttt{pow$\times$(bb+ga)} & 1.50(3) & - & - & 0.044(4) & 0.8(2) & 0.0006(2) & 1.1(1) & 3(2) & - & 43 & 0.69 & -10.09(7) & 5.5(8) \\
D07 & 92704-04-02-04 & 54251.848 & \texttt{pow$\times$bb} & 1.50(5) & - & - & 0.044(6) & - & - & 1.6(1) & 0.5(2) & 0.99(7) & 74 & 1.18 & -10.2(1) & 5(1) \\
D07 & 92704-03-22-00 & 54254.925 & \texttt{pow$\times$bb} & 1.57(6) & - & - & 0.043(7) & - & - & 1.2(3) & 1(2) & - & 45 & 1.19 & -10.3(2) & 4(1) \\
D07 & 92704-03-24-00 & 54256.302 & \texttt{pow$\times$bb} & 1.51(8) & - & - & 0.036(7) & - & - & 1.4(2) & 0.6(4) & - & 45 & 1.05 & -10.3(2) & 5(2) \\
D07 & 92704-04-03-00 & 54256.612 & \texttt{pow$\times$bb} & 1.48(5) & - & - & 0.031(4) & - & - & 1.7(1) & 0.38(8) & 0.89(6) & 74 & 1.06 & -10.2(1) & 6(1) \\
\hline
R10 & 95409-01-12-00 & 55281.592 & \texttt{pow$\times$(bb+ga)} & 1.53(1) & - & - & 0.75(2) & 0.87(7) & 0.019(1) & 0.85(4) & 204(50) & - & 43 & 1.57 & -8.64(3) & 9.8(7) \\
R10 & 95409-01-12-04 & 55286.738 & \texttt{pow$\times$(bb+ga)} & 1.59(1) & - & - & 1.00(3) & 0.80(6) & 0.021(2) & 0.78(4) & 349(104) & - & 43 & 1.85 & -8.56(4) & 10.1(9) \\
R10 & 95409-01-13-03 & 55288.380 & \texttt{pow$\times$(bb+ga)} & 1.575(9) & - & - & 0.97(2) & 0.96(5) & 0.026(2) & 0.80(3) & 367(82) & - & 43 & 1.98 & -8.51(3) & 11.2(8) \\
R10 & 95409-01-13-00 & 55289.628 & \texttt{pow$\times$(bb+ga)} & 1.587(9) & - & - & 1.02(2) & 0.95(5) & 0.026(2) & 0.81(3) & 340(78) & - & 43 & 1.87 & -8.51(3) & 10.9(8) \\
R10 & 95409-01-13-05 & 55292.796 & \texttt{pow$\times$(bb+ga)} & 1.61(1) & - & - & 1.18(4) & 0.93(7) & 0.031(2) & 0.82(4) & 346(100) & - & 43 & 1.49 & -8.49(4) & 10.6(9) \\
R10 & 95409-01-13-01 & 55293.095 & \texttt{pow$\times$(bb+ga)} & 1.625(9) & - & - & 1.25(3) & 0.80(6) & 0.025(2) & 0.88(4) & 212(63) & - & 43 & 1.80 & -8.57(4) & 8.8(7) \\
R10 & 95409-01-13-06 & 55294.129 & \texttt{pow$\times$(bb+ga)} & 1.624(9) & - & - & 1.24(3) & 0.97(5) & 0.032(2) & 0.85(3) & 264(63) & - & 43 & 1.33 & -8.54(3) & 9.4(7) \\
R10 & 95409-01-14-00 & 55295.008 & \texttt{pow$\times$(bb+ga)} & 1.66(1) & - & - & 1.32(4) & 0.82(6) & 0.026(2) & 0.80(4) & 352(106) & - & 43 & 1.54 & -8.51(4) & 10.2(9) \\
R10 & 95409-01-14-02 & 55297.883 & \texttt{pow$\times$(bb+ga)} & 1.852(8) & - & - & 2.03(4) & 0.96(4) & 0.037(2) & 0.84(3) & 391(73) & - & 43 & 1.89 & -8.39(2) & 12.5(7) \\
R10 & 95409-01-14-06 & 55299.787 & \texttt{pow$\times$(bb+ga)} & 2.07(1) & - & - & 3.01(8) & 0.95(4) & 0.039(2) & 0.77(2) & 1045(153) & - & 43 & 1.85 & -8.11(2) & 19.7(9) \\
R10 & 95409-01-18-04 & 55327.044 & \texttt{pow$\times$(bb+ga)} & 2.43(6) & - & - & 2.0(3) & 0.96(7) & 0.028(2) & 0.768(8) & 3461(198) & - & 43 & 1.61 & -7.59(1) & 57.3(2) \\
\hline
D11 & 96409-01-02-02 & 55573.469 & \texttt{pow$\times$(bb+ga)} & 2.34(8) & - & - & 0.35(8) & 0.76(6) & 0.0057(6) & 0.61(1) & 2662(271) & - & 43 & 1.96 & -8.10(2) & 71(1) \\
D11 & 96409-01-04-02 & 55586.502 & \texttt{pow$\times$(bb+ga)} & 2.3(1) & - & - & 0.31(8) & 0.8(1) & 0.0044(6) & 0.62(2) & 1765(285) & - & 43 & 1.86 & -8.27(2) & 65(1) \\
D11 & 96409-01-04-07 & 55587.503 & \texttt{pow$\times$(bb+ga)} & 2.2(1) & - & - & 0.2(1) & 0.8(2) & 0.0038(9) & 0.60(2) & 1640(422) & - & 43 & 1.08 & -8.34(3) & 68(3) \\
D11 & 96409-01-04-08 & 55588.554 & \texttt{pow$\times$(bb+ga)} & 2.3(2) & - & - & 0.18(8) & 0.8(1) & 0.0031(6) & 0.57(2) & 2307(590) & - & 43 & 1.35 & -8.28(4) & 75(4) \\
D11 & 96409-01-05-04 & 55590.441 & \texttt{pow$\times$(bb+ga)} & 2.4(1) & - & - & 0.3(1) & 0.6(1) & 0.0031(6) & 0.58(3) & 1787(610) & - & 43 & 1.29 & -8.38(4) & 58(3) \\
D11 & 96409-01-05-05 & 55592.739 & \texttt{pow$\times$(bb+ga)} & 2.1(1) & - & - & 0.15(5) & 0.94(9) & 0.0046(6) & 0.63(2) & 1024(196) & - & 43 & 1.93 & -8.47(2) & 68(3) \\
D11 & 96409-01-05-03 & 55594.902 & \texttt{pow$\times$(bb+ga)} & 2.11(6) & - & - & 0.20(3) & 0.75(7) & 0.0040(4) & 0.66(2) & 571(97) & - & 43 & 1.86 & -8.63(2) & 53(2) \\
D11 & 96409-01-06-00 & 55597.264 & \texttt{pow$\times$(bb+ga)} & 1.94(6) & - & - & 0.17(3) & 1.0(1) & 0.0045(6) & 0.76(3) & 163(40) & - & 43 & 1.40 & -8.94(3) & 35(2) \\
D11 & 96409-01-06-01 & 55598.677 & \texttt{pow$\times$(bb+ga)} & 1.90(5) & - & - & 0.14(2) & 0.79(7) & 0.0041(4) & 0.74(2) & 198(34) & - & 43 & 1.24 & -8.89(2) & 41(2) \\
D11 & 96409-01-06-02 & 55601.891 & \texttt{pow$\times$(bb+ga)} & 1.66(4) & - & - & 0.09(1) & 1.05(9) & 0.0041(5) & 0.83(3) & 63(14) & - & 43 & 0.72 & -9.19(3) & 25(2) \\
D11 & 96409-01-07-00 & 55603.992 & \texttt{pow$\times$(bb+ga)} & 1.67(6) & - & - & 0.10(2) & 0.9(2) & 0.0020(6) & 0.8(1) & 28(23) & - & 43 & 0.88 & -9.52(9) & 13(3) \\
D11 & 96409-01-07-03 & 55604.912 & \texttt{pow$\times$(bb+ga)} & 1.63(4) & - & - & 0.088(9) & 0.7(2) & 0.0014(4) & 0.76(7) & 45(31) & - & 43 & 0.93 & -9.50(7) & 14(2) \\
\hline
\captionsetup{width=1.25\textwidth,font={scriptsize}}
\caption{Best-fit parameters and $1\sigma$ standard uncertainty in
  parentheses for all observations where a disc was deemed required.  The
  spectral models employed for the disc and continuum were \texttt{diskbb}
  and \texttt{powerlaw} (or \texttt{bknpower}), respectively.  The Fe
  K$\alpha$ line at 6.4 keV is fit with the \texttt{gauss} model.  The
  entire model is absorbed using \texttt{phabs} with $N_{\rm H} = 5.7
  \times 10^{21}~{\rm cm^{-2}}$ (Miller et al. 2009b) and an
  energy-independent constant is applied to the HEXTE data to account for
  offset relative to the PCA.  A 0.6\% systematic error is added to all PCA
  energy channels.  From {\it left} to {\it right}, the columns give the
  state transition ID, observation ID, MJD for the midpoint of the PCA
  observation, the best-fit \texttt{XSPEC} model, power-law index, break
  energy$^{a}$, high-energy power-law index$^{a}$, power-law normalisation,
  Gaussian line width, Gaussian line normalisation, disc colour temperature
  as returned by \texttt{diskbb}, disc normalisation, constant applied to
  the HEXTE data $^{b}$, number of degrees of freedom, reduced $\chi^{2}$,
  disc flux over the 0.1-10 keV band, and disc fraction.  Dashes indicate
  that a particular parameter was not applicable in the fit.  $^{a}$Applies for fits using the
  \texttt{bknpower} model; $^{b}$Applies for fits using the HEXTE
  data.  The full version of this table for all fits used in this work is available as an electronic supplement.} \label{tab:bestfits_sample} \\
\end{longtable}
\end{landscape}
\end{tiny}

%===========================================================================

\end{document}